%% file: mscthesis.tex
\documentclass[msc,10pt,noupper,appendixpart,oneside]{ubcthesis}
\usepackage{hyperref}

\usepackage{graphicx}

\usepackage{lscape}

\input babarsym.tex

\institution{The University Of British Columbia}
\institutionaddress{Vancouver, Canada}
\department{Physics}
\numberofsignatures{0}      

\previousdegree{B.Sc. (hons), University of Canterbury, 2005}

\title{B Counting at \babar}
\author{Grant D. McGregor}
\copyrightyear{2008}
\submitdate{August 2008}
\renewcommand\thepart         {\Roman{part}}
\renewcommand\thechapter      {\arabic{chapter}}
\renewcommand\thesection      {\thechapter.\arabic{section}}
\renewcommand\thesubsection   {\thesection.\arabic{subsection}}
\renewcommand\thesubsubsection{\thesubsection.\arabic{subsubsection}}
\renewcommand\theparagraph    {\thesubsubsection.\arabic{paragraph}}

\begin{document}
\newcommand{\Matrix}[1]{\boldsymbol{#1}}
\newcommand{\ud} {\mathrm{d}}

\frontmatter

\maketitle
\begin{abstract}
In this thesis we examine the method of counting \BB events produced in the 
\babar\ experiment. The original method was proposed in 2000, but improvements to track reconstruction 
and our understanding of the detector since that date make it appropriate to revisit the \B Counting method.
We propose a new set of cuts designed to minimize the sensitivity to  time-varying backgrounds. We find the
new method counts \BB events with an associated systematic uncertainty of $\pm 0.6\%$.   

\end{abstract}

\tableofcontents
\listoftables
\listoffigures

\acknowledgements

This work was only possible thanks to the guidance of my supervisor, Chris Hearty. Thanks also to the other members of the UBC \babar\ group ---
David Asgeirsson, Bryan Fulsom, Janis McKenna and Tom Mattison --- and to the Trigger/Filter/Luminosity 
Analysis Working Group, especially Al Eisner and Rainer Bartoldus. Finally, thanks to the the Canadian 
Commonwealth Scholarship scheme and the Canadian Bureau of International Education (CBIE) whose funding made 
this work possible. 

\mainmatter

\input{chapter1.tex}
\input{chapter2.tex}

\input{chapter3.tex}

\input{chapter4.tex}

\input{chapter5.tex}
\input{chapter6.tex}

\input{chapter7.tex}

\input{chapter8.tex}

\input{chapter9.tex}

\input{chapter10.tex}

\appendix

\input{appendix.tex}


\end{document}

%% file: chapter1.tex
\chapter{Introduction}

The \babar\ experiment is located at the Stanford Linear Accelerator Center (SLAC). Data from \epem annihilations at a centre-of-mass 
(CM) energy   $\sqrt{s}\approx 10.58$ GeV are taken at the SLAC  PEP-II storage-rings with the \babar\ detector. This chapter provides
a brief introduction to the experiment.

PEP-II is a  `$B$-Factory' --- it is designed to produce a high number of $B$ mesons. This is achieved through the process $\epem \to \FourS \to \BB$. 
Beams of 9.0 GeV electrons and 3.1 GeV  positrons collide with a CM energy of 10.58 GeV, which is the peak of the \FourS resonance. Running at this energy is desirable
because the \FourS meson (a bound \bbbar state) decays more than 96\% of the time to a \BB state. The design luminosity of PEP-II  
($3 \times 10^{33}$ cm$^{2}$s$^{-1}$)   was exceeded in 2001 and by  2006, peak luminosities above  
$12 \times 10^{33}$ cm$^{2}$s$^{-1}$ were recorded.

\begin{figure}[htb] 
  \begin{center}
    \includegraphics[width=4.5in]{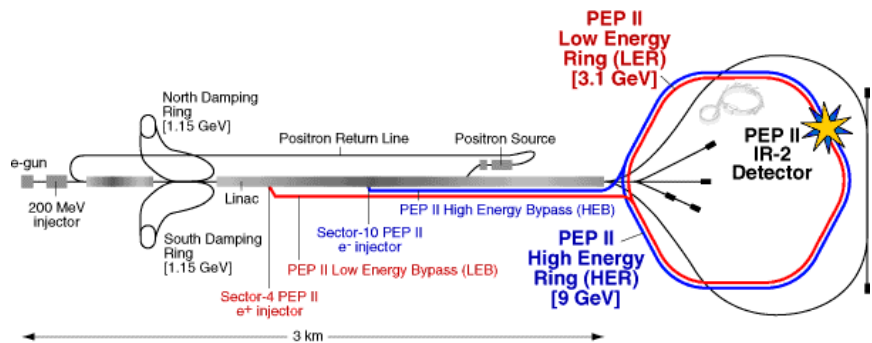}
    \caption{Diagram of PEP-II linac and storage-rings.}\label{fig:pepii}
  \end{center}
\end{figure}

Asymmetric beam energies cause the decay products to have a Lorentz boost of $\beta\gamma = 0.56$ relative to  the laboratory. This is one of the most
important design features of the \babar\ experiment.

 In the rest frame of each \FourS particle, the \B mesons are created almost at rest. By boosting along the $z$ axis, the \B mesons travel a measurable distance
inside the detector before decaying. The mean lifetime of  a $B$ meson is very short (around 1.5 ps), but by reconstructing the decay vertices
of each \B it  possible to calculate how far each travelled and hence the lifetimes. 

For some analyses (such as studies of $CP$ violation in the \BB system) the relative  difference in decay times is especially important. The 
 \babar\ detector is specifically designed to enable precise measurements of this quantity. 

PEP-II is run at the \FourS resonance most of the time, but it is important to occasionally run `off-peak'. Typically just over 10\% of the time, 
PEP-II is run around 40 MeV below the resonance peak.  This is achieved by lowering the energy of the electron beam, which reduces the boost by less than
one percent.

\begin{figure}[htbp] 
  \begin{center}\label{fig:upsilon_spectrum}
    \includegraphics[width=4.0in]{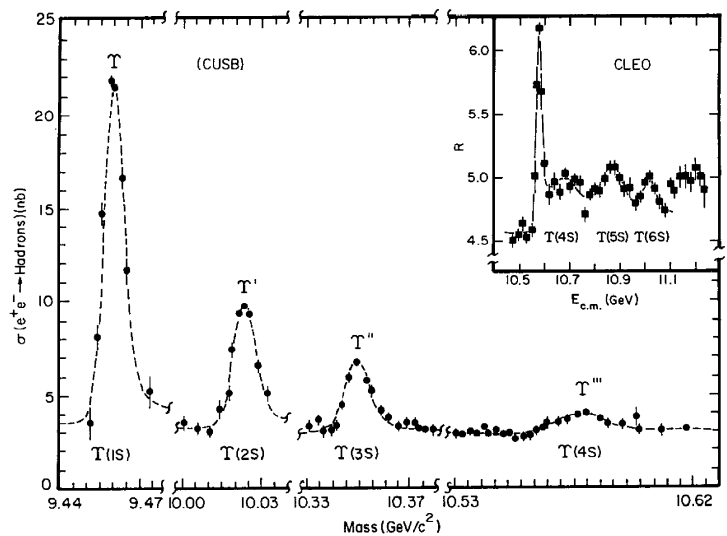}
    \caption[The spectrum of hadron production near 10.58 GeV.]{The spectrum of hadron production near PEP-II's operational CM energy. 
    The curve shows the cross-section for inclusive production of hadrons (vertical axis) as a function of CM energy. The peak at the \FourS resonance
    is clearly visible. The plot
    is originally from \emph{Upsilon Spectroscopy} by Besson and Skwarnicki \cite{Besson:1993mm}.}
  \end{center}
\end{figure}

\emph{Continuum} is the name used to describe all  non-$\BB$ events produced in the detector. The continuum contains many types of events, by far the
 most common of which is $\epem \to \epem$ (Bhabha events) which have a cross-section close to 40 nb. For each type of quark $q$ (apart from $t$), 
events of the type $\epem \to \qqbar$ have a cross-section of order 1 nb. The other leptons (\mumu or \tautau) are  produced with cross-sections
also close to 1 nb. 

These cross-sections all scale in a known way  when PEP-II runs at the decreased off-peak energy, while 
  \FourS (and hence  $\BB$) production is `switched off'. The off-peak data can then be used to understand on-peak backgrounds and when scaled in the correct
way can be used to count \B events in on-peak data. Any $B$ physics analysis at
\babar\ will make use of the off-peak data, and it is an integral part of \B Counting.

\section{CP Violation}
The \babar\ experiment was designed to achieve a number of physics goals. The primary motivation for the experiment was to study
$CP$ violation in the decays of $B$ mesons \cite{Harrison:1998}. A significant result~\cite{Aubert:2001a} demonstrating the existence of $CP$ violation in this 
sector was published in 2001, two years after running began. 

$CP$ is the product of two quantum mechanical operators.  Charge conjugation ($C$)  interchanges particles with their anti-particles and parity ($P$) 
changes the `handedness' of a co-ordinate system, i.e. $P$ sends $(t, \mathbf{x}) \to (t, -\mathbf{x})$. 

Independently, $C$ and $P$ are not symmetries of nature and
in particular, neither is a symmetry of the weak nuclear force. For example, left-handed neutrinos are transformed to right-handed neutrinos by the operation of $P$, and to
left-handed anti-neutrinos by $C$. Neither of these states is observed to participate in the weak interaction.

Cronin and Fitch  were awarded  the 1980  Nobel Prize in Physics for demonstrating the combination $CP$ was also not a symmetry 
of nature \cite{Christenson:1964}. They observed this in  the decay of neutral kaons, specifically through the mode $\KL \to \pipi$ which would not
occur if $CP$ were conserved. Following this, $CP$ violation was also anticipated in $B$ meson decays, and was expected to be larger and more varied than in the kaon system. 
Investigating this was one of the main reasons for running  the  \babar\ experiment.

There are several reasons for studying $CP$ violation with an experiment like $\babar$. Measurements of $CP$ violation are important to test the Standard Model
 (SM) of particle physics. Any results outside SM predictions by definition indicate \emph{new} physics.   A second reason relates to the matter-antimatter asymmetry 
of the universe. $CP$ violation is one of the conditions proposed by Sakharov \cite{Sakharov:1988} to explain how the observed asymmetry could occur. 

In the Standard Model, $CP$ violation is governed by the weak quark mixing matrix, known as the Cabibbo-Kobayashi-Maskawa (CKM) matrix \cite{Cabibbo:1963,Kobayashi:1973}. 
 This takes the form:

\begin{equation}
   V
   =\left( \begin{array}{ccc}
     V_{ud} & V_{us} & V_{ub} \\
     V_{cd} & V_{cs} & V_{cb} \\ 
     V_{td} & V_{ts} & V_{tb} 
   \end{array} \right)
\end{equation}
where each $V_{ij}$ is complex, and the probability of a transition between two types of quark $i$ and $j$ is proportional to $| V_{ij} |^2$. 
Here $u$, $d$, $s$, $c$, $t$ and $b$ represent up, down, strange, charm, top and bottom quarks respectively. 

By studying a variety of different decay channels, \babar\ can be used to measure  or put limits on several components of the CKM matrix. In particular,
many channels seen at \babar\ allow the parameter
\begin{equation}
\beta = \textrm{Arg}\left(-\frac{V_{cd}V^\ast_{cb}}{V_{td}V^\ast_{tb}}\right)
\end{equation}
to be measured. Other searches for CP-violation, such as in gluonic penguin decays  allow the measurement of different CKM matrix elements 
(see for example \cite{Lingel:1998}) including:
\begin{equation}
\gamma = \textrm{Arg}\left(-\frac{V_{ud}V^\ast_{ub}}{V_{cd}V^\ast_{cb}}\right).
\end{equation}

\section{Other Physics at \babar }

The \babar\ experiment has multiple uses apart from searching for $CP$ violation in $B$ decays. It is a high luminosity experiment, and as well as producing many
$B$ mesons (enabling study of rare $B$ processes),  collisions create many other species including charmonium states (such as \jpsi), other charmed mesons 
(such as $D$), $\tau$ leptons
 and two-photon states. 

Data from \babar\ are used to study each of these to refine and test the Standard Model. The cross-section for \ccbar production in the
 \babar\ detector is slightly larger than for \bbbar, while the \tautau cross-section is slightly smaller. 

The charm physics research at \babar\ includes  measuring $D$ meson and charm baryon lifetimes, rare $D$ meson decay and $D$ mixing.
 An important
 result demonstrating evidence for \DzDzb mixing observed in $\Dz \to \kaon \pi$ decays was published in 2007 \cite{Aubert:2007}.  

An analysis group is specifically devoted to studying charmonium  (\ccbar) states.  This group  studies several types of physics.
 This includes initial state radiation (ISR) production (such as $\epem \to \gamma_{I\!S\!R}\pipi \jpsi $), fully reconstructed 
$B$ events  with charmonium final states (for example $B \to \jpsi X$ or $B \to \psitwos X$ where $X$ represents other unspecified 
particles) and studies of exotic states like Y(4260). 

Other groups study charmless hadronic decays of $B$ mesons (e.g. $B \to \pi \pi$), leptonic $c$  and $b$ decays (e.g. $D, B  \to 
\ell \nul$, where $\ell = e, \mu$ or $\tau$)  
and radiative penguins. The last of these describes  processes involving radiative loops such as $b \to s \g$. Radiative penguin 
events provide  one way
to test theories beyond the  Standard Model  because new virtual particles could appear in the loop with detectable effects. 

Also, research at \babar\ includes  measurements of the properties of the $\tau$ lepton produced through $\epem \to \tautau$.
An important area of research in the $\tau$ sector is the search for lepton flavour violation through decays such as
$\tau^+ \to \mup \gamma$ or $\tau^+ \to \epem\ep$.

%% file: chapter2.tex
\chapter{The \babar\   Detector}

This chapter provides a brief introduction to the hardware of the   \babar\ detector.  Detailed  descriptions of the detector~\cite{Aubert:2001} 
and also of PEP-II~\cite{PEP2:1993} are available elsewhere.

It will be useful to briefly explain the (right-handed) co-ordinate system used in the \babar\ experiment. 
\begin{itemize}
\item The $+z$ axis is the direction of the high-energy (electron) beam. 
\item The $+y$ axis is vertically upwards.
\item The azimuthal angle $\phi$ lies  in the $xy$ plane. It is zero on the $+x$ axis, and increases towards the $+y$ axis.
\item The polar angle $\theta$ is measured from the $z$ axis. It is zero on the $+z$ axis and $\pi$ on the $-z$ axis. 

\end{itemize}

\begin{figure}[htb] 
  \begin{center}
    \includegraphics[width=4.5in]{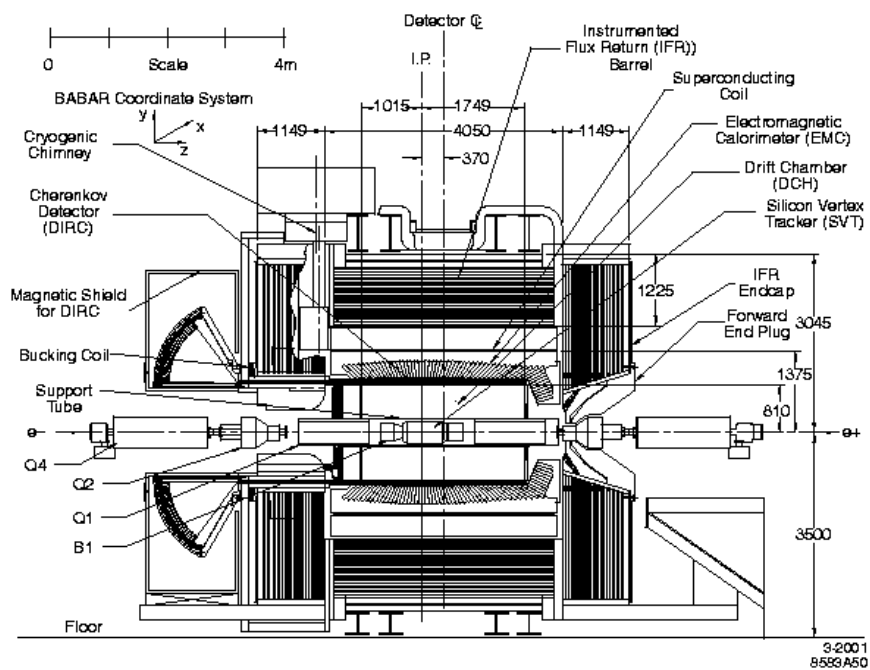}
    \caption{Cross-section of the \babar\ detector.}\label{fig:babarside}
  \end{center}
\end{figure}

The beam energy asymmetry of PEP-II means the optimal detector shape is also asymmetric. A schematic of the detector is shown in Fig.~\ref{fig:babarside}. 
Full details of the detector and its sub-systems can be found in \cite{TecDesRep:1995} and \cite{Aubert:2001}. The detector was designed to meet a large number
of requirements which include:
\begin{itemize}
\item  performing tracking on particles with transverse momentum ($p_{t}$) between $\sim 60$ MeV/c and  $\sim 4$ GeV/c;   
\item detecting photons and neutral pions with energies between $\sim 20$ MeV and  $\sim 5$ GeV, and; 
\item distinguishing electrons, muons, pions, kaons and protons over a wide range of momenta.

\end{itemize}

To meet these requirements, the detector was constructed with five major subsystems.  The innermost components
 are a five-layer  double-sided silicon vertex tracker (SVT), which provides precision tracking information and a 40-layer
 drift chamber (DCH), which provides momentum 
information for charged tracks. Outside these is a particle identification system, a detector of internally-reflected \v{C}erenkov
 light (DIRC).  Surrounding 
all these is an electromagnetic calorimeter (EMC) built from more than six thousand CsI(Tl) crystals. The EMC records energy
deposited by electrically charged particles and photons and also has some sensitivity to neutral hadrons such as neutrons and
\KL. The EMC is able to detect neutral pions by recording the photons from $\piz \to \gamma\gamma$. Finally, outside 
the superconducting coil is the muon detection system, called the Instrumented Flux Return (IFR)  consisting mostly of 
Limited Streamer Tube (LST) modules.

\section{The Silicon Vertex Tracker}

The silicon vertex tracker (SVT) is the component of the \babar\ detector closest to the interaction point (IP). It provides the most precise angular measurements
of tracks because further from the IP, track precision is limited by multiple scattering. The SVT  consists of five cylindrical layers of double-sided silicon-strip
detectors. 

In total, there are 340 silicon detectors within the SVT, with around 150,000 readout channels. The individual silicon detectors vary in size between $43\times 42$ mm$^2$ 
($z \times \phi$) and  $43\times 42$ mm$^2$. Each detector is double-sided, with strips to measure $\phi$ running parallel to the beam axis ($\phi$ strips) and strips to 
measure $z$ ($z$ strips) running orthogonally on the opposite side. The total area covered by the five layers of silicon detectors is around one square metre.

The ends of each strip are held at a potential difference of around 35--45 V.  When a charged particle passes through the silicon, it creates free electrons and
electron-hole pairs. This charge
is collected by the applied voltage,  and after being amplified  can be read out as a `hit' signal.  The track position is calculated by combining all the hits from
 individual silicon detectors.

The resolution of the SVT depends on track angle and varies between layers. For tracks at  normal incidence, the resolution of inner layers is approximately 
15 $\mu$m in $z$ and 10  $\mu$m in $\phi$. For outer layers the resolutions are approximately  35 $\mu$m and 20 $\mu$m respectively.

Some very low momentum particles such as slow pions in \Dstar decay have insufficient energy to reach outer components of the detector. For these particles,  the SVT 
provides the only tracking information.

\section{The Drift Chamber}
The multi-wire drift chamber (DCH) provides precise tracking information and to a lesser extent is used to perform some particle identification (PID). It was constructed on the campus of the University of British Columbia at TRIUMF. 
The DCH is cylindrical and surrounds the SVT. It is filled with a gas mixture of helium and isobutane, which is kept 4 mb above atmospheric pressure. 

\begin{figure}[htb] 
  \begin{center}
    \includegraphics[width=2.0in]{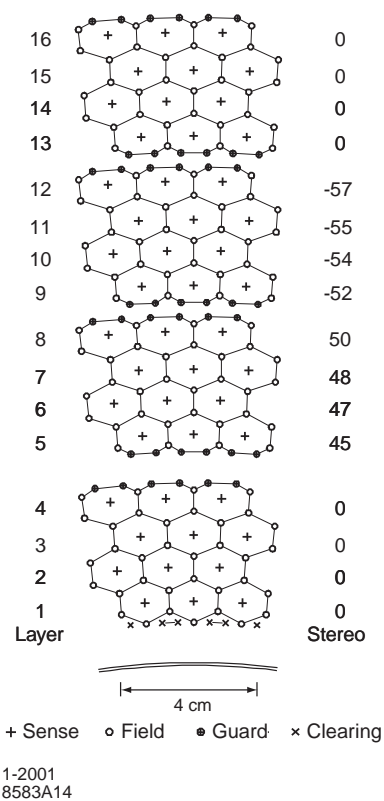}
    \caption[A schematic drawing of DCH drift cells.]{A schematic drawing of 16 layers of DCH drift cells. The lines connecting wires are included for 
visualisation purposes only. Numbers in the `Stereo' column indicate the angle in mrad of the stereo layers.}\label{fig:DCH}
  \end{center}
\end{figure}

The DCH is strung with nearly thirty thousand wires arranged into 40 
layers (and ten `superlayers') of hexagonal cells. The layout of these cells for 16 layers is 
shown in Fig.~\ref{fig:DCH}. 
A tungsten-rhenium `sense'  wire in the centre of each cell is held at a positive high voltage, of around 1960 V. The surrounding 
gold-coated aluminium `field' wires are 
at ground potential. Cells on the boundary of a superlayer have two gold-coated aluminium `guard' wires. These are held at 340 V and improve the performance of the cells
and ensure the gain of boundary cells is the same as that of inner ones.

Of the ten superlayers, four are `stereo'. The wires of these layers are strung  at a slight angle (between $\pm 45$ and $\pm 76$ mrad) relative to the $z$ axis. 
This is important for tracking as it enables  $\theta$ co-ordinates to be measured. 

A charged particle moving through the drift chamber ionizes the gas. The ionized gas molecules within a cell  move towards a ground wire and the electrons released from these
  molecules are accelerated towards the sense wire. In turn these electrons collide with and  ionize other gas molecules. This causes an `avalanche' of negative charge
 (up to a gain of 50,000) to arrive at the sense wire. For a particle energetic enough to exit the drift chamber, up to 40 DCH `hits' of this sort (one per cell) can be
 recorded. 

The drift chamber is contained within the 1.5 T field of the super-conducting coil. Measurements of track curvature in the drift chamber allow momenta to be determined. The
momentum resolution of the drift chamber for a particle of transverse momentum $p_t$    (i.e. the component of momentum
perpendicular to the $z$ axis)  in units of GeV/c is:
\begin{equation}
\sigma_{p_t} = (0.13 \pm 0.01)\% \cdot p_t + (0.45 \pm 0.03)\% .
\end{equation}
To be measured in the drift chamber (i.e. to travel past the inner radius of the DCH and trigger enough wires to provide a good measurement), a track must have
 transverse momenta of at least 100 MeV/$c$. 

\begin{figure}[htb] 
  \begin{center}
    \includegraphics[width=3.0in]{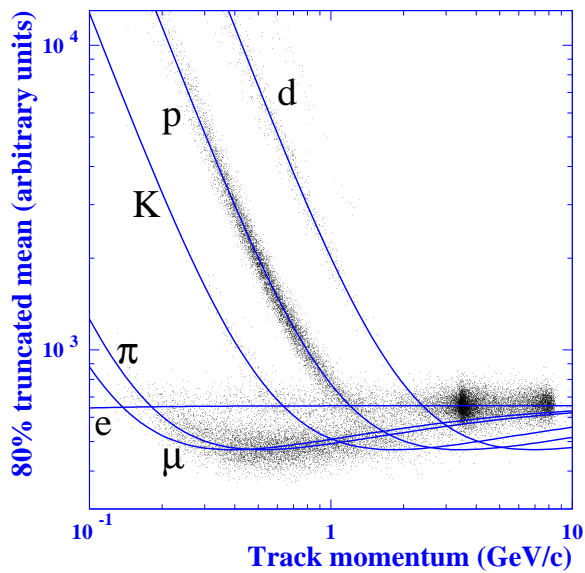}
    \caption[Drift Chamber $\ud E /  \ud x$ as a function of particle momentum.]{Drift Chamber $\ud E /  \ud x$ as a
 function of particle momentum. The scatter points are from beam scan data and the curves are appropriately parameterized Bethe-Bloch 
curves. The letters have their usual meanings, labelling the curves muon, electron, pion, kaon, proton and deuteron respectively. }\label{fig:dedx}
  \end{center}
\end{figure}

Different species of particles have a characteristic rate of energy loss with distance ($\ud E /  \ud x$), and the drift chamber  can  provide PID for 
low-momentum tracks by measuring this as a particle moves through the chamber. The $\ud E /  \ud x$ distributions for different
particles as a function of momentum are shown in Fig.~\ref{fig:dedx}.    Particle ID is achieved by recording the 
total charge deposited in each cell. Corrections are made to account  for several sources of bias including 
signal saturation, variations in gas pressure and differences  in cell geometry.

For low-momentum  tracks (i.e. those  below 1 GeV/c), the overall  $\ud E /  \ud x$ resolution is around 7\%. The final PID combines this information with
that gathered  from the DIRC and other subsystems.

\section{The DIRC}
The function of the detector of internally-reflected \v{C}erenkov light (DIRC) is to perform particle identification. In particular, it especially separates kaons and 
pions of momenta between 2.0 and  4.0 GeV/c.  It records the \v{C}erenkov radiation of 
charged particles passing through a ring of fused-silica bars located outside the drift chamber. This type of light is emitted whenever a charged particle passes through
a medium with a velocity greater than the speed of light in that medium. 

For a medium of refractive index $n$ (for fused-silica, $n\approx 1.47$), a particle moving with velocity $v=\beta c$ (where
$c$ denotes the speed of light in a vacuum) emits a characteristic cone of \v{C}erenkov light at an angle $\cos(\theta_c) = 1/(\beta n)$.  When the \v{C}erenkov angle
 $\theta_c$ of a particle is measured in the DIRC, momentum information from the drift chamber allows the mass 
(and hence type) of the particle to be determined.

\begin{figure}[htb] 
  \begin{center}
    \includegraphics[width=3.0in]{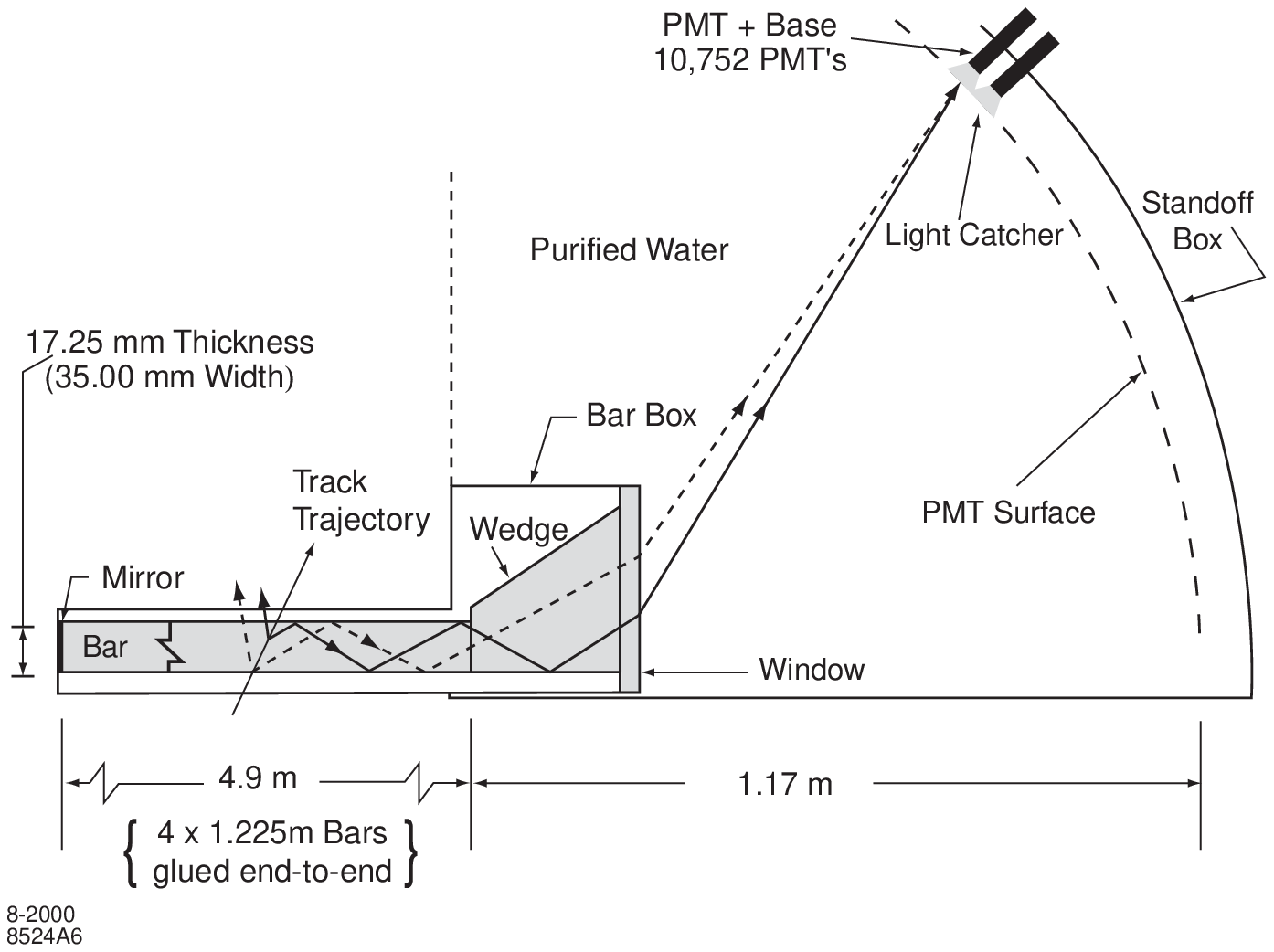}
    \caption[A schematic drawing of the DIRC.]{A schematic drawing of a DIRC radiator bar (left) and imaging region. }\label{fig:DIRC}
  \end{center}
\end{figure}

Ring-imaging \v{C}erenkov counters have commonly been used in particle detectors since the 1980s, but the \babar\   DIRC differs from all previous ones by 
utilising total internal reflection.  The DIRC radiator comprises 144  synthetic quartz bars of length 4.9m which function both as light guides and
 as a medium for \v{C}erenkov radiation. 
As shown in Fig.~\ref{fig:DIRC}, light is carried towards DIRC electronics in the imaging region  --- located at the backward-end of the detector --- 
by successive internal reflections.
A mirror is placed at the forward-end of the radiator bars to ensure forward-travelling photons are reflected back to arrive at the DIRC electronics. 

At the DIRC imaging region, the light expands into the `Standoff Box' which is filled with six cubic metres of  purified water of a similar refractive index to the 
quartz. The photons are detected at the rear of this by almost 11,000 photomultiplier tubes (PMTs), each of diameter 2.82 cm. The position and  arrival-time
of PMT signals are then used to calculate the \v{C}erenkov angle of particles passing through the DIRC radiator. 

At the high-momentum end  of the DIRC's functional window (around 4.0 GeV/c), the difference in the  \v{C}erenkov angle of pions and kaons is 3.5 mr. At this
momentum, the DIRC is able to separate the two species to  around $3\sigma$. This increases for lower momenta, up to a separation of as much as  $10\sigma$ 
at 2.0 GeV/c.

\section{The Electromagnetic Calorimeter}
The Electromagnetic Calorimeter (EMC) measures electromagnetic showers 
in the energy range 20 MeV to 9 GeV. It comprises 6580 thallium-doped caesium
iodide (CsI(Tl)) crystals which sit in 56 rings in a cylindrical barrel and forward end-cap which together cover a solid angle of 
 $-0.775 \leq \cos{\theta} \leq 0.962$ in the laboratory frame. This coverage 
region was chosen because few photons travel in the extreme backward direction. The barrel part of the EMC is located between the DCH and the magnet cryostat.

Thallium-doped CsI is used because of its favourable properties. It has a high light yield, a small Moli\`ere radius and  a short radiation length. 
These allow excellent resolution of both energy and angle within the compact design of the detector.

\begin{figure}[htb] 
  \begin{center}
    \includegraphics[width=3.5in]{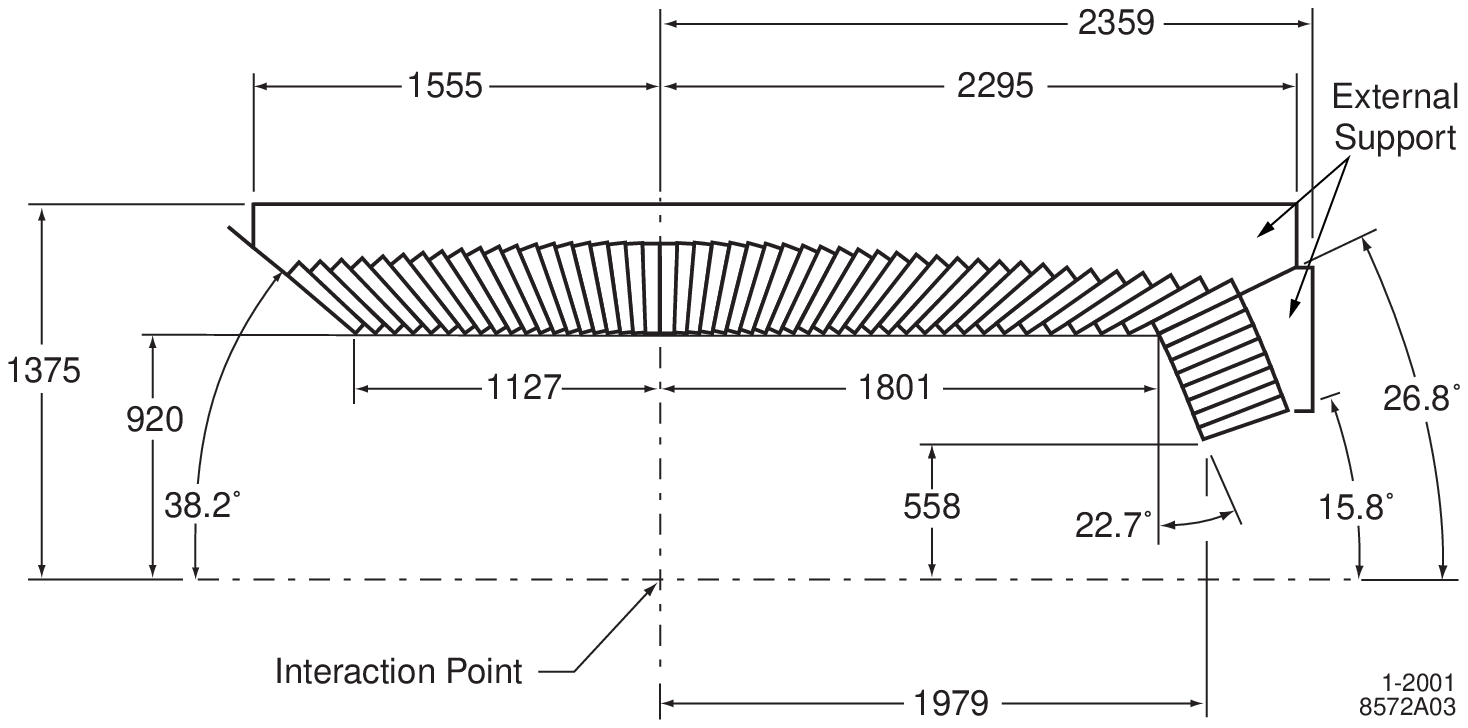}
    \caption[A cross-section of the EMC (top half).]{A cross-section of the EMC (top half). The  EMC is symmetric in $\phi$. Dimensions are given in mm.}
    \label{fig:EMC}
  \end{center}
\end{figure}

Altogether, the EMC weighs more than 26 tonnes. The crystals are held from the rear by an aluminium support system.  Similarly, cooling equipment and cables 
are also located at the rear.  This configuration minimises  the amount of material between the interaction point (IP) and the crystals. The support structure
is in turn attached to the coil cryostat. The layout of the aluminium support  and the 56 crystal rings can be seen in  Fig.~\ref{fig:EMC}.

High-energy  electrons and photons create electromagnetic showers when travelling through the EMC crystals. The shower size depends on the 
properties of the particle. The crystal absorbs the shower photons, and re-emits them as visible scintillation light. When tested with a source of 1.836 MeV photons, the 
average light yield per crystal is 7300 emitted photons per MeV of energy. The scintillation photons are read out with a package consisting  of two silicon 
photodiodes and two preamplifiers mounted on each crystal. 

There is a degradation in time due to radiation damage inside the EMC, and the crystals are constantly monitored to gauge their individual performance. This
is achieved by recording the EMC response both to known radioactive sources and to real events (i.e. from colliding beams). The latter involves
calibrating the system by measuring  particles of a known type, energy and position with the EMC. A variety of events are used for this calibration,
including \epem (Bhabhas), \gaga and $\mumu$.

\section{The Muon Detection System}
The IFR consists of three sections --- a central cylindrical barrel and two end caps. When \babar\ was commissioned, the IFR was filled
with Resistive Plate Counters (RPCs). Due to degradation and diminishing efficiency over the course of running, it became necessary to replace 
the barrel RPCs between 2004 and 2006  with a different technology (LSTs). RPCs are  still used in the  end caps.

The IFR is designed to detect deeply penetrating particles such as muons and  neutral hadrons (mostly \KL and neutrons).
 The system covers a large solid angle and was designed to have high efficiency and high background rejection. 

\begin{figure}[htb] 
  \begin{center}
    \includegraphics[width=4.5in]{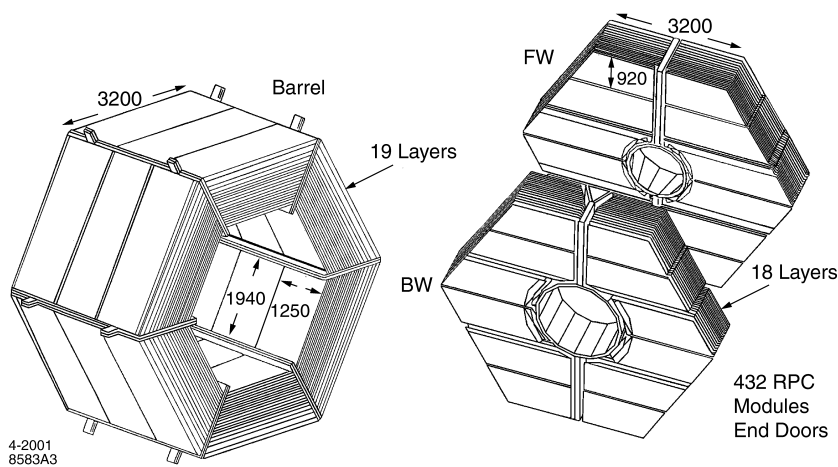}
    \caption[The IFR barrel and forward and backward endcaps.]{The IFR barrel and forward (FW) and backward (BW) endcaps.}\label{fig:IFR}
  \end{center}
\end{figure}

The IFR structure, shown in Fig.~\ref{fig:IFR}, is made from large quantities of iron, segmented into 18 plates of thickness 
varying between 2cm (nearest the beam) and 10cm (the outermost
layer).  The total thickness of the barrel is 65 cm, and the end caps 60 cm.   The iron (which also functions as the superconducting magnet's flux return)
 is optimized to be a muon filter and neutral hadron absorber allowing the best identification of muons and $\KL$s. 
  The iron layers are separated by a 3.2 cm gap  where up to 19 RPCs, LSTs, or  layers of brass can be housed.

RPCs and LSTs are different designs which achieve the same goal --- to provide accurately timed, precise two-dimensional positions of charges passing through. An 
RPC is filled with a gas mixture of argon, Freon and isobutane. Two 
graphite-coated surfaces located on either side of the gas  are held at 
a potential difference of
6700-7600 V. Ionising particles create a streamer between these two surfaces and their presence is read-out by pairs of capacitive strips running  parallel 
and perpendicular to the beam. 

Linseed oil was used to coat inner surfaces, and degradation of this and the graphite surfaces contributed to efficiency losses as the 
experiment progressed. By 2002, the muon identification efficiency had dropped from 87\% (at the start of running) to 78\% at a pion 
misidentification rate of 4\%. For this reason, LSTs  were installed to replace all RPCs in the barrel.
 Two sextants were fully replaced in summer 2004, and the remaining four 
sextants in summer 2006. The RPCs in the forward end cap
were replaced with new, more robust RPCs in fall 2000 \cite{Anulli:2005}.

LSTs function slightly differently from RPCs. An LST consists of either seven or eight
gas-filled cells. Through the length of each cell runs a wire held at high voltage (HV) of around 5500 V. To stop the wire sagging, several 
wire-holders are fitted along the cell's length. The gas mixture in the LST cells is predominantly carbon dioxide, and also contains smaller 
amounts of isobutane and argon.

A charged particle passing through the cell ionizes the gas and creates a streamer which can be
read out from the wire. This provides the $\phi$ co-ordinate. The streamer also induces a charge on
 a plane below the wire. Running parallel to the wires are a series of conducting strips ($z$-planes) which detect this charge and provide
the $z$ co-ordinate. 

For the upgrades in 2004 and 2005, the nineteen layers of RPCs in each barrel sextant were replaced by twelve layers of LSTs and six of brass. The outermost layer
of RPCs was inaccessible, so no LSTs could be installed there. The brass was installed in every second layer starting with the fifth both to compensate for the loss of
absorption between the two outer layers and to increase the total absorption length.

%% file: chapter3.tex
\chapter{Particle Lists}\label{particlelists}

The \babar\ database of events (including more than 400 million \BB events)  is very large and several stages of processing exist to reduce the time and storage 
requirements of analysts. The final selected data are stored in ntuples, and analysts must make a number of decisions regarding the types of information
that should be stored. For example, it is often desirable to store only events of a certain type, and necessary to define what properties a track
candidate should have before it is considered relevant.

As well as containing  well-measured particles (charged tracks and neutrals) from genuine physics events, raw data includes a lot of other (`junk')
information which must be filtered as much as possible.  This includes particles from cosmic rays, and beam-gas and beam-wall interactions. The 
software used by the \babar\ Collaboration is flexible enough to give analysts control over this filtering. 

The term \emph{particle list} means a list of all charged or neutral candidates in a particular event  which pass certain criteria. These criteria are varied
and depending on the type of list can include limits on track angle, momentum, distance of the track from the interaction point and many other quantities. 
A \emph{candidate} is a  collection of hits within
the detector (for neutral particles this takes the form of energy deposits in certain detector components) which  after reconstruction
 have been identified  as a likely particle. Candidates can be either charged or neutral, and there is no guarantee a candidate is in fact
anything other than background noise --- it only has to appear to be a genuine particle.

A charged candidate is a track associated with a particle moving inside the detector which (because of its charge) curves in the magnetic field. 
A neutral candidate is  any  EMC cluster  which has no matching track.

\section{Lists Used in \B Counting}

The most basic lists, on which others are based are ChargedTracks and CalorNeutral. These contain every charged and  neutral candidate 
after event reconstruction.  

The lists used in the past for \B Counting are   ChargedTracksAcc (for charged candidates) and GoodNeutralLooseAcc (neutrals).
 There is also an additional requirement based on the number of charged candidates on the GoodTracksAccLoose list.

The lists used in this analysis are slightly different to these and are instead based  on the established lists GoodTracksLoose (for charged candidates) and 
CalorNeutral with some customized additions. The requirements for GoodTracksLoose, ChargedTracksAcc and GoodNeutralLooseAcc are shown in Table \ref{tab:lists1}. 
\begin{table}[h]       \caption{Criteria of some existing particle lists used in \B Counting.} 
  \centering
  \begin{tabular}{|c|c|c|}  

    \hline
    GoodTracksLoose & \ ChargedTracksAcc & \ GoodNeutralLooseAcc \\
    \hline
     $p_T > 0.05$ GeV/c       & \   $0.41 < \theta < 2.54$      & \  Raw energy $> 0.03$ GeV \\
      DOCA in xy $< 1.5 $   cm     & \                                  & \    $0.41 < \theta < 2.409$  \\
      DOCA along z $< 2.5 $ cm    & \      & \ \\
    $ | \mathbf{p} | < 10$ GeV/c  & \ & \  \\

  $ V_{0}$ daughters excluded & \ & \  \\
    \hline
  \end{tabular}
    \label{tab:lists1}
\end{table}

 The addition of `Acc' (within acceptance region) to a list name
denotes angular requirements which keep the track within the detector. The list GoodTracksAccLoose is the same as GoodTracksLoose with the additional requirement  
$0.41 < \theta < 2.54$ for each track.

DOCA stands for the Distance of Closest Approach of the track to the interaction
point. By rejecting tracks which do not closely approach the IP, a large number of background tracks can be removed from the event. 

Tracks clearly identified as $V_{0}$ daughters are removed from the GoodTracksLoose list. These are tracks which originate at a vertex far from the IP, and are
obviously secondary decay products.

\section{Charged Candidates}
The list of charged candidates used in this thesis is based on the standard GoodTracksLoose with $V_{0}$ daughters included. For many analyses
it is convenient not to include them in the main charged list so they can later be identified and  added only if desired. For example, it
is beneficial for some analyses to reconstruct \piz mesons in this way.  For \B Counting,
this sort of flexibility is not necessary. The quantities of importance for labelling events \emph{hadronic} or \emph{mu-like} mostly rely on information
from the entire event such as the total event energy, the total number of tracks, or the sphericity of the event. The sphericity is
measured by a quantity called  R2, which is the ratio of the event's second and zeroth Fox-Wolfram moments \cite{Fox:1978}. 

In addition to this, some angular requirements are imposed to keep tracks within the detector's acceptance region. These are determined by
comparing data and Monte-Carlo simulated data. At very high or very low angles,  MC is less able to simulate tracks, so the maximum and minimum
angular values are chosen to ensure data and MC agree closely in the selected region.

At very high or very low angles there are fewer layers of the detector (especially the drift chamber)  which particles
can traverse. Tracks passing near the edges of detector will  be more difficult to reconstruct than other tracks, and some
may be lost. This is difficult to  simulate accurately with Monte Carlo. The angular cuts chosen are almost identical to those used in  'Acc' charged
lists.

\begin{figure}[htb] 
  \begin{center}
    \includegraphics[width=4.5in]{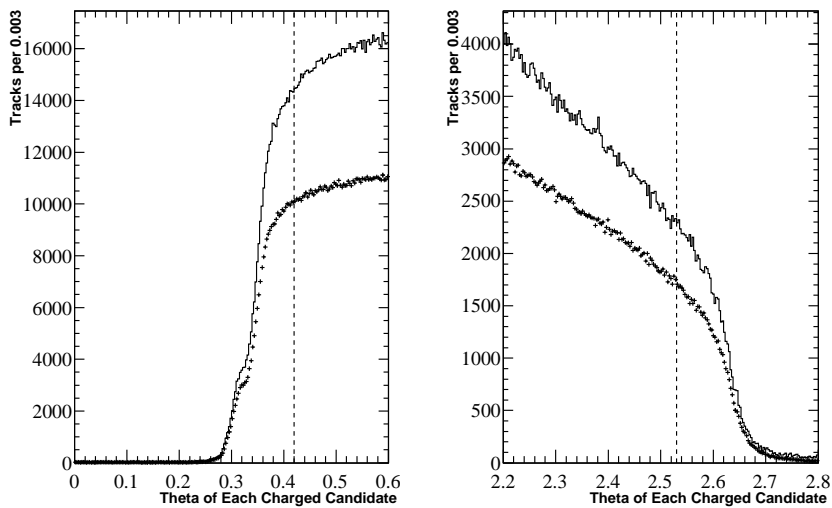}
    \caption[High and low theta regions the GoodTracksLoose list.]{High and low theta regions for tracks on the GoodTracksLoose list.
      The solid line represents on-peak data, and the points (+) off-peak data.}\label{fig:chargedangles}
  \end{center}
\end{figure}

Figure \ref{fig:chargedangles} shows  the high and low $\theta$ regions for tracks on the GoodTracksLoose list. The solid curve represents
 on-peak data, and the dotted curve off-peak data. In both regions
the agreement between MC and data is good even beyond the cuts, but sharp drop in data events at just below 0.4 and above 2.6 indicate the regions where 
the effects of reaching the detector's edges become apparent. The cuts are chosen on either side of these drops to ensure the selected region
is both modelled well by MC and doesn't include regions where edge effects appear.

For later convenience, this new list is given the name ChargedTracksBC.

\section{Neutral Candidates}
The list of neutral candidates used in this analysis is based on the standard CalorNeutral list. It has additional angular requirements (again to ensure
the region studied is well-described by Monte Carlo), and a minimum energy requirement of 100 MeV. 

\begin{figure}[htb] 
  \begin{center}
    \includegraphics[width=4.5in]{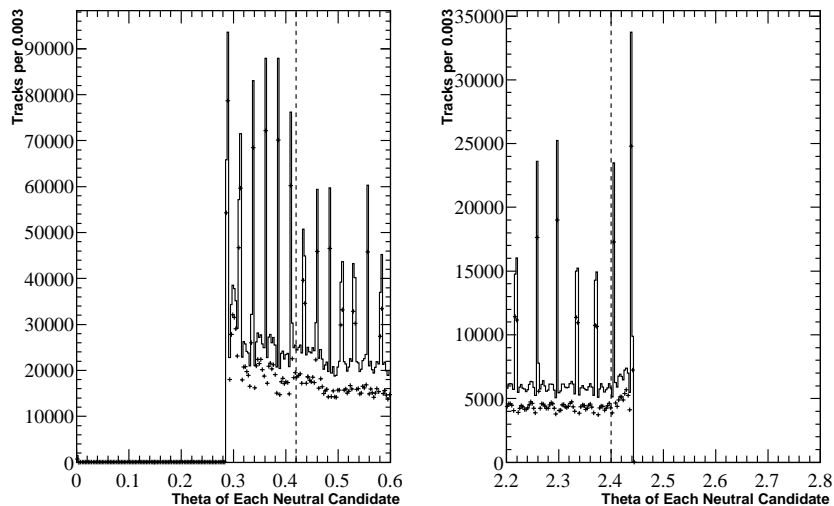}
    \caption[High and low theta regions for the CalorNeutral list.]{High and low theta regions for candidates on the CalorNeutral list.
 The solid line represents on-peak data, and the points (+) off-peak data.}\label{fig:neutralangles}
  \end{center}
\end{figure}

The reasoning behind choosing these particular angular cuts
($0.42 < \theta < 2.40$) is similar to that of  the charged tracks list. The high and low theta regions are shown in Fig.~\ref{fig:neutralangles}.
The acceptance region for neutral particles depends on the angular range of the EMC (see Fig.~\ref{fig:EMC}), so is different than for charged tracks. 
The shape of Fig.~\ref{fig:neutralangles} clearly shows the EMC construction --- the highest peaks are at the centre of EMC crystals.

\begin{figure}[htb] 
  \begin{center}
    \includegraphics[width=4.5in]{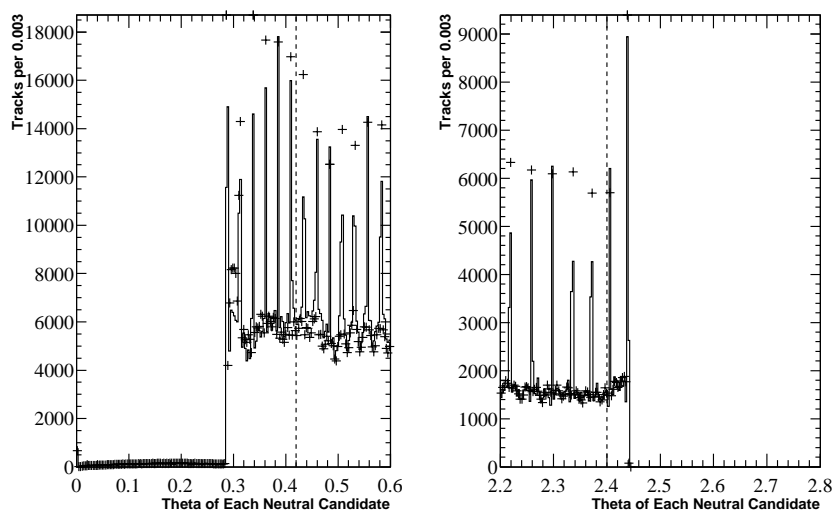}
    \caption[\BB data and MC for high and low neutral angles.]{\BB data and MC for high and low neutral angles. 
      Candidates are on the CalorNeutral list. The solid line represents \BB data the points (+) \BB MC.}\label{fig:neutralangles2}
  \end{center}
\end{figure}

The angular cut for high theta values is made to exclude the two most extreme crystals, ensuring neutrals are well within the detector. 
For the low theta region, the cut is based on the agreement between MC
and data. Although EMC crystals are in place for theta values as low as 0.3, the best agreement between MC and data occurs above 0.42. Within reasonable limits, 
the exact choice of angular cut is 
arbitrary, but the range $0.42 < \theta < 2.40$ is made for the neutral list for the above reasons.

 \begin{figure}[htb] 
  \begin{center}
    \includegraphics[width=4.5in]{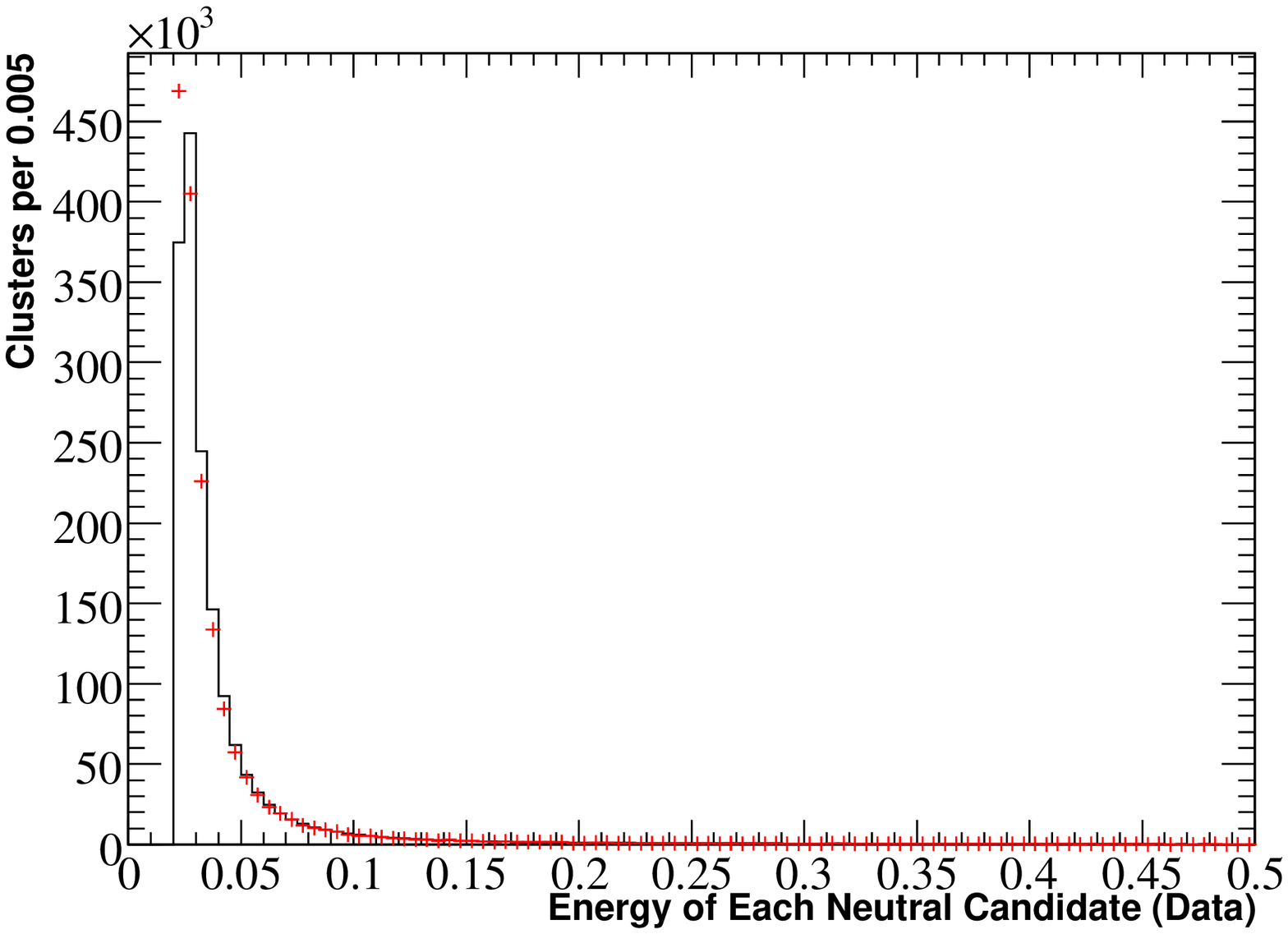}
    \caption[Neutral backgrounds in back-to-back mu-pair events.]{Neutral backgrounds in Run 6 back-to-back mu-pair events. The solid curve represents data, and
the points (+) mu-pair MC. The energy of each neutral cluster in an event is plotted for  mu-like events where the muons are within 0.02 radians
of  back-to-back.}\label{fig:neutralbkg}
  \end{center}
\end{figure}

To reduce  backgrounds, a minimum energy requirement of 100 MeV per cluster is also imposed on the neutral list. Erroneous neutral clusters
due to beam gas or electronic noise have a random distribution throughout any event. The level of this background can be determined by examining
muon-pair events where the muons are back-to-back. These $\epem \to \mumu$ events are very clean --- the back-to-back requirement
implies the muons have not radiated photons, so within certain limits the muons are the only genuine physics tracks in the event. 

Figure \ref{fig:neutralbkg} shows the distribution of energy per neutral cluster  for such back-to-back muon events (solid curve) and compares this
to mu-pair Monte Carlo. The events plotted pass isBCMuMu and additionally the two highest-momentum tracks are within 0.02 radians ($1.1^{\circ}$) of back-to-back. 
 A cut at 0.1 GeV removes the vast majority of neutral energy from this event at the expense of little physics, so this requirement
is included in the \B Counting neutral list. This new list is given the name NeutralsBC.

\begin{center}
\begin{table}[b]      \caption{Criteria of the lists used in this thesis.}
  \centering
  \begin{tabular}{|c|c|}  

    \hline
    ChargedTracksBC &\  NeutralsBC \\
    \hline
    On ChargedTracks list &\ On CalorNeutral list \\
    $p_T > 0.05$ GeV/c      &\   Cluster energy $> 0.1$ GeV \\
   
      DOCA in xy $< 1.5 $   cm &\   $0.42 < \theta < 2.40$ \\
   
      DOCA along z $< 2.5 $ cm  &\  \\
  
     $ | \mathbf{p} | < 10$ GeV/c  &\   \\

    $0.42 < \theta < 2.53$ &\   \\
    \hline
  \end{tabular}
    \label{tab:mylists}
\end{table}
\end{center}

%% file: chapter4.tex
\chapter{\B Counting}\label{chapt:BCount}

As its name suggests, \emph{B Counting} involves determining the number of \BB events in a particular data sample. It is necessary to know this to measure
any $B$ meson branching ratio for a particular final state. For example,  $\BR (\Bz \to D^{-} \pi^{+}  ) $ is approximately 0.3\%. This value is equal
to the number of $\Bz \to D^{-} \pi^{+}$ decays in  a particular sample divided by the total number of \Bz events in the sample. 

\section{Method for \B Counting}\label{BCountMethod}

The $B$ mesons produced at \babar\ are from the decay of the \FourS meson (a \bbbar state). Due to inefficiencies, the ratio of  \BzBzb 
production to \BpBm production from \FourS  decay is difficult to measure, but the currently accepted value  is consistent with 1. 
The \babar\ Collaboration has published the result \cite{Aubert:2004bb}:

\begin{eqnarray}\label{eqn:FourSRatio}
R^{+\!/0} & = & \BR(\FourS \to \BpBm) / \BR(\FourS \to \BzBzb) \nonumber \\
 & = & 1.006 \pm 0.036 \ \textrm{(stat.)} \pm 0.031 \ \textrm{(syst.)}.
\end{eqnarray}
It is common to assume this number is exactly unity. Hence, to calculate a particular $B$ branching ratio in a sample, it is only necessary to count the total 
number of \BB events in the sample, regardless of charge.   

The current method used to count \B mesons by the \babar\ Collaboration has been in place since 2000 \cite{Hearty:2000}.  In a sample of on-peak data, the number 
of \BB events is equal
to the total  number of hadronic events ($N^{0}_{H}$)  less the number of non-\BB hadronic events ($N^{0}_{X}$). 
\begin{equation}\label{eqn:BProduction}
N^{0}_{B} = N^{0}_{H} - N^{0}_{X}
\end{equation}
In this sense,  \emph{hadronic} means  events which look like $\epem \to$ hadrons. The superscript ``0'' in each case indicates these quantities are
the actual numbers produced of each event type. In reality, every quantity has an associated efficiency since nothing can be counted perfectly. In the \B 
Counting code, we can only use the number of each quantity  \emph{counted}, and the number of \B mesons produced is calculated from the number
of \B mesons counted by dividing by the \BB efficiency $\varepsilon_{B}$.

The number of on-peak continuum events can be found by scaling (by luminosity) an off-peak sample where, without \FourS production, all events are non-\BB. 
The small decrease in energy from on-peak to off-peak data-taking changes all continuum production rates slightly, but almost all of these events
scale in the same way as muon-pair events (i.e. $\epem \to \mumu$). Hence, the ratio of muon-pairs in the on- and off-peak samples is approximately equal to 
the ratio of the luminosities of the two samples. \B Counting relies on the fact that the ratio of muon-pair to continuum (i.e. non-\BB) events in a
sample is the same (to an excellent approximation) regardless of the CM energy:
\begin{equation}\label{eqn:EqualRatio}
\frac{N^{0}_{X}}{N^{0}_{\mu} } \approx \frac{N'^{0}_{X}}{N'^{0}_{\mu} }.
\end{equation}
In this thesis, unless stated otherwise,  primed symbols represent off-peak values.

\section{Derivation of the \B Counting Formula}\label{sec:BCountingDerivation}

Suppose we wish to calculate the number of \B mesons in a sample of on-peak data of luminosity $\mathcal{L}$, using an off-peak
data sample of luminosity $\mathcal{L}'$. For simplicity, off-peak quantities are primed, and hadronic, mu-pair, \BB  and continuum  quantities have
the subscripts $H$, $\mu$, $B$  and $X$ respectively. The numbers produced of each quantity (as opposed to those counted by the \B Counting selectors) 
have the superscript ``0''.

In a sample of on-peak data, the number of \BB events is equal to the total  number of hadronic events ($N^{0}_{H}$)  
less the number of non-\BB hadronic events ($N^{0}_{X}$).  
\begin{equation}\label{eqn:BBProduction}
N^{0}_{B} = N^{0}_{H} - N^{0}_{X}
\end{equation}

The number of on-peak continuum events can be found by scaling (by luminosity) an off-peak sample (in which all events are non-\BB). 
The small decrease in energy from on-peak to off-peak data-taking changes all continuum production rates slightly, but almost all of these events
scale in similar ways with luminosity. 

For any particular type of event, the number counted is equal to the number produced multiplied by the efficiency ($\varepsilon$). So for example,
\begin{equation}\label{eqn:lumi1}
N_{\mu} =  \varepsilon_{\mu}N^0_{\mu} =  \varepsilon_{\mu}\sigma_{\mu}\mathcal{L}
\end{equation}
and
\begin{equation}\label{eqn:lumi2}
N'_{\mu} = \varepsilon'_{\mu}N'^0_{\mu} = \varepsilon'_{\mu}\sigma'_{\mu}\mathcal{L}'.
\end{equation}

For off-peak data, there is no \FourS production, so we can assume that all hadrons are from continuum events (and hence the symbols
$N'_H$ and $N'_X$ are equivalent):
\begin{equation}\label{eqn:lumi3}
N'_H = N'_{X}  =  \varepsilon'_{X}N'^0_{X} = \varepsilon'_{X}\sigma'_{X}\mathcal{L}'.
\end{equation}

We define:
\begin{eqnarray}\label{eqn:Appkappadef}
\kappa & \equiv & \frac{\varepsilon'_{\mu}\sigma'_{\mu}}{\varepsilon_{\mu}\sigma_{\mu}}\cdot \frac{\varepsilon_X \sigma_X}{\varepsilon'_X \sigma'_X } \\
       & \equiv & \kappa_{\mu} \cdot \kappa_X.
\end{eqnarray}

We combine (\ref{eqn:lumi2}), (\ref{eqn:lumi3}) and (\ref{eqn:Appkappadef}) to give:
\begin{equation}\label{eqn:bigratio}
\frac{N'_X}{N'_{\mu}}\kappa = \frac{\varepsilon'_{X}\sigma'_{X}\mathcal{L}'}{ \varepsilon'_{\mu}\sigma'_{\mu}\mathcal{L}'}\kappa 
= \frac{ \varepsilon_{X}\sigma_{X}}{ \varepsilon_{\mu}\sigma_{\mu}}.
\end{equation}

The hadronic events in the on-peak sample consist of continuum and  \BB  events:
\begin{equation}
N^0_{H} = N^0_X + N^0_{B}
\end{equation}
and the number of \emph{counted} hadronic events is
\begin{equation}
N_{H} = \varepsilon_X N^0_X + \varepsilon_{B} N^0_{B}.
\end{equation}

Hence from (\ref{eqn:bigratio}), we can write
\begin{eqnarray}
 \varepsilon_{B} N^0_{B} & = & N_{H} - \varepsilon_X N^0_X  \\
 & = &  N_{H} -  \varepsilon_{X}\sigma_{X}\mathcal{L} \\
 & = &  N_{H} - \frac{N'_H}{N'_{\mu}}\cdot\kappa\cdot \varepsilon_{\mu}\sigma_{\mu}\cdot \frac{N_{\mu}}{ \varepsilon_{\mu}\sigma_{\mu}} \\
 & = & N_{H} - N_{\mu}\cdot\frac{N'_H}{N'_{\mu}}\cdot\kappa.
\end{eqnarray}

Hence, the number of \BB mesons produced in the on-peak sample is given by:
\begin{equation}\label{eqn:nb}
 N^0_{B} = \frac{1}{\varepsilon_{B}}( N_{H} -  N_{\mu}\cdot R_{o\!f\!\!f} \cdot \kappa ),
\end{equation}
where
\begin{equation}
R_{o\!f\!\!f} \equiv  \frac{N'_X}{N'_{\mu}}.
\end{equation}
In general, $\kappa$ is close to unity. The exact values and uncertainties  of $\kappa_{\mu}$ and $\kappa_{X}$ are discussed in detail
later.

\section{\B Counting Goals}

\begin{figure}[tbp] 
  \begin{center}
    \includegraphics[width=3.5in]{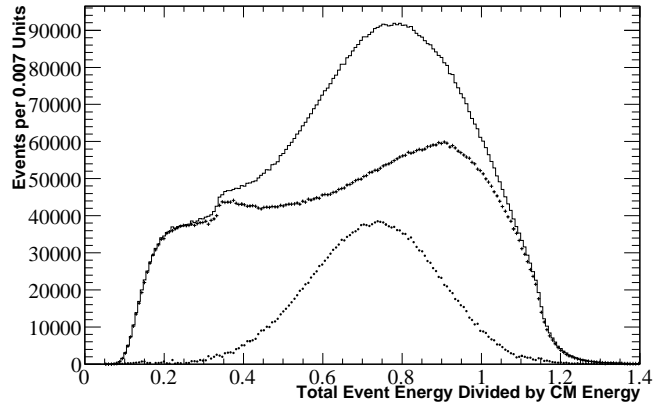}
    \caption[ETotal/eCM for events passing hadronic cuts.]{Total energy divided by CM energy (ETotal/eCM) 

for events passing all \B Counting hadronic cuts
 apart from one on ETotal for a sample of data from Runs 5 and 6. The bin width is 0.007. The solid ({\scriptsize +}) histogram  represents on-peak (off-peak) data. The 
off-peak data is scaled by luminosity to the on-peak data, and the dotted histogram is the difference between the two. \label{fig:ETotalDemo}}
  \end{center}
\end{figure}

Figure \ref{fig:ETotalDemo} demonstrates the scaling of on- and off-peak  data by the number of muon-pairs in each sample. The figure is a plot of total 
 energy for events passing all hadronic cuts,
except for one on ETotal itself (the existing \B Counting cut is set at 4.5 GeV). Samples of on- and off-peak data from the same time period are plotted 
 --- in this case from `Run 4', 
which took place between September 2003 and July 2004. The off-peak data is luminosity-scaled (by the ratio of muon pairs in each sample) to match the on-peak
data. 

\begin{figure}[tbp] 
  \begin{center}
    \includegraphics[width=3.5in]{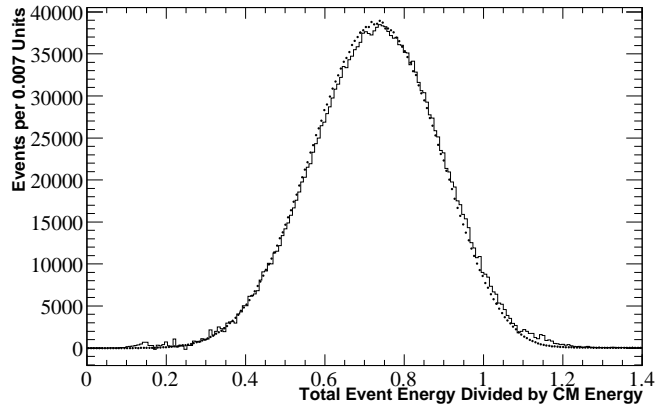}
    \caption[ETotal/eCM on- and off-peak subtraction, overlaid with MC.]{ETotal/eCM on- and off-peak subtraction, overlaid with \BB Monte Carlo. The solid 
histogram  represents data (off-peak subtracted from on-peak in Fig.~\ref{fig:ETotalDemo}) and the dotted histogram 
is from MC. The histograms are normalised to have the same area from 0.0--1.0.}\label{fig:ETotalDemoBB}
  \end{center}
\end{figure}

When correctly-scaled off-peak data is subtracted from on-peak data, the remaining events are those from \FourS decay, the vast majority of which
are \BB events. This \BB data is shown in Fig.~\ref{fig:ETotalDemo} and again in Fig.~\ref{fig:ETotalDemoBB} where it is overlaid with \BB Monte Carlo (MC)
generated events. The agreement between  data and MC is generally very good.

Ideally the \B Counting hadronic selection efficiency should be:
\begin{itemize}
\item high for $B$ meson events,
\item insensitive to changing background conditions (i.e. stable in time), and
\item in good agreement with simulation \BB  Monte Carlo.

\end{itemize}

Similarly the \B Counting muon selection should also be insensitive to backgrounds and be in good agreement with MC.  The specific selection requirements 
are described in later chapters.

More than 400 million \BB pairs have been created in \babar\ during its running, so  \B Counting is a very high statistics task and uncertainties
are predominantly systematic. The main aim of the work described in this thesis is to find selection criteria optimising these requirements and minimising
the final systematic error in the number of \B pairs counted.

\section{TrkFixup}
This thesis proposes the first major update to \B Counting since  the original method was devised in 2000. Since then,  many improvements have been made to 
both hardware and software within \babar\ which have improved the precision of many analyses. One of the most significant for \B Counting, named 
\emph{TrkFixup} was proposed in 2005 and fully implemented by mid-2007.

A systematic uncertainty present in any \babar\ analysis comes from the tracking precision. There are many reasons the tracking information of charged particles 
passing through the detector can be incorrect. For example, a collection of drift chamber and SVT hits due to noise could be incorrectly reconstructed
as a charged track. Similarly low-momentum particles which loop around within the detector  can be misidentified as more than one track. 

TrkFixup improves the tracking precision through several methods. Firstly, it improves track resolution
by modifying pattern recognition algorithms. It also increases the track finding probability and removes unwanted (`junk') tracks  such as duplicate (e.g. looping)
 tracks and backgrounds. This is achieved by re-examining `questionable' tracks with more sophisticated algorithms than previously used.

These improvements are important to \B Counting for several reasons. Reducing sensitivity to beam backgrounds is a key part of this thesis and the addition
of TrkFixup to track reconstruction improves this even before any adjustment to the \B Counting cuts is made. Some other improvements are  obvious, 
such as increased precision in measurements of track momentum and position. This is important for border-line events which narrowly pass or fail a particular
cut. 

When counting muon-like events, a requirement of the \B Counting code is that the event have at least two `good' tracks. TrkFixup affects
\B Counting directly through this cut, as events previously with only one good track could now be found to have two or more. Similarly, 
 an event previously labelled as mu-like could be found after TrkFixup to have fewer than two good tracks and thus 
fail the selection.

 \section{Original \B Counting Requirements}

The original requirements (i.e.\ those in place before the research described herein)  for an event to be labelled as mu-like
 or hadronic for \B Counting purposes were proposed in 2000 \cite{Hearty:2000}. An event passing
these criteria during processing receives a label (known as a `tag') \emph{isBCMuMu} or \emph{isBCMultiHadron} respectively. 
The criteria are summarized in Table \ref{table:isBC}. 

During reconstruction, events can receive one or more BGF tag.  They have very simple criteria and are designed to quickly accept all possibly useful physics events. 
The BGF tags enable some time to be saved during reconstruction, as  an event must have at least one of them to be fully processed. BGFMultiHadron and BGFMuMu are two 
of these which are used in \B Counting. 

BGFMultiHadron has the very simple selection  requirements of  at least three charged tracks, and R2 of  less than 0.98 in the CM frame. R2 is the second 
Fox-Wolfram Moment divided by the zeroth Fox-Wolfram Moment and is a measure of the event's sphericity \cite{Fox:1978}.  A pair of back-to-back particles
have R2 approaching one, while very spherical events (i.e. ones with many tracks distributed evenly around the event) have R2 approaching zero.   Thus the events
which do not receive the BGFMultiHadron tag are mostly $\epem \to \ellell$.  The three track requirement  also filters out several types of two-photon and
Initial State Radiation (ISR) events.

\begin{center}
\begin{table}[ht]   
  \caption[Criteria of isBCMuMu and isBCMultiHadron.]{
    \label{table:isBC}Criteria of the isBCMuMu and isBCMultiHadron \B Counting tags. The definitions of each quantity are given in the text.}
  \centering
  \begin{tabular}{|c|c|}
    \hline
    isBCMuMu Tag &\  isBCMultiHadron Tag \\
    \hline
    BGFMuMu &\ BGFMultiHadron \\
    masspair $> 7.5$ GeV/$c^2$    &\   R2All $\leq 0.5 $ \\

     acolincm $< 0.17453$ \  ($10^{\circ}$) &\   nGTL $\geq 3 $ \\

     nTracks $\geq 2$  &\ ETotal $\geq 4.5 $ GeV \\

     maxpCosTheta $< 0.7485 $  &\   PrimVtxdr $< 0.5$ cm \\

     maxpEmcCandEnergy $> 0 $ GeV  &\ $|$PrimVtxdz$| < 6.0$ cm  \\
     maxpEmcLab $ < 1 $ GeV  &\   \\ 
    \hline
  \end{tabular}

\end{table}
\end{center}

The criteria of BGFMuMu are also very broad, and are designed to quickly reject events which cannot be mu-pair events. To be labelled BGFMuMu, the two tracks of
 highest momenta must have  momenta above 4 GeV/c and 2 GeV/c respectively and be approximately back-to-back ($2.8 < \theta_{1} + \theta_{2} < 3.5$). Also,
 the energy deposited in the EMC by the two highest-momentum tracks must add to less than 2 GeV. 

The isBCMuMu tag has many criteria which depend on the properties of the two highest-momenta tracks --- if the event is a genuine mu-pair event, these will
almost certainly be the two muons. The quantity \emph{masspair} is the invariant mass of the pair and \emph{acolincm} is the acolinearity (the angular departure
from back-to-back)  of the pair in the  centre-of-mass (CM) frame. Also in the CM frame, both tracks must have $| \cos (\theta )| < 0.7485 $. The maximum 
energy deposited in the EMC by one of these tracks must be 
less than 1 GeV in the laboratory  frame, and at least one must leave some energy in the EMC. The former condition rejects approximately 99.7\% of all Bhabha events, 
which on average deposit just over 5 GeV  of energy per electron track. The final isBCMuMu quantity is \emph{nTracks}, 
the number of reconstructed charged tracks on the TaggingList list.

The criteria of isBCMultiHadron mostly involve quantities related to the entire event. \emph{R2All} is R2 calculated only using charged and neutral
candidates from within the detector acceptance region (i.e. from the ChargedTracksAcc and GoodNeutralLooseAcc lists). There are also  requirements on the 
number of tracks on the GoodTracksLoose list (\emph{nGTL}) and the total  event energy recorded in the laboratory frame (\emph{ETotal}). 

A possible source of background events is beam-wall interactions. These are  showers  caused by a particle from the beam interacting with the 
walls of the beam pipe and can resemble genuine physics events.  The primary vertex is the point calculated after reconstruction to be the most
 likely vertex for all tracks in the event. To reject beam-gas interactions, isBCMultiHadron has requirements on the distance between the primary vertex and the beamspot
in the $xy$ plane (\emph{PrimVtxdr}) and in $z$ (\emph{PrimVtxdz}).

The number of \B mesons in a particular on-peak sample can  be calculated using these tags and (\ref{eqn:nb}) by equating the number of isBCMuMu events 
with $N_{\mu}$ and the number of isBCMultiHadron events with $N_{H}$. 

The value of  $R_{o\!f\!\!f}$ depends on beam conditions
and backgrounds and so changes with time.  Off-peak running occurs only for short periods around larger blocks of on-peak 
running, so when calculating the number of \B mesons in any particular on-peak sample, the  value of  $R_{o\!f\!\!f}$ is based on the closest blocks
 of off-peak running.  

Many effects including changing background conditions cause long-term time-variations in selector (hadronic and muon-pair) efficiencies and this
is discussed in the following chapters. The association of
on-peak data with only the most recent off-peak running period ensures this variation has a minimal effect on \B Counting.

%% file: chapter5.tex
\chapter{Muon Selection}\label{chap:muonselection}

The motivation for the work in this thesis is to improve the understanding of \B Counting at \babar\  and if possible to modify the method to reduce overall
uncertainty. With more than 400 million \BB pairs recorded, statistical effects are often negligible, so improvements 
must come through reducing systematic
uncertainty. This chapter describes proposed modifications to the mu-pair part of the \B Counting code.

\section{Monte Carlo Simulation}\label{sec:MuonMC}
Muon-pair events are simulated with the KK2F MC generator \cite{Jalach:2000}. Before any detector simulation is applied, the physics cross-sections for the process 
$\epem \to \mumu$ within the detector can be calculated using a \babar\ application called  \emph{GeneratorsQAApp}. Based on 10 million generated on- and 
off-peak events, the cross-sections are $(1.11853 \pm 0.00011)$ nb and $(1.12647 \pm 0.00011)$ nb respectively. 

For detector simulation of all MC events, the  \babar\ Collaboration uses GEANT 4 \cite{Agostinelli:2002}. After GEANT has been applied to the simulated
events, they are made to correspond to a particular time by adding background frames from data. These frames are random `snapshots' of  detector occupancy
and are recorded at a rate of 1 Hz. The events are then reconstructed in the same way as data. This background mixing increases the realism of the 
Monte Carlo simulation by more accurately reflecting the processes occurring simultaneously within the detector.

 The detector conditions (such as the number of non-functioning EMC crystals), the levels of beam-background,  luminosity and other effects vary with time 
and cause a time-dependence in the efficiency of the mu-pair selector.  
The efficiency  of isBCMuMu  (and almost any other quantity) for a sample of MC depends on the time at which it was `taken', i.e. on the time
of the mixed-in backgrounds. Here, efficiency means the proportion of generated mu-pair events which pass the mu-pair \B Counting selector. In this way 
MC simulates the same effect in data.

The efficiency of isBCMuMu for simulated mu-pair MC is approximately 43\%. As described in Chapter \ref{chapt:BCount},  the number of mu-pair events in a data
 sample is used to scale the sample by luminosity. In modifying the muon selection, the goal is to reduce the total uncertainty when counting \B  mesons, so the
 absolute value of
the efficiency is not as important its variation  in time. 

\section{Cut on Muon Pair Invariant Mass}\label{FloatMasspair}
Ideally the muon selection will function identically on both on- and off-peak data. The total change in energy between the two is small (around 40 MeV), but 
any cut which varies depending on CM energy should be altered to  minimize the effect.  

Of the isBCMuMu cuts listed in Table  \ref{table:isBC}, the one altered the most by this effect is masspair --- the invariant mass of the two highest momentum tracks. 
This cut is designed to reject tau pairs  but accept almost all genuine muon events. The tau pairs are very short-lived and decay into multiple particles before 
reaching the beam-pipe walls. These events fail because the two highest-momentum tracks do not carry the majority of the energy and their invariant mass is
comparatively low.  For muon-pair events, masspair peaks near the beam energy.  

A simple method of removing sensitivity to CM energy (eCM) is to define a new quantity, masspair scaled by the total energy  ---  masspair/eCM --- and cut
instead on this. If the cut value in Table \ref{table:isBC}  is scaled by the on-peak beam energy of 10.58 GeV, the revised cut is:
\begin{equation}
  \frac{\textrm{masspair}}{\textrm{eCM}} > \frac{7.5 \ \textrm{GeV}/c^2}{10.58 \  \textrm{GeV}} \approx 0.709.
\end{equation}
In this thesis,  we set $c =1$ for convenience unless otherwise stated.

\section{Minimum EMC Energy Requirement}
As described in Chapter \ref{chapt:BCount}, to receive the isBCMuMu tag an event must meet certain EMC requirements. Of the two highest-momenta tracks, the maximum
energy deposited in the EMC by either one must be non-zero and less than 1 GeV. It was found during this research that depending on the time-of-running, up to 2\% of 
events  failed muon selection because both tracks deposited no energy in the EMC, i.e. events which would otherwise pass the selection were rejected because
the quantity maxpEmcCandEnergy was zero. 

In the original \B Counting code, the maximum EMC energy cut was calculated in the lab frame, while the minimum was calculated in the CM. However, any zero measurement
in the EMC is equivalent regardless of frame, so for simplicity in this thesis, the minimum and maximum allowed EMC deposits are both calculated in the lab frame. 

\begin{figure}[htb] 
  \begin{center}
    \includegraphics[width=4.0in]{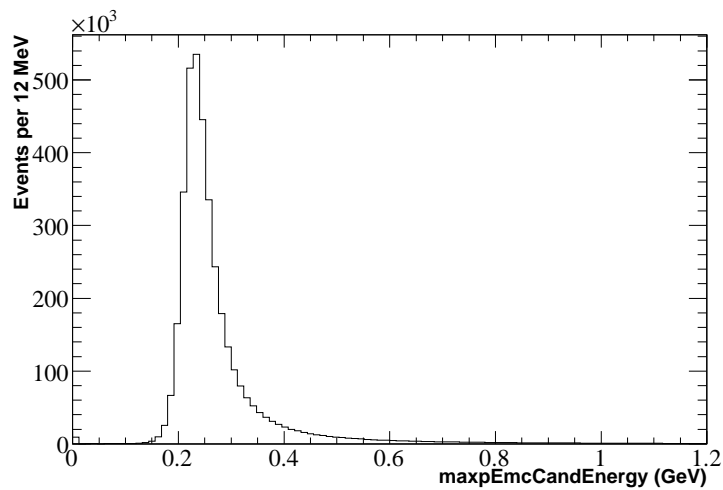}
    \caption[Distribution of maxpEmcCandEnergy.]
	    {Distribution of maxpEmcCandEnergy for on-peak Run 6 data passing all other muon-pair selection cuts.}
            \label{fig:maxpEmcPlot}
  \end{center}
\end{figure}

This effect was studied in samples of at least a million on-peak data  events with the BGFMuMu tag from each of the six major running periods (Runs 1--6). Some
of the quantities of interest for events passing all other muon selection requirements are shown in Tables \ref{table:maxpemc1} and \ref{table:maxpemc2}. 
The value of p1EmcCandEnergy is the amount of energy left
in the EMC by the highest momentum track, and p2EmcCandEnergy the amount for the second highest momentum track. The maximum of the two is named maxpEmcCandEnergy.
The distribution of this quantity for a sample of Run 6 on-peak  data passing all other muon-selection cuts is shown in Fig.~\ref{fig:maxpEmcPlot}. 

Any track has a non-zero chance of leaving no energy in the EMC. This could be due to any of several reasons including reconstruction errors or the particle hitting
 a gap between two EMC crystals. Assuming the individual probabilities are independent, the probability of both of the two highest momentum tracks 
leaving no EMC energy should be approximately the square of this number.  However, as can be seen in Tables \ref{table:maxpemc1} and \ref{table:maxpemc2},
 the number of events with both EMC clusters equal to zero 
(i.e. where maxpEmcCandEnergy=0) is much higher than would be expected. 

\begin{figure}[htb] 
  \begin{center}
    \includegraphics[width=4.0in]{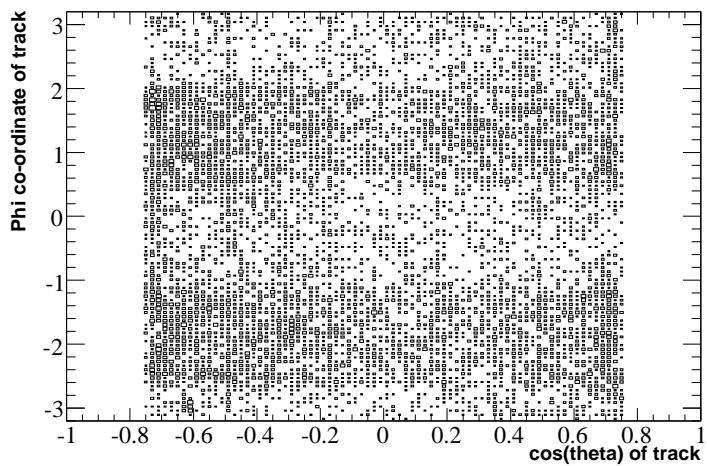}
    \caption[Angular distribution of muons with maxpEmcCandEnergy = 0.]
	   {Angular distribution of muons from  Run 6 data  with maxpEmcCandEnergy of zero. The value of $\phi$ and $\cos(\theta)$ in the CM frame are plotted
for the two highest-momentum tracks in events which  pass all other muon-selection requirements. The number of events in each bin of this $100 \times 100$ 
grid is proportional to the area of the box at the bin-centre.}
            \label{fig:maxpEmcScatterPlot}
  \end{center}
\end{figure}

This effect does not appear to be due to  EMC failures in a particular region. Figure \ref{fig:maxpEmcScatterPlot} shows the angular 
distribution in $\cos(\theta)$ and $\phi$ of the muons in a sample of
 Run 6 data events which pass all other muon-selection cuts, but which have maxpEmcCandEnergy of
zero. Note that the distribution is fairly randomly spread in  $\theta$ and $\phi$, although some broad bands are visible at constant $\phi$ due to detector
geometry.

The high proportion of events with maxpEmcCandEnergy of zero indicates some other effect is present --- it is especially noticeable in Runs 2 and 3 where 
the probability of both tracks leaving nothing in the EMC
is greater than the probability that the highest momentum track alone deposits nothing. This was examined by Rainer Bartoldus and the \babar\ Trigger, Filter and 
Luminosity Analysis Working Group (AWG). The predominant cause was found in events where the DCH and EMC event times differed.

The Drift Chamber and EMC have different timing systems, so can occasionally disagree on the event time ($t_0$). The EMC $t_0$ is calculated in two stages. First,
 the average arrival-time of energy in a wide time-window around the event is found. The window is then narrowed around this average to select the EMC hits
to be reconstructed. This method is susceptible to calculating incorrect $t_0$ if background events such as stale
Bhabhas arrive at an inopportune time.  Stale Bhabhas are unwanted Bhabha events which happen to occur at almost the same time as another physics event.

Such an event passing through the EMC at either end of the event window could cause the average arrival time of energy to shift to a time
when no energy is in the EMC. Through this process events that would otherwise pass muon selection can have no energy visible in the EMC
and be rejected. The proportion of events with EMC information lost in this way is complicated to simulate, but Tables~\ref{table:maxpemc1} and 
\ref{table:maxpemc2} show reasonable  agreement (usually within a factor of 2) between data and Monte Carlo.

The majority of these events can be regained by changing the minimum maxpEmcCandEnergy to:
\begin{itemize}
\item Either maxpEmcCandEnergy $ > 0$,
\item or at least one of the two highest momentum tracks is identified as a muon in the IFR.
\end{itemize}

\newpage
\begin{landscape}
  \begin{center}
    \begin{table}        
      \caption[Effect of maxpEmcCandEnergy (Runs 1--3)]{
	\label{table:maxpemc1}Effect of the minimum maxpEmcCandEnergy requirement on samples from Runs 1--3. }               
      \begin{tabular}{|l||cc|cc|cc|}
	\hline
        &\  \bf{Run 1} &\     &\ \bf{Run 2} &\     &\ \bf{Run 3} &\   \\
        &\  Data       &\ MC  &\ Data  &\  MC   &\ Data &\ MC  \\
	\hline
	\# BGFMuMu Events  &\ $1.001 \times 10^6$   &\ $0.807 \times 10^6$  &\ $1.542 \times 10^6$ &\ $2.580 \times 10^6$ &\ $4.622 \times 10^6$ &\ $1.718 \times 10^6$   \\
	\hline
	Pass all cuts  but minimum &\ $0.681 \times 10^6$   &\ $0.576 \times 10^6$  &\ $1.055 \times 10^6$ &\ 
        $1.837 \times 10^6$   &\ $3.155 \times 10^6$ &\ $ 1.223 \times 10^6$  \\
	EMC energy ($\dag$)   &\    &\     &\  &\  &\ &\ \\
	\hline
	\% of $\dag$ with  &\ 0.52  &\ 0.32  &\ 1.34 &\ 0.41  &\ 1.09 &\ 0.38  \\
	p1EmcCandEnergy=0 &\    &\     &\  &\  &\ &\ \\
	and p2EmcCandEnergy$\neq 0$ &\    &\     &\  &\  &\ &\ \\
	\hline
        \% of $\dag$ with  &\ 0.91  &\ 0.70  &\ 1.71 &\ 0.83  &\ 1.44 &\ 0.79 \\
	p2EmcCandEnergy=0 &\    &\     &\  &\  &\ &\ \\
	and p1EmcCandEnergy$\neq 0$ &\    &\     &\  &\  &\ &\ \\
	\hline
        \% of $\dag$ with  &\ 0.04  &\ 0.14  &\ 1.70 &\ 1.91  &\ 1.37 &\ 1.66 \\
	maxpEmcCandEnergy=0 ($\ddag$) &\    &\     &\  &\  &\ &\ \\
	\hline
       \% of $\ddag$ with  at&\ 23.3  &\ 85.4 &\ 83.3 &\ 93.4  &\ 80.6 &\ 91.3 \\
       least one track identified  &\    &\     &\  &\  &\ &\ \\
       as a muon in IFR  &\    &\     &\  &\  &\ &\ \\
	\hline
      \end{tabular}

    \end{table}
  \end{center}
\end{landscape}

\newpage
\begin{landscape}
  \begin{center}
    \begin{table}       
      \caption[Effect of maxpEmcCandEnergy (Runs 4--6)]{
	\label{table:maxpemc2}Effect of the minimum maxpEmcCandEnergy requirement on samples from Runs 4--6. }                
      \begin{tabular}{|l||cc|cc|cc|}
	\hline
        &\  \bf{Run 4} &\     &\ \bf{Run 5} &\     &\ \bf{Run 6} &\   \\
        &\  Data       &\ MC  &\ Data  &\  MC   &\ Data &\ MC  \\
	\hline
	\# BGFMuMu Events  &\ $2.516 \times 10^6$   &\ $3.556 \times 10^6$  &\ $2.694 \times 10^6$ &\ $3.045 \times 10^6$ &\ $3.151 \times 10^6$ &\ $1.003 \times 10^6$   \\
	\hline
	Pass all cuts  but minimum &\ $1.724 \times 10^6$   &\ $2.542 \times 10^6$  &\ $1.874 \times 10^6$ &\ 
        $2.186 \times 10^6$   &\ $2.154 \times 10^6$ &\ $ 0.711 \times 10^6$  \\
	EMC energy ($\dag$)   &\    &\     &\  &\  &\ &\ \\
	\hline
	\% of $\dag$ with  &\ 0.66  &\ 0.45  &\ 0.65 &\ 0.43  &\ 0.63 &\ 0.36  \\
	p1EmcCandEnergy=0 &\    &\     &\  &\  &\ &\ \\
	and p2EmcCandEnergy$\neq 0$ &\    &\     &\  &\  &\ &\ \\
	\hline
        \% of $\dag$ with  &\ 1.07  &\ 0.84  &\ 1.08 &\ 0.83  &\ 1.04 &\ 0.76 \\
	p2EmcCandEnergy=0 &\    &\     &\  &\  &\ &\ \\
	and p1EmcCandEnergy$\neq 0$ &\    &\     &\  &\  &\ &\ \\
	\hline
        \% of $\dag$ with  &\ 0.19  &\ 0.27 &\ 0.16 &\ 0.31  &\ 0.21 &\ 0.42 \\
	maxpEmcCandEnergy=0 ($\ddag$) &\    &\     &\  &\  &\ &\ \\
	\hline
       \% of $\ddag$ with at &\ 65.1  &\ 93.4 &\ 89.8 &\ 97.9  &\ 87.6 &\ 93.5 \\
       least one track identified  &\    &\     &\  &\  &\ &\ \\
       as a muon in IFR &\    &\     &\  &\  &\ &\ \\
	\hline
      \end{tabular}

    \end{table}
  \end{center}
\end{landscape}

A requirement on the IFR has not been used previously in B Counting. This is because in general its muon identification efficiency is too low and variability with
time too large to provide stable identification. However, in this case the IFR is used only to recover genuine muon-pair events that would otherwise be lost, so enables
a net efficiency gain and improved stability in time.

Muons are identified in the IFR by a Neutral Net muon selector \cite{Mohapatra:2004}, specifically by muNNVeryLoose. 
The efficiency of this 
selector varies with time, but in muon-pair MC the probability that at least one of the two highest-momentum tracks passes is typically over 90\%.  

The final row of Tables \ref{table:maxpemc1} and \ref{table:maxpemc2} shows the percentage of events recovered by adding this IFR requirement. In Run 6 data for example,
87.6\% of events which pass all other isBCMuMu requirements but which have maxpEmcCandEnergy of zero  are recovered.  

\section{Optimization of Cuts}\label{Sec:MuOpt}

The muons counted by isBCMuMu selection are used to measure the luminosity of a sample of events. Since on- and off-peak data are taken at different times, 
the two have different background conditions. Hence, to correctly perform the \B Counting background subtraction it is important the muon-pair selection be as
insensitive as possible  to changing backgrounds. The variation is reduced to some extent by  changing to the particle lists described in 
Chapter \ref{particlelists} and by altering the
masspair and maxpEmcCandEnergy cuts as described above. 

This sensitivity can be measured by the changes in muon-pair MC efficiency with time. Ideally the efficiency should be identical regardless of run number --- a sample
of generated mu-pair events with Run 1 backgrounds should have the same efficiency as the same events with Run 6 backgrounds. In practice this is not the case, but
by choosing different cut values for the requirements outlined in Table \ref{table:isBC} it can be reduced. 

The requirement of BGFMuMu is essential as it quickly filters out large amounts of events which cannot be muon-pairs. Likewise there is little flexibility
in the requirement of at least two well-identified charged tracks. The selector should identify events with well-tracked muons, and all events of this type
will have at least two  tracks on the charged list. The modification of the minimum maxpEmcCandEnergy requirement described above helps veto 
Bhabha events and is not further adjusted. 

The four remaining isBCMuMu requirements have slightly more flexibility in their cut values. The justification for the current cuts is given 
in \cite{Hearty:2000}, but  the addition of TrkFixup and the different particle lists (ChargedTracksBC and NeutralsBC) used in this thesis
are reason enough that they should be re-evaluated at this stage. Additionally, the method used in this thesis to find an optimal cut-set is 
improved and more robust. 

To optimise the isBCMuMu cutset, the month-by-month efficiency for muon-pair MC in both on- and off-peak running was calculated for different cutset choices. 
Up to and  including Run 6, there are 63 months with  on-peak muon-pair MC and 27 months with off-peak MC. For convenience, each  month from the time of running is
given an  integer label --- from January 1999 (month 0) to December 2007 (month 107). 

The total number of cutsets trialled in the optimization procedure was $13 \times 17 \times 20 \times 14 = 61880$. 
The weighted mean efficiencies of these cutsets
are between 0.262 and 0.450.

\begin{table}[ht]   
  \caption[Cut optimization of isBCMuMu.]{
    \label{table:mucutvariation}Cut optimization of isBCMuMu. }
  \centering
  \begin{tabular}{|c|c|c|}
    \hline
    Quantity  &\  No. of increments  &\ Range \\
    \hline
     masspair/eCM &\ 13 &\ $(6.0$ -- $9.0)/10.58$\\
       maxpCosTheta  &\  17  &\  0.650 -- 0.765 \\ 
       acolincm &\ 20  &\ 0.01 -- 0.20 \\
        maxpEmcLab &\  14  &\ 0.7 -- 2.0 \ GeV \\
    \hline
  \end{tabular}

\end{table}

The forward and backward ends of the detector have different acceptance regions, and the different angular cuts on ChargedTracksBC and NeutralsBC reflect this.
When varying maxpCosTheta (the maximum value of $|\cos{\theta_{C\!M}}|$ for the two highest momentum tracks), some care must be taken to ensure the cut is 
not varied into a region excluded by the list's own angular cuts. 

To find the $\theta_{C\!M}$ angles corresponding to the angular cuts  of ChargedTracksBC for a typical muon  we  boost into the laboratory  frame from the 
CM by $\beta\gamma = -0.56$ along $z$.  The method for calculating the transformation of the angles of a velocity can be found in many
relativistic kinematics texts, for example \cite{Hagedorn:1963}. We find:

\begin{equation}\label{eqn:lorentz}
\tan{\theta_{C\!M}} = \frac{v \sin{\theta}}{\gamma(v \cos{\theta} + \beta c)}
\end{equation}
where $v$ is the velocity of a particle travelling at angle $\theta$ relative to the $z$ axis in the laboratory frame. Practically all muons from $\epem \to \mumu$ 
have CM momenta between 5.0 and 5.5 GeV/$c$, which corresponds to having velocities of 0.9999$c$ in the laboratory frame (the two are identical 
to four significant figures). Using (\ref{eqn:lorentz}) and this velocity, the minimum and maximum angles of ChargedTracksBC (0.42 and 2.53) boost 
to 0.698 ($\cos{\theta_{C\!M}} = 0.766$) and 2.776 ($\cos{\theta_{C\!M}} = 0.934$) in the laboratory frame respectively. 

Hence, when varying maxpCosTheta, values above 0.766 should be ignored to agree with the angular requirements of the charged list. In practice, cut values
up to and including 0.765 are trialled.

\begin{figure}[p] 
  \begin{center}
    \includegraphics[width=3.8in]{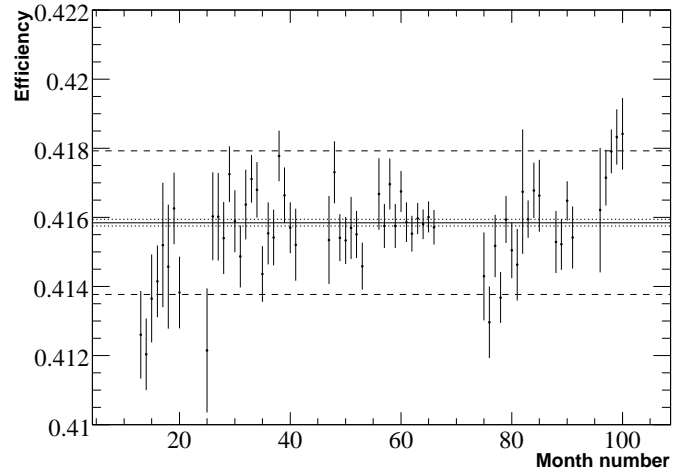}
    \caption[On-peak mu-pair MC efficiency variation.]{On-peak mu-pair MC efficiency variation. Details of cutset choice and error bars are in text. 
      The $\chi^2 $ value is 124.04 and $M$ (the number of degrees of freedom) is 63. The dashed lines show $\pm 0.5\%$ from the mean.}
      \label{fig:OnPeakMuVar}
  \end{center}
\end{figure}

\begin{figure}[p] 
  \begin{center}
    \includegraphics[width=3.8in]{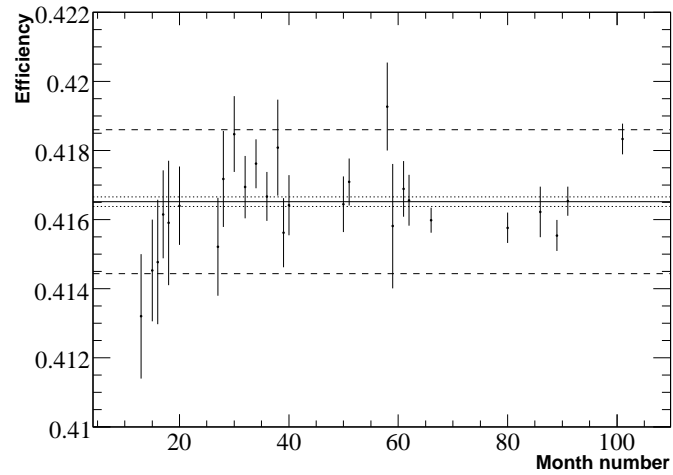}
    \caption[Off-peak mu-pair MC efficiency variation.]{Off-peak mu-pair MC efficiency variation for the same cutset choice as on-peak distribution.
      The $\chi^2 $ value is 47.94 and $M$ (the number of degrees of freedom) is 27. The dashed lines show $\pm 0.5\%$ from the mean. }\label{fig:OffPeakMuVar}
  \end{center}
\end{figure}

The on- and off-peak muon-pair MC efficiency variation is shown in Fig.~\ref{fig:OnPeakMuVar} and \ref{fig:OffPeakMuVar}. The cutset shown has 
fairly typical time-variation. It has an acolincm cut of 0.07, masspair/eCM of \  $(8.00/10.58)$, maxpCosTheta of 0.765 and a maximum maxpEmcCandEnergy 
cut of 1.0 GeV.

In each month $i$, the statistical error $\Delta(\varepsilon_i)$ is assumed to be binomial with
\begin{equation}\label{eqn:effstatvar}
\Delta(\varepsilon_i) =  \sqrt{\varepsilon_i  (1- \varepsilon_i) /N_{i}}
\end{equation}
where $\varepsilon_i$ is the efficiency  and $N_i$ the total number of generated events for that month. If we  assume the efficiencies come
from a statistical distribution which is constant in time, the weighted mean  $\hat{\varepsilon}$  (the solid horizontal line on the plots) is given by
\begin{equation}\label{eqn:wmean}
\hat{\varepsilon} = \frac{1}{w} \sum_{i=1}^{M} w_i \varepsilon_i
\end{equation}
where $M$ is the number of months with MC events, $w_i = 1/(\Delta(\varepsilon_i))^2$ and $w = \sum w_i$. The statistical standard deviation (shown by dashed horizontal lines)
of $\hat{\varepsilon}$ is $1/\sqrt{w}$ \ \cite{PDG}. The variation around the weighted mean  is obviously not random, 
and this  shows the presence of some non-statistical systematic effects. 

\begin{figure}[p] 
  \begin{center}
    \includegraphics[width=4.3in]{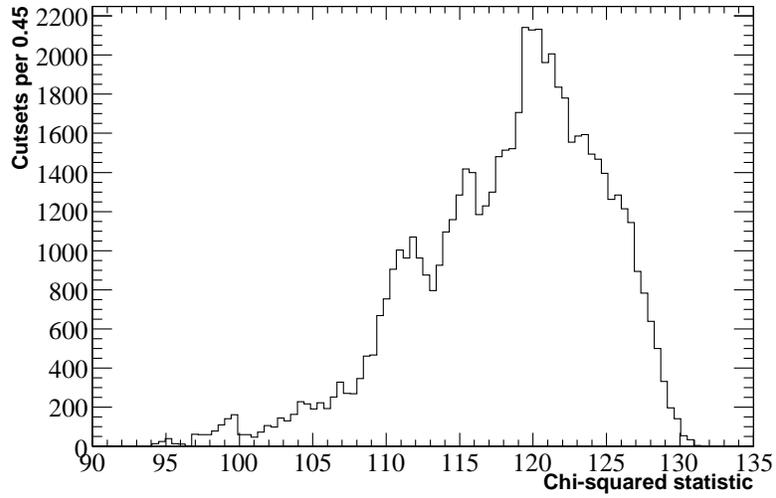}
    \caption[Distribution of $\chi^2$ statistic (on-peak).]{Distribution of $\chi^2$ statistic for trialled cutsets 
with on-peak mu-pair MC.
$M$ (the number of degrees of freedom) is 63.}\label{fig:X2Muonpeak}
  \end{center}
\end{figure}

\begin{figure}[p] 
  \begin{center}
    \includegraphics[width=4.3in]{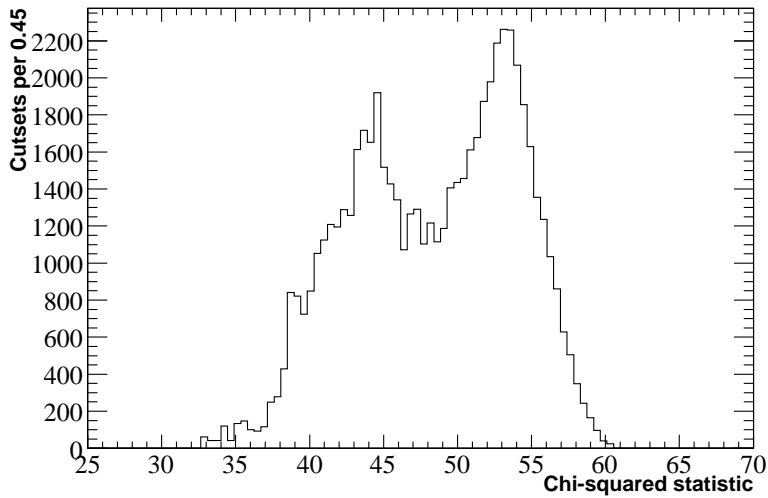}
    \caption[Distribution of $\chi^2$ statistic (off-peak).]{Distribution of $\chi^2$ statistic for trialled cutsets 
with off-peak mu-pair MC.
$M$ (the number of degrees of freedom) is 27.}\label{fig:X2Muoffpeak}
  \end{center}
\end{figure}

The amount of systematic variation present can be measured by calculating the $\chi^2$ statistic 
\begin{equation}\label{eqn:chisquared}
\chi^2 = \sum_{i=1}^{M} \frac{ ( \varepsilon_i - \hat{\varepsilon} )^2}{ (\Delta(\varepsilon_i))^2}
\end{equation}%
for each cutset. If the variation was entirely statistical, an average $\chi^2$ value of $M -1$ would be expected. The value of $M-1$ is 62 for on-peak MC 
 and 26 for off-peak MC. For the cutsets trialled, $\chi^2$
is much higher than this: for example 124.04 and 47.94 in Fig.~\ref{fig:OnPeakMuVar} and \ref{fig:OffPeakMuVar}. The on- and off-peak distributions
 of $\chi^2$ values for the 61880 cutsets are shown in Fig.~\ref{fig:X2Muonpeak} and \ref{fig:X2Muoffpeak}.

All cutsets trialled have a lower $\chi^2$ statistic than the existing \B Counting muon-pair selection. For the same 
samples of on- and off-peak MC, the
 $\chi^2$ statistics of the isBCMuMu tag were 1016.5 and 402.25 respectively.

\section{Estimating Systematic Uncertainty}\label{sec:MuSystematics}
From Fig.~\ref{fig:X2Muonpeak} and \ref{fig:X2Muoffpeak} it is clear that all cutsets have $\chi^2$ values indicating variation beyond statistical 
fluctuations. There is however, a broad range of $\chi^2$ values --- this indicates that some choices of cutset are objectively better (i.e. less
sensitive  to systematic effects) than others.

To gauge the size of the systematic uncertainty we look for $d_0 $ in a modified $\chi^2$  statistic

\begin{equation}\label{eqn:modchi}
\chi'^{2}(d)  =  \sum_{i=1}^{M} \frac{ ( \varepsilon_i - \hat{\varepsilon} )^2}{ (\Delta(\varepsilon_i))^2 + d^2} \qquad (d\geq 0)
\end{equation}
such that 
\begin{equation}\label{eqn:modchi2}
\chi'^{2}(d_0) = M-1.
\end{equation}

Equation (\ref{eqn:modchi}) is equivalent to a polynomial of order $2M$ in $d$. Clearly for $d=0$, $\chi'^{2}(0) = \chi^2 >M -1$ and for large $d$, $\chi'^{2}(d) \to 0$. 
So, since (\ref{eqn:modchi}) is continuous,  by the Intermediate Value Theorem there is at least one real $d_0$ satisfying  (\ref{eqn:modchi2}). 

In practice, $d_0$ is found numerically by incrementing $d$ from zero in steps of   $1\times 10^{-6}$ until $\chi'^{2}(d)< M-1$. The value of $d$ at this point is taken
to be equal to $d_0$.  The ranges of  $d_0$ values for different cutsets are  $(5.57$ -- $9.27)\times  10^{-4}$ in on-peak MC and $(2.93$ -- $7.89)\times  10^{-4}$
in off-peak MC. 

The total uncertainty for a particular cut-set is then the sum in quadrature  of statistical uncertainty ($1/\sqrt{w}$) and systematic uncertainty ($d_0$). The `optimal'
set of cuts is the one which causes the lowest uncertainty in the total number of \B mesons. 

It is important to note the difference between the 
uncertainty used for the purposes of optimization and the value of the systematic uncertainty in data. The optimization procedure seeks to find the cutsets
(muon-pair and hadronic) which cause the lowest uncertainty in \B Counting due to time-variation of MC.  This amounts to finding the cutsets which are
most stable in time. 

The systematic uncertainty for MC does not correspond directly to the same uncertainty in data. Steps are taken in data to reduce the effect of varying efficiencies, and
the association of on-peak data with only the most recent  $R_{o\!f\!\!f}$  value reduces this effect to a large degree. The estimation of systematic uncertainties in 
data is discussed in Chapter \ref{chap:systuncert}.

\section{Variation in  $\kappa_{\mu}$}
The results above can be used to evaluate the value of $\kappa_{\mu}$ as defined in (\ref{eqn:Appkappadef}).  In past \B Counting code, $\kappa$, the product of 
$\kappa_{\mu}$ and $\kappa_{X}$ has been set equal to unity.  $\kappa_{\mu}$ is equal to the ratio of off- to on-peak muon-pair cross-sections, multiplied
by the respective efficiencies. 

The theoretical cross-section for electron annihilation to a muon-pair  $\epem \to \mumu$ at CM energy squared of $s$ is approximately \cite{Peskin:1995}:
\begin{equation}
\sigma_{\mu} = \frac{4 \pi \alpha^2}{3s} \approx \frac{86.8 \  \textrm{nb}}{s \ (\textrm{in GeV}^2)}.
\end{equation}
So, since the muon-pair cross-section scales inversely with $s$, the expected value of $\kappa_{\mu}$ is approximately:
\begin{equation}\label{eqn:kappamuapprox}
  \kappa_{\mu}=  \frac{\epsilon'_{\mu}\sigma'_{\mu}}{ \epsilon_{\mu}\sigma_{\mu}}  \approx \frac{10.58^2}{10.54^2} \approx 1.0076.
\end{equation}

To find the time-variation in $\kappa_{\mu}$ some care needs to be taken when choosing
time bins. On- and off-peak running occur at different times, and often the gap between off-peak running periods is quite large. To manage this,
the running time is divided into 24 unequal bins such that each contains periods of both off- and on-peak running. This makes it possible to calculate a value for
 $\kappa_{\mu}$ in each time bin.

Variation in  $\kappa_{\mu}$ is calculated using the same method used to calculate variations in mu-pair efficiency.
The statistical uncertainty in $\kappa_{\mu}$ includes the effects of the independent statistical uncertainties of on- and off-peak mu-pair MC efficiencies
which are given by (\ref{eqn:effstatvar}).  There is also a small contribution from the uncertainty in the MC cross-sections, $\Delta(\sigma_{\mu})$ and
 $\Delta(\sigma'_{\mu})$.

\begin{figure}[ht] 
  \begin{center}
    \includegraphics[width=4.3in]{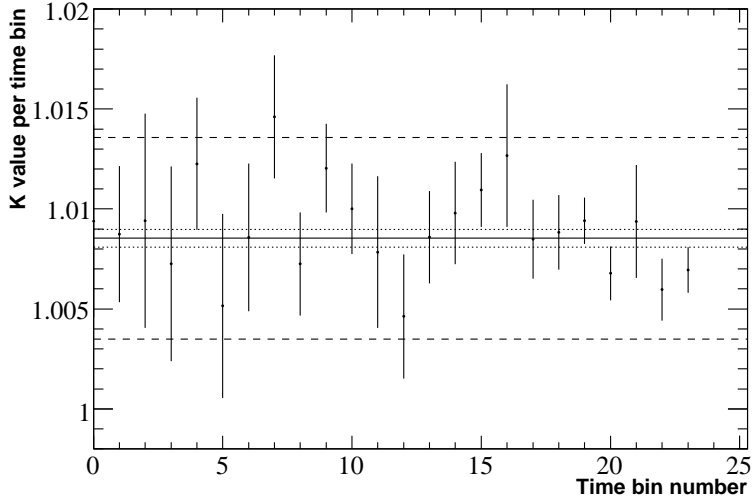}
    \caption[Time-variation of $\kappa_{\mu}$.]{Time-variation of  $\kappa_{\mu}$ for the cutset shown in Fig.~\ref{fig:OnPeakMuVar} and \ref{fig:OffPeakMuVar}.
     The $\chi^2 $ value is 21.01, $M$ is 24 and the weighted average is 1.00854. The dashed lines show $\pm 0.5\%$ from the mean.}
      \label{fig:KappaMuVar}
  \end{center}
\end{figure}

\begin{equation}\label{eqn:bionomial}
\Delta(\varepsilon_i \sigma_{\mu} ) = [ (\Delta(\varepsilon_i) \sigma_{\mu})^2 + (\varepsilon_i \Delta(\sigma_{\mu}))^2 ]^{\frac{1}{2}}
\end{equation}
and similarly for off-peak MC. The method used for  calculating  propagation of uncertainties is given in Appendix \ref{appendix:Uncert}. The statistical 
uncertainty  in $\kappa_{\mu}$ for time-period $i$ is then given by (\ref{app:KappaMu}):
\begin{equation}
\Delta(\kappa_{\mu,i}) = \Bigg[ \left( \frac{ \Delta(\varepsilon'_i \sigma'_{ \mu})}{\varepsilon_i \sigma_{\mu}  } \right)^2 + 
  \left( \frac{ \varepsilon'_i \sigma'_{\mu}  \cdot \Delta(\varepsilon_i \sigma_{\mu} ) }{ ( \varepsilon_i \sigma_{\mu} )^2}  \right)^2  \Bigg]^{\frac{1}{2}}.
\end{equation}

The shape of the time distribution for the same cutset considered in Section \ref{Sec:MuOpt} is shown in Fig.~\ref{fig:KappaMuVar}. It is much 
closer to a constant than for  either the on- or off-peak efficiencies, and the variation in time seems to be mostly consistent with only statistical
variation.  This is an indication that  much of the variation in mu-pair MC is due to systematic effects which affect on-peak and off-peak MC similarly. 

\begin{figure}[ht] 
  \begin{center}
    \includegraphics[width=4.3in]{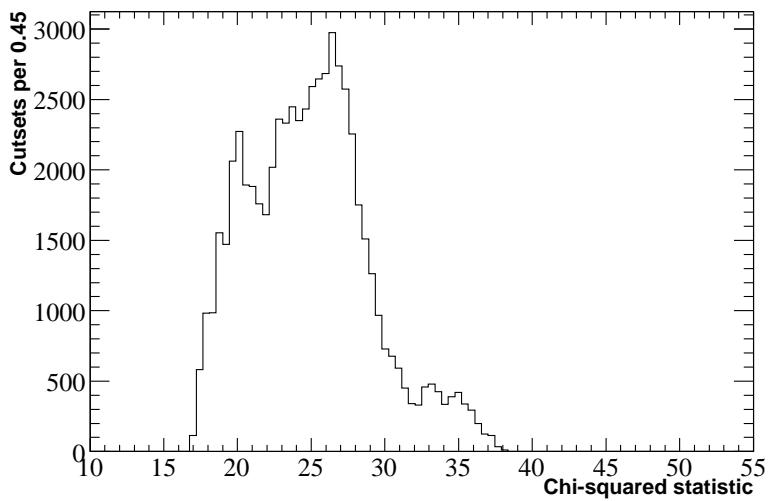}
    \caption[Distribution of $\chi^2$ statistic ($\kappa_{\mu}$).]{Distribution of $\kappa_{\mu}$ $\chi^2$ statistic 
for all trialled cutsets. $M$ 
(the number of degrees of freedom) is 24.}
      \label{fig:X2MuKappa}
  \end{center}
\end{figure}

With 24 periods of time considered, the value of $M-1$ for $\kappa_{\mu}$ is 23. The $\chi^{2}$ statistic can again be calculated, and for the pictured  cutset
 it has a value of 21.01. The weighted mean value of $\kappa_{\mu}$ for this cutset is 
\begin{equation}\label{eqn:kappamuest}
\hat{\kappa}_{\mu} = 1.00854 \pm 0.00044
\end{equation}
which within two (statistical) standard deviations of the theoretical expectation. The 
distribution of $\chi^{2}$ values for all cutsets is shown in Fig.~\ref{fig:X2MuKappa}. Across all cutsets, the mean value of $\chi^{2}$ is 24.88. 

As before, the amount of systematic uncertainty present in each cutset's $\kappa_{\mu}$ variation can be estimated by finding $d_0$ satisfying a modified
(\ref{eqn:modchi2}). We do not allow $d^2$ to be negative, so for cutsets with $\chi^2$ less than 23, $d_0$ is set to zero. Following the same method as above,
the range of $d_0$ values is between zero and 0.00172. This is added in quadrature with the statistical uncertainty to give the total uncertainty
on each cutset's $\kappa_{\mu}$ value.

%% file: chapter6.tex
\chapter{Hadronic Selection}

The criteria for good hadronic selection requirements were mentioned in Section \ref{BCountMethod}. This chapter describes proposed modifications to the
hadronic part of the \B Counting code. 

The change to the NeutralsBC list with a  minimum neutral energy requirement reduces the sensitivity to backgrounds before any optimization is performed 
on the hadronic selection. Several of the other cut quantities are changed to match the new lists. Instead  of requiring at least three tracks
on the GoodTracksLoose list,  the cut is instead based on ChargedTracksBC. Previously,  the values of ETotal and R2All were based on the 
the GoodNeutralLooseAcc and ChargedTracksAcc lists. Now, they are  calculated by combining all candidates on the ChargedTracksBC and NeutralsBC lists. 

Similarly, PrimVtxdr and PrimVtxdz (the distance of the event's primary vertex from the beamspot in $r$ and $z$) were formerly calculated by finding 
the vertex of all tracks on ChargedTracksAcc. In this thesis, these quantities are calculated from tracks on an intermediate charged list, which
is equivalent to  ChargedTracksBC before any cuts on DOCA (in either $xy$ or $z$)  are made. 

\section{Monte Carlo Simulation}
The method for optimizing hadronic selection is similar to the mu-pair selection optimization discussed in the previous chapter. 
Hadronic continuum events are simulated with $\epem \to \uubar / \ddbar/\ssbar$ and $\epem \to \ccbar$
on-peak and off-peak MC using the JETSET MC generator \cite{Sjostrand:1995}. As with muon-pair MC, once the MC events are generated,  detector response is simulated with 
GEANT 4 and background frames are added.  

In data, there are additional sources of hadronic continuum events --- Initial State Radiation (ISR) and two-photon events. These are small in comparison
to the other continuum events and their effects are discussed in later  chapters. The present optimization procedure seeks to 
minimize the time-sensitivity of  hadronic selection for $uds$ and $\ccbar$ Monte Carlo only.  

The production cross-sections for the continuum MC used in the optimization  procedure are given in Table \ref{table:contXS}. These values are
based on 1 million  on- and off-peak events of each type simulated by the KK2F generator.   \BB events are simulated in a similar way to continuum MC, 
but with the EvtGen MC generator  \cite{Ryd:2003}. The efficiency of the isBCMultiHadron tag for 
\BB MC is approximately 96\%.

 \begin{table}[ht] 
   \caption[Continuum MC cross-sections.]{
    \label{table:contXS}Continuum MC cross-sections used in optimization. } 
   \centering
  \begin{tabular}{|c|c|c|}
    \hline
    Continuum  &\ On-Peak $\sigma$ &\  Off-Peak $\sigma$  \\
    MC Type  &\    &\  \\
    \hline
     \uubar &\ $1.5735 \pm 0.0013$ nb  &\ $ 1.5847 \pm 0.0013$ nb \\
      \ddbar  &\  $0.3931 \pm 0.0003$ nb  &\ $0.3959 \pm 0.0003$ nb \\ 
      \ssbar &\ $0.3751 \pm 0.0003$ nb  &\ $0.3778 \pm 0.0003$ nb \\
      \ccbar &\  $1.2929 \pm 0.0004$ nb &\  $1.3021 \pm 0.0004 $ nb\\
    \hline
  \end{tabular}

\end{table}

\section{Cut on Total Energy}
As in Section \ref{FloatMasspair}, hadronic selection should function in the same
way, regardless of CM energy.  Of the isBCMultiHadron cuts listed in Table  \ref{table:isBC}, the one affected by this is the cut on total event
energy in the laboratory frame. This dependence is removed by scaling by eCM. If the cut value is scaled by the on-peak beam energy, 
the adjusted requirement is:
\begin{equation}
  \frac{\textrm{ETotal}}{\textrm{eCM}} > \frac{4.5 \ \textrm{GeV}}{10.58 \  \textrm{GeV}} \approx 0.4253.
\end{equation}

The cut on total lab energy is primarily in place to reject beam-gas and two-photon events. The adjusted cut slightly increases the isBCMultiHadron efficiency 
for off-peak continuum events.

\section{Cut on Highest-Momentum Track}
Events of the type $\epem \to \ellell$ where $\ell$ is either $e$ or $\mu$ typically have at least one track with approximately half the beam
energy in the CM frame. This does not happen for \B decays, as the \B mesons very quickly decay into other particles well within the detector.
Typically each \B meson decays into two tracks, each  with approximately a quarter of the beam energy. 
One possible way to increase the Bhabha and mu-pair rejection  in the hadronic selection without affecting  \BB efficiency is to  
reject events where the highest momentum track  has close to half the beam energy. 

Some poorly-tracked particles  have an artificially  high momentum (for example a momentum of $4 \pm 2$ GeV/$c$), and 
SVT-only tracks
are especially prone to this. At short distances chord and arc length are very similar, so to precisely measure a track's curvature (and thus momentum) 
it is usually necessary to make measurements outside the SVT, some distance from the beamspot. A minimum requirement for this is at least one
hit in the drift chamber. 

In hadronic selection optimization, one variable is a new proposed cut on p1Mag, the CM momentum
of the highest-momentum track. If the highest-momentum track has at least one hit in the drift chamber and its momentum is
greater than the cut value, the event is rejected. 

\begin{figure}[ht] 
  \begin{center}
    \includegraphics[width=4.0in]{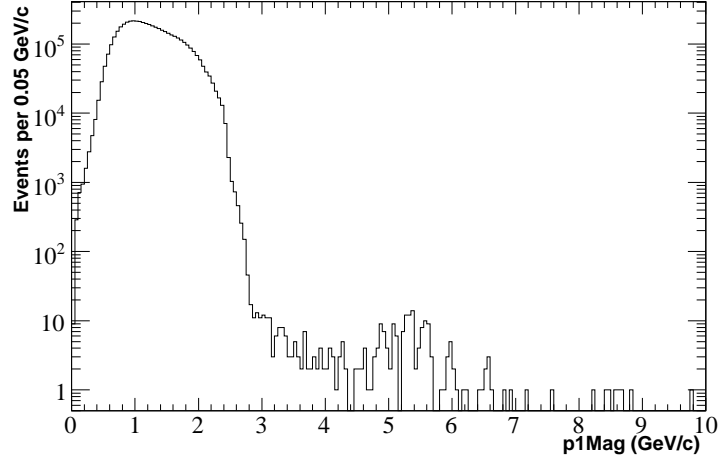}
    \caption[Distribution of p1Mag for a sample of \BB Monte Carlo.]{Distribution of p1Mag for a sample of \BB Monte Carlo events where
the highest-momentum track has at least one DCH hit.}
      \label{fig:p1MagBB}
  \end{center}
\end{figure}

Figure \ref{fig:p1MagBB} shows the distribution of p1Mag for \BB MC events where the highest-momentum track has at least one drift chamber hit. The
sample contains 1.8 million Run 6 MC events.  Clearly, above 3.0 GeV/$c$ the number of events is negligible.

In the optimization procedure, the cut on p1Mag (with the requirement on  DCH hits) is varied between 3.0  and 5.0 GeV/$c$ or 
is turned off altogether. 

\section{Optimization of Cuts}
Similarly to mu-pair selection, there are certain cuts necessary for isBCMultiHadron which allow little flexibility. These include the cuts on
PrimVtxdr and PrimVtxdz. The reasons for the current cut values on these quantities are given in \cite{Hearty:2000}. Both cuts are designed to reject backgrounds,
while maintaining high \BB efficiency.  They especially reject events which have a vertex at the beam-wall. 

\begin{figure}[p] 
  \begin{center}
    \includegraphics[width=4.3in]{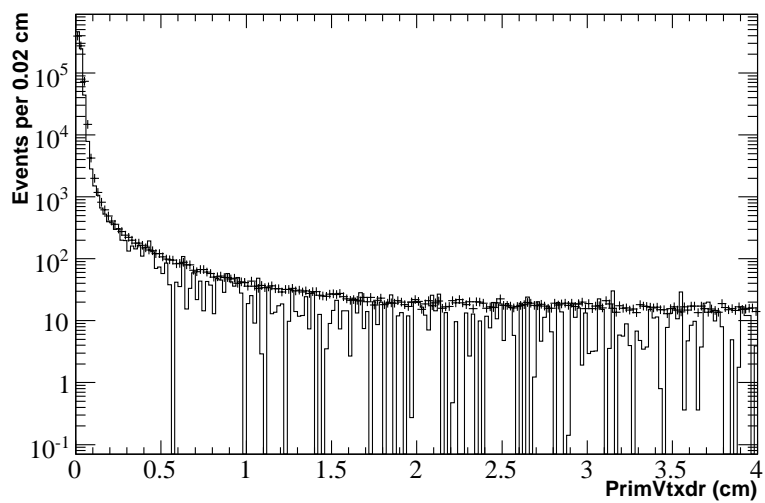}
    \caption[PrimVtxdr for a data sample from Run 6, overlaid with  MC.]
	    {PrimVtxdr for a data sample from Run 6 (solid line), overlaid with \BB MC.}
      \label{fig:PrimVtxdrDataMC}
  \end{center}
\end{figure}

\begin{figure}[p] 
  \begin{center}
    \includegraphics[width=4.3in]{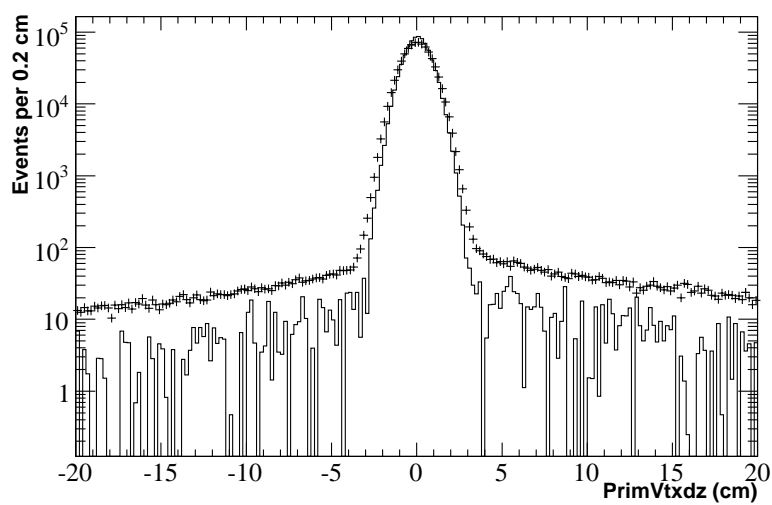}
    \caption[PrimVtxdz for a data sample from Run 6, overlaid with  MC.]
	    {PrimVtxdz for a data sample from Run 6 (solid line), overlaid with \BB MC.}
      \label{fig:PrimVtxdzDataMC}
  \end{center}
\end{figure}

Figures \ref{fig:PrimVtxdrDataMC} and \ref{fig:PrimVtxdzDataMC} show the distributions of these two quantities for a sample of \BB data 
(i.e.  off-peak subtracted from on-peak)  from Run 6 overlaid with \BB MC. The data sample contains the \BB events 
from a sample of 2.81 million on-peak events which pass all isBCMultiHadron requirements apart from those on PrimVtxdr and PrimVtxdz.  The cuts used
in hadronic selection (PrimVtxdr $<$ 0.5 cm, $|$PrimVtxdz$| <$ 6.0 cm) are far from the sharp peaks of these distributions and movements
around these cut values have an almost  negligible effect on the MC time-variation or \BB efficiency.


To ensure the selected events are hadronic, the requirements of BGFMultiHadron and having at least three good tracks are necessary. There are two remaining
cut quantities (in addition to p1Mag) which can be optimized: R2All and ETotal. The total number of cutsets trialled 
during the optimization
procedure is $36 \times 26 \times 12 = 11232$.
\begin{table}[ht]  
  \caption[Cut optimization of isBCMultiHadron.]{
    \label{table:MHcutvariation}Cut optimization of isBCMultiHadron. }    
  \centering
  \begin{tabular}{|c|c|c|}
    \hline
    Quantity  &\  No. of increments  &\ Range \\
    \hline
    R2All &\ 36  &\ 0.45 -- 0.80 \\
    ETotal/eCM  &\  26  &\  $\big( (2.5$ -- $5.0)/10.58 \big)$ \\ 
    p1Mag ($\geq 1$  DCH hit) &\ 12 &\ 3.0 -- 5.0 GeV/$c$, or no cut \\
    \hline
  \end{tabular}

\end{table}

The optimization procedure is very similar to the mu-pair optimization. There are 66 months of $uds$ and $\ccbar$ on-peak MC, 27 of $uds$ off-peak MC and
26 of $\ccbar$ off-peak MC. The statistical uncertainties are again treated binomially and are given by (\ref{eqn:effstatvar}). The weighted mean
efficiency and $\chi^2$ statistic are given by (\ref{eqn:wmean}) and  (\ref{eqn:chisquared}) respectively. 

\begin{figure}[p] 
  \begin{center}
    \includegraphics[width=3.8in]{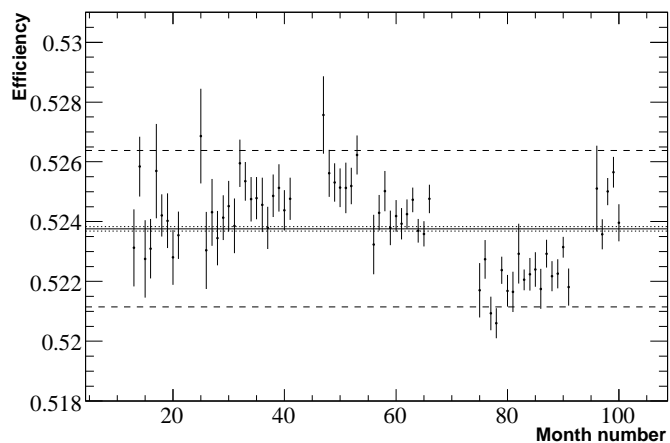}
    \caption[Time-variation of on-peak $uds$ MC.]{Time-variation of on-peak $uds$ MC efficiency.  Details of cutset choice are in text. 
    The $\chi^2$ value is 372.46 and $M$ (the number of degrees of freedom) is 66. The weighted mean efficiency is  $0.52376 \pm 0.00008$. 
The dashed lines show $\pm 0.5\%$ from the mean. }
      \label{fig:udsOnMCVar}
  \end{center}
\end{figure}

\begin{figure}[p] 
  \begin{center}
    \includegraphics[width=3.8in]{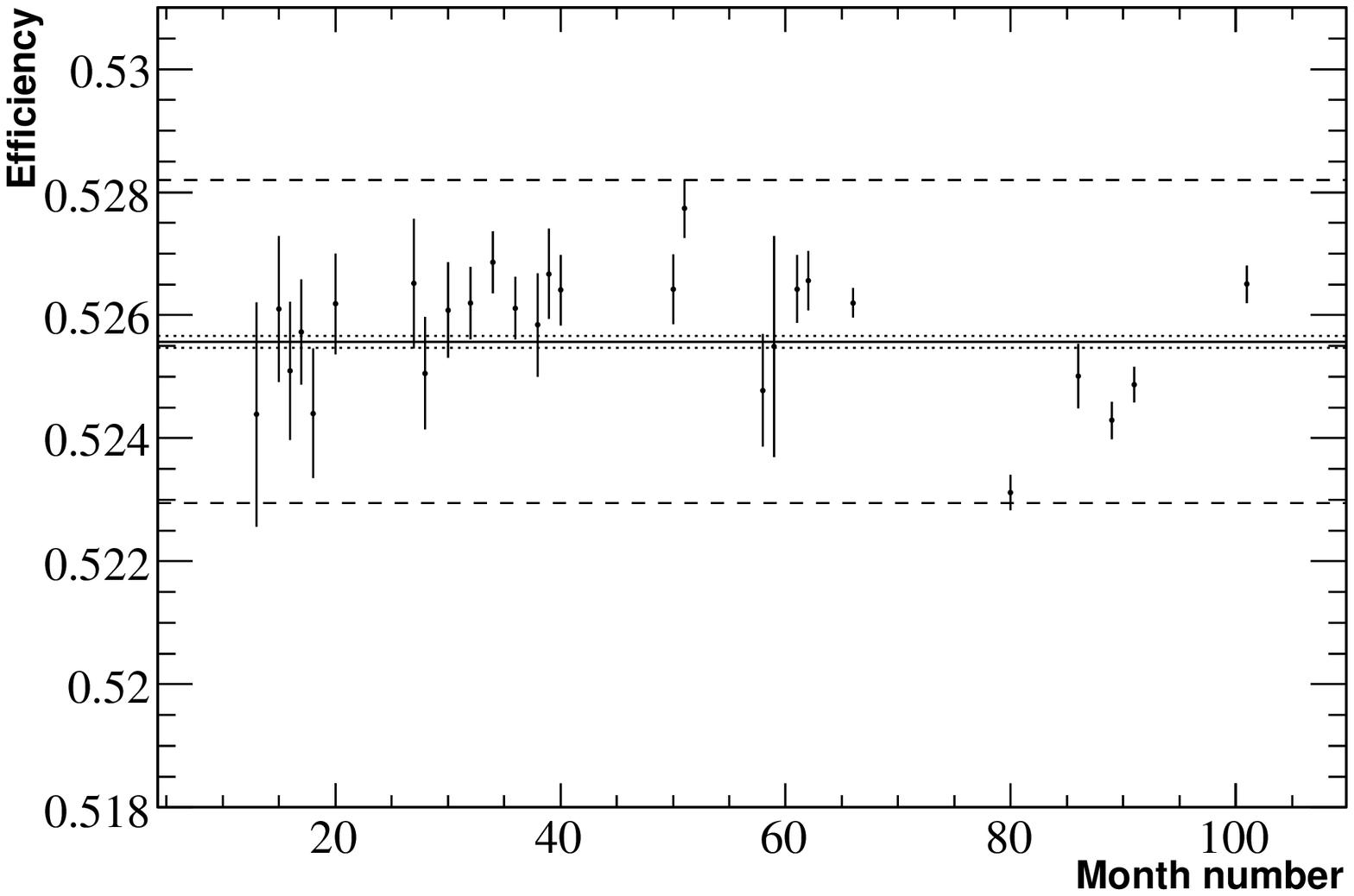}
    \caption[Time-variation of off-peak $uds$ MC.]{Time-variation of off-peak $uds$ MC efficiency. Details of cutset choice are in text. 
    The $\chi^2$ value is 160.71  and $M$ (the number of degrees of freedom) is 27. The weighted mean efficiency is $0.52557 \pm 0.00010$. 
The dashed lines show $\pm 0.5\%$ from the mean.}
      \label{fig:udsOffMCVar}
  \end{center}
\end{figure}

\begin{figure}[p] 
  \begin{center}
    \includegraphics[width=3.8in]{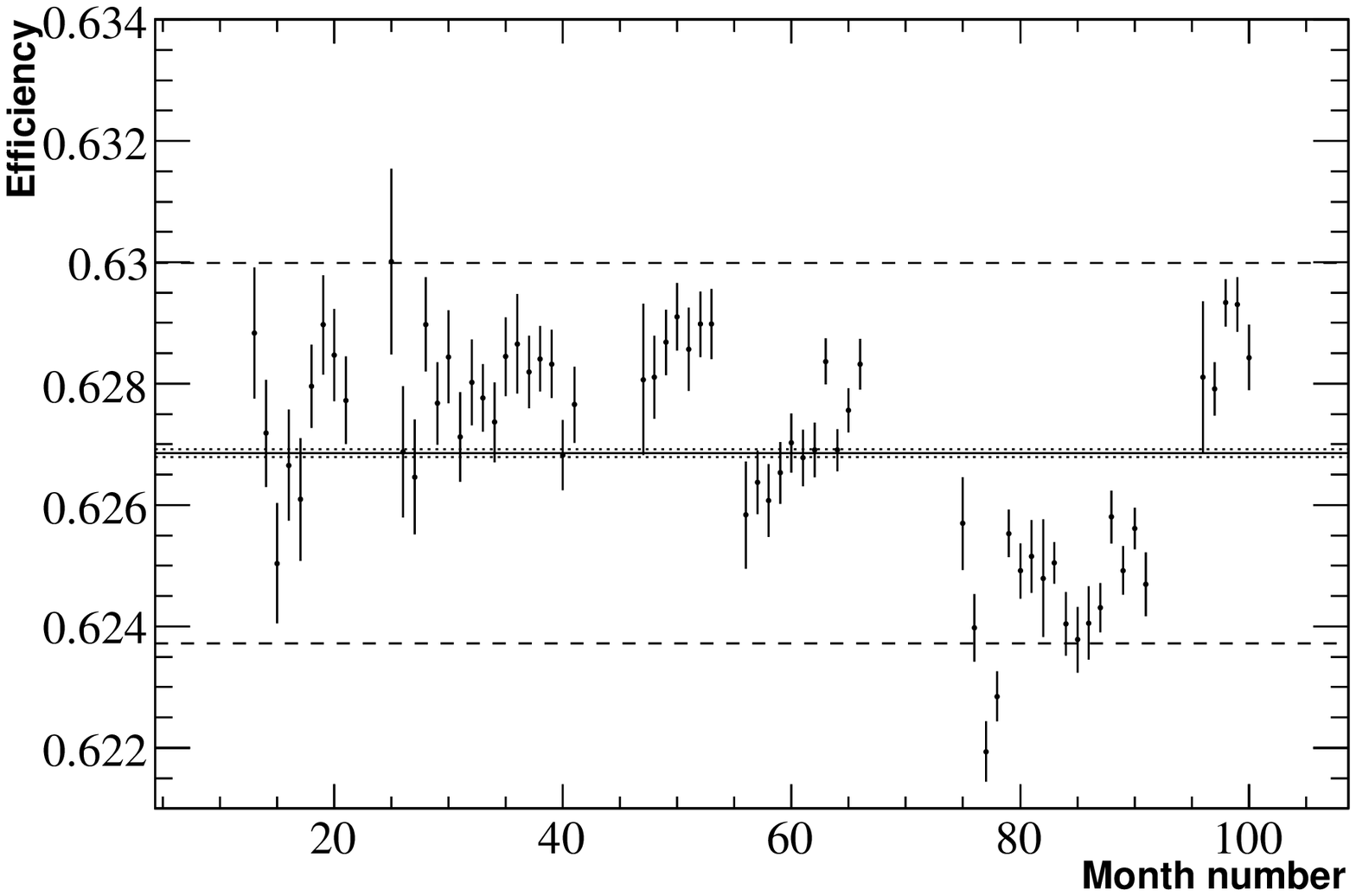}
    \caption[Time-variation of on-peak \ccbar MC.]{Time-variation of on-peak \ccbar MC efficiency.  Details of cutset choice are in text. 
    The $\chi^2$ value is 842.20 and $M$ (the number of degrees of freedom) is 66. The weighted mean efficiency is $0.62686 \pm 0.00007$.
The dashed lines show $\pm 0.5\%$ from the mean.}
      \label{fig:ccbarOnMCVar}
  \end{center}
\end{figure}

\begin{figure}[p] 
  \begin{center}
    \includegraphics[width=3.8in]{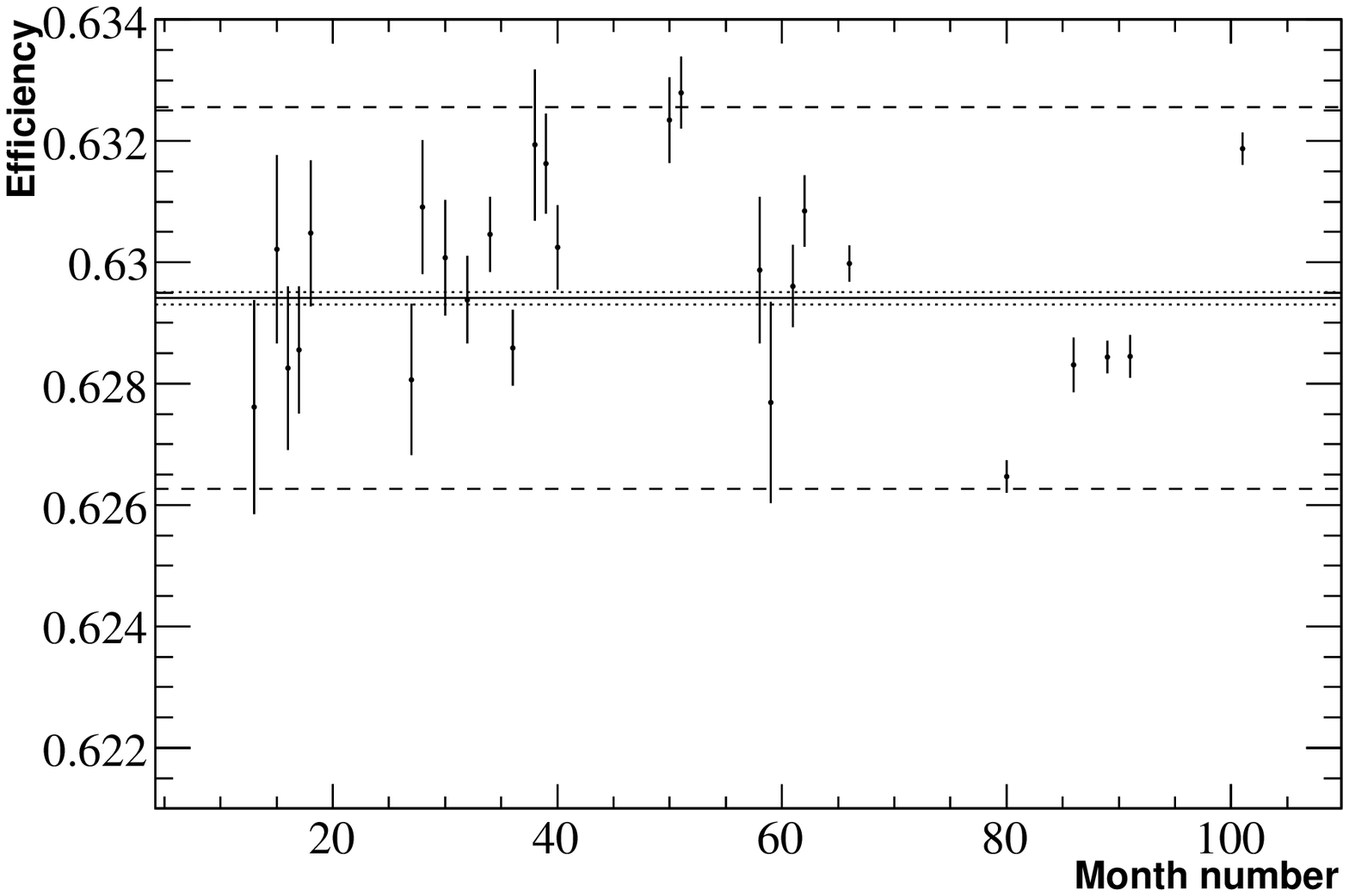}
    \caption[Time-variation of off-peak \ccbar MC.]{Time-variation of off-peak \ccbar MC efficiency. Details of cutset choice are in text. 
    The $\chi^2$ value is 319.67 and $M$ (the number of degrees of freedom) is 26.  The weighted mean efficiency is  $0.62941 \pm 0.000010$.
The dashed lines show $\pm 0.5\%$ from the mean.}
      \label{fig:ccbarOffMCVar}
  \end{center}
\end{figure}

The time variations of these four  types of MC are shown in Fig.~\ref{fig:udsOnMCVar} --  \ref{fig:ccbarOffMCVar}. The cutset pictured has fairly typical
time-variation: R2All $< 0.50$, (ETotal/eCM) $> 0.42533$ and p1Mag $<3.8$ GeV/$c$. These plots all display obvious systematic time-variation. The 
shapes of the distributions are reasonably similar across all four plots and they also resemble the muon-pair efficiency variation 
of Fig.~\ref{fig:OnPeakMuVar} and \ref{fig:OffPeakMuVar}. 

\begin{figure}[p] 
  \begin{center}
    \includegraphics[width=4.0in]{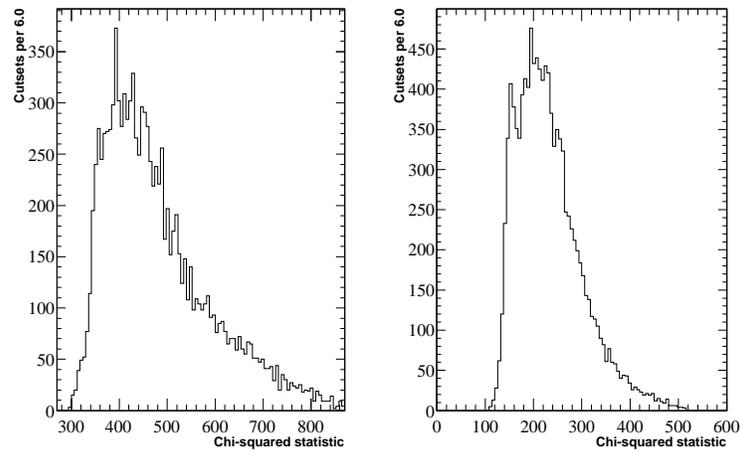}
    \caption[Distribution of $\chi^2$ statistic for $uds$ MC.]{Distribution of $\chi^2$ statistic for trialled 
cutsets (left) for $uds$ on-peak MC and (right) for $uds$ 
off-peak MC. The mean values of these distributions are 479.86 and 232.47 respectively. $M$ is 66 and 27 respectively. }
      \label{fig:X2uds}
  \end{center}
\end{figure}

\begin{figure}[p] 
  \begin{center}
    \includegraphics[width=4.0in]{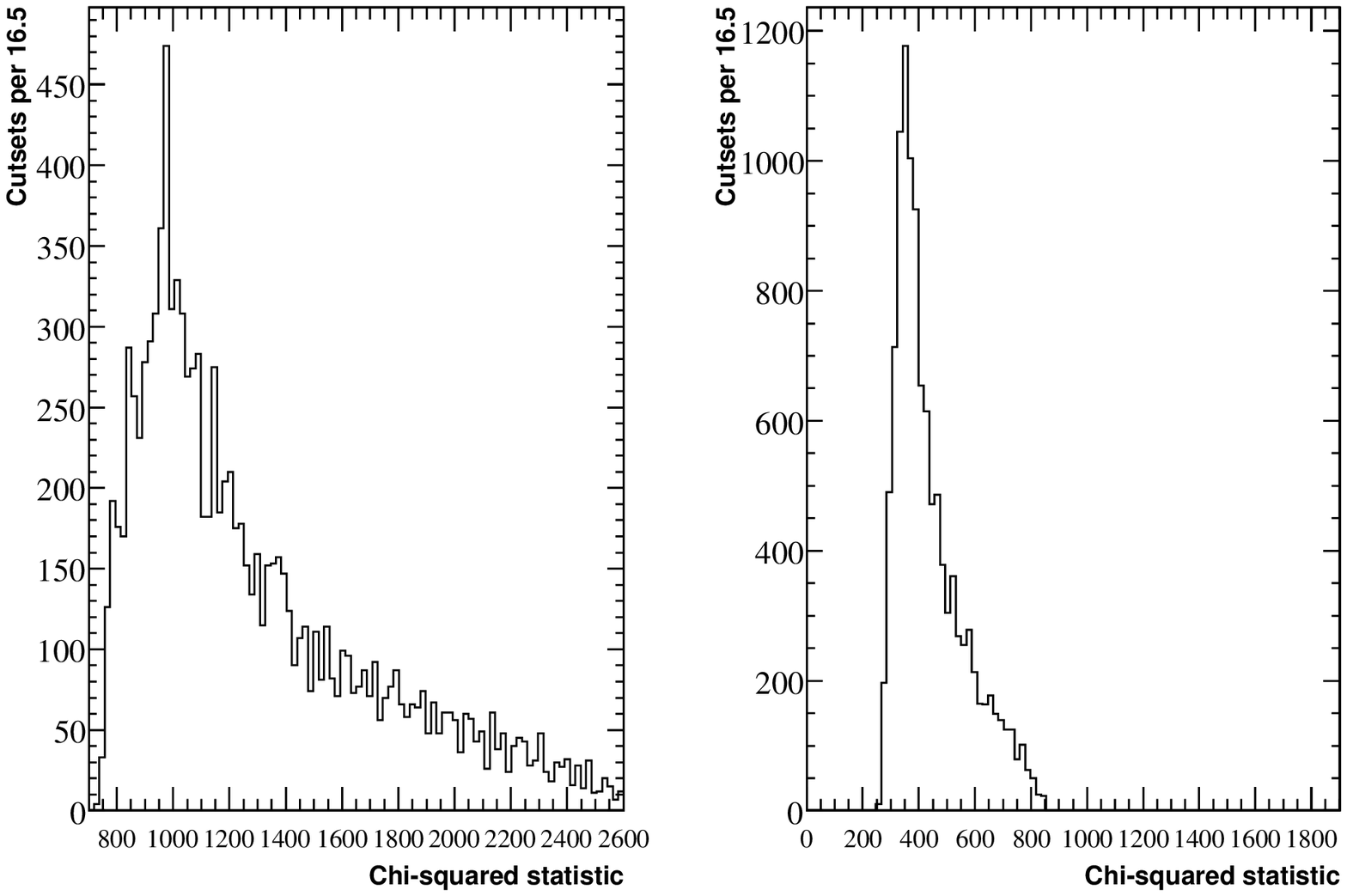}
    \caption[Distribution of $\chi^2$ statistic for \ccbar MC.]{Distribution of $\chi^2$ statistic for trialled 
cutsets (left) for \ccbar on-peak MC and
(right) for \ccbar off-peak (right) MC.
    The mean values of these distributions are 1292.82 and 441.28 respectively. $M$ is 66 and 26 respectively.}
      \label{fig:X2ccbar}
  \end{center}
\end{figure}

The distributions of $\chi^2$ for all 11232 cutsets are shown in Fig.~ \ref{fig:X2uds} and \ref{fig:X2ccbar}. These are all clearly well above the
$M-1$ values for each type of MC. 

A  comparison between the new cutsets and the existing \B Counting selector, isBCMultiHadron can be made. The on-peak
(off-peak) $\chi^2$ statistic for this tag is 1081.82 (527.27) for $uds$ MC  and 2669.32 (1304.13) for \ccbar MC.  Almost all  cutsets for each MC
type have $\chi^2$ values below these which indicates an improvement on average.



\section{Estimating Systematic Uncertainty}
To estimate the systematic uncertainties in continuum MC time-variation, we  use the procedure outlined in Section \ref{sec:MuSystematics}.
We look for $d_0 $ in the modified $\chi^2$  statistic of (\ref{eqn:modchi}) such that
\begin{equation}\label{eqn:modchi2a}
\chi'^{2}(d_0) = M-1.
\end{equation}

Again, $d_0$ is found numerically by incrementing $d$ from zero in steps of   $10^{-6}$ until $\chi'^{2}(d)< M-1$. The 
ranges of  $d_0$ values for different cutsets are  $(1.07$ -- $1.69)\times  10^{-3}$ in on-peak $uds$  MC, $(0.67$ -- $1.59)\times  10^{-3}$
in off-peak $uds$ MC,  $(1.60$ -- $2.13)\times  10^{-3}$ in on-peak \ccbar  MC and  $(1.29$ -- $2.16)\times  10^{-3}$
in off-peak \ccbar MC.  

As with muon-pairs, the total uncertainty for a particular cutset is then the sum in quadrature  of statistical uncertainty ($1/\sqrt{w}$) and
 systematic uncertainty ($d_0$). The `optimal' set of hadronic cuts is determined at the same time as the `optimal' set of muon-pair cuts by finding the
combination  which causes the lowest uncertainty in the total number of \B mesons.

\section{Variation in  $\kappa_{X}$}\label{sec:kappacont}

Using the efficiencies and cross-sections found above, we can evaluate the continuum part of $\kappa$, defined in (\ref{eqn:Appkappadef}). The cross-sections
 of $\epem \to \uubar / \ddbar/\ssbar/\ccbar$ \ all scale inversely with $s$ (the CM energy squared): 
\begin{equation}\label{eqn:kappacontest}
\kappa_{X} =   \frac{ \varepsilon_{X}\sigma_{X}}{ \varepsilon'_{X}\sigma'_{X}} \approx \frac{10.54^2}{10.58^2} \approx 0.9925. 
\end{equation}
Note that in this approximation, the product of $\kappa_{X}$ and $\kappa_{\mu}$ (defined in  (\ref{eqn:kappamuapprox})) is exactly unity.

There are other  processes in data which can pass the hadronic selection and can affect  $\kappa_{X}$ but are not included in this approximation. 
These include two photon events, Initial State Radiation, cosmic rays, Bhabhas and beam gas events. The hadronic selection is designed to minimize
many of these (especially cosmic rays and beam gas events) and the effects of these sorts of contamination are discussed in Chapter \ref{chap:contamination}.

To ensure there is on- and off-peak $uds$ and \ccbar MC in every time bin, the time-variation of $\kappa_{X}$ is divided into 23 unequal time
periods. As with $\kappa_{\mu}$, the statistical uncertainty of $\kappa_{X}$ in each time bin includes  the  statistical uncertainties 
of each type of  MC given by (\ref{eqn:effstatvar}) (e.g. $\Delta(\varepsilon_{uds,i})$) and a contribution from the uncertainty in the MC cross-sections 
(e.g. $\Delta(\sigma_{uds})$).

To an excellent approximation, 
\begin{equation}
\kappa_{X} =   \frac{ \varepsilon_{uds}\sigma_{uds} + \varepsilon_{\ccbar}\sigma_{\ccbar} }{ \varepsilon'_{uds}\sigma'_{uds} + 
\varepsilon'_{\ccbar}\sigma'_{\ccbar}} 
\end{equation}
and the statistical uncertainty in $\kappa_{X}$ for each time period is given by (\ref{app:KappaCont}):

\begin{eqnarray}
\Delta(\kappa_{X, i}) & = & \Bigg[  
\left( \frac{ \Delta(\varepsilon_{uds,i} \sigma_{uds})}{\varepsilon'_{uds,i} \sigma'_{uds} + \varepsilon'_{\ccbar ,i} \sigma'_{\ccbar}} \right)^2 +{}
\left( \frac{ \Delta(\varepsilon_{\ccbar ,i} \sigma_{\ccbar})}{\varepsilon'_{uds,i} \sigma'_{uds} + \varepsilon'_{\ccbar ,i} \sigma'_{\ccbar}} \right)^2  +  \nonumber\\
& & {}+ \left( \frac{  \Delta(\varepsilon'_{uds,i} \sigma'_{uds})\cdot (\varepsilon_{uds,i} \sigma_{uds} + \varepsilon_{\ccbar ,i} \sigma_{\ccbar})      }
     {(\varepsilon'_{uds,i} \sigma'_{uds} + \varepsilon'_{\ccbar ,i} \sigma'_{\ccbar)})^2} \right)^2 + \nonumber\\
& & {}+  \left( \frac{  \Delta(\varepsilon'_{\ccbar ,i} \sigma'_{\ccbar})\cdot (\varepsilon_{uds,i} \sigma_{uds} + \varepsilon_{\ccbar ,i} \sigma_{\ccbar})      }
   {(\varepsilon'_{uds,i} \sigma'_{uds} + \varepsilon'_{\ccbar ,i} \sigma'_{\ccbar)})^2} \right)^2 \Bigg]^{\frac{1}{2}}
\end{eqnarray}
where, following (\ref{eqn:bionomial}), 
\begin{equation}
\Delta(\varepsilon_{uds,i} \sigma_{uds} ) = [ (\Delta(\varepsilon_{uds,i}) \sigma_{uds})^2 + (\varepsilon_{uds,i} \Delta(\sigma_{uds}))^2 ]^{\frac{1}{2}}
\end{equation}
and similarly for other MC types. 

\begin{figure}[htp] 
  \begin{center}
    \includegraphics[width=3.8in]{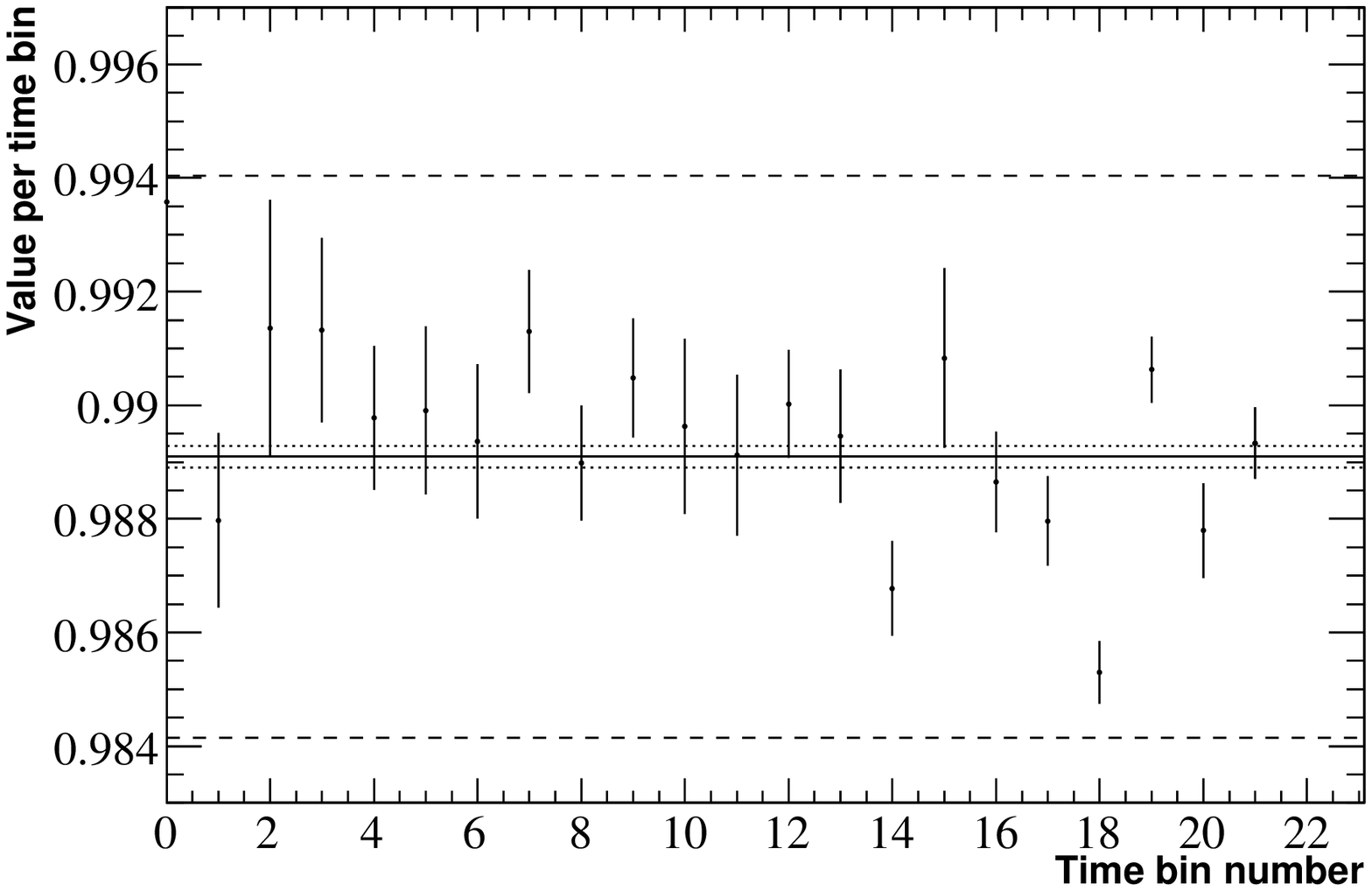}
    \caption[Time-variation of $\kappa_{X}$.]{Time-variation of  $\kappa_{X}$ for the cutset shown in Fig.~\ref{fig:udsOnMCVar} -- \ref{fig:ccbarOffMCVar}.
     The $\chi^2 $ value is 93.82, $M$ is 23  and the weighted average is 0.98909. The dashed lines show $\pm 0.5\%$ from the mean.}
      \label{fig:KappaMHVar}
  \end{center}
\end{figure}

\begin{figure}[ht] 
  \begin{center}
    \includegraphics[width=4.0in]{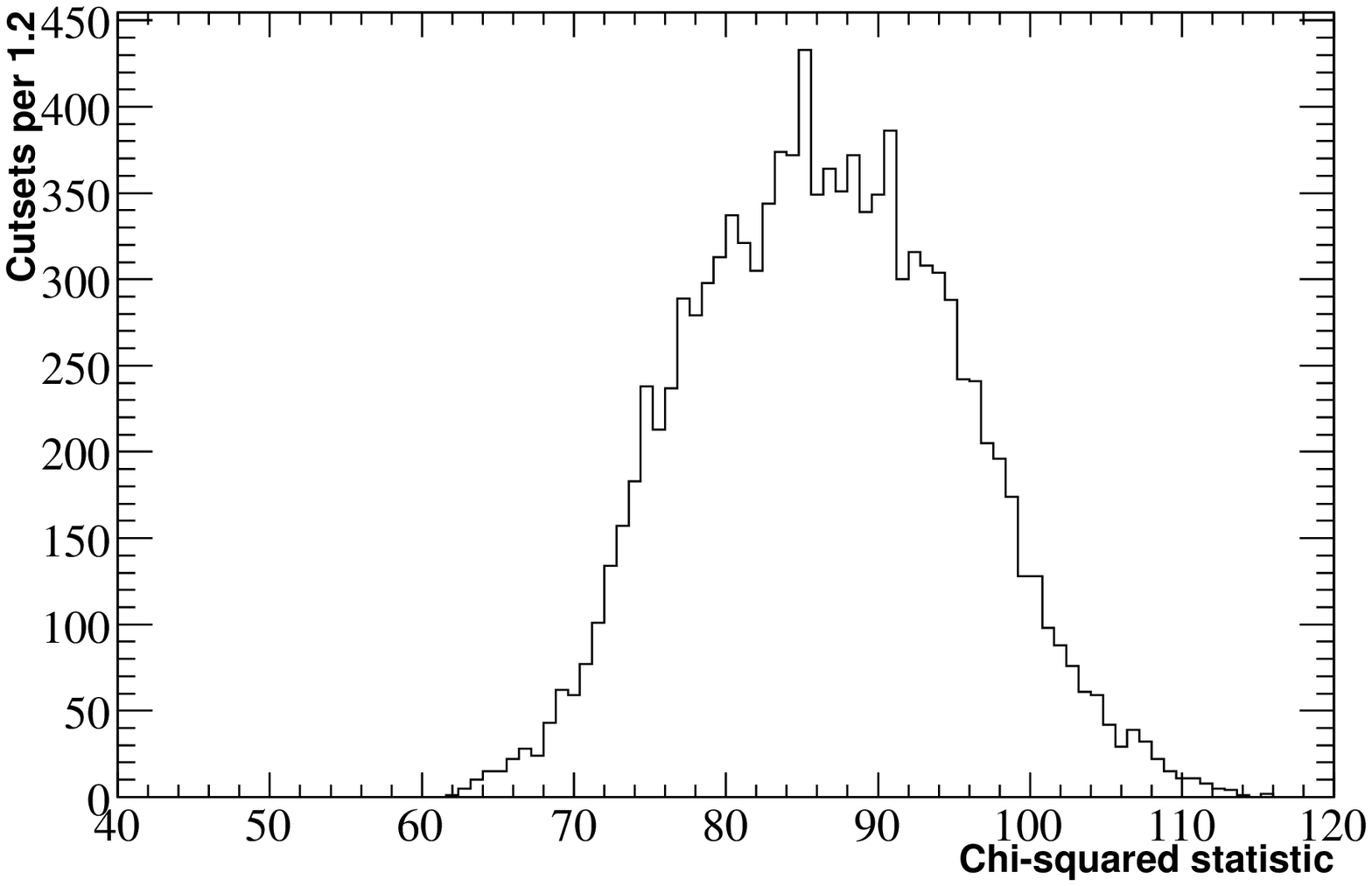}
    \caption[Distribution of $\chi^2$ statistic ($\kappa_{X}$).]{Distribution of $\kappa_{X}$ $\chi^2$ statistic for all trialled cutsets. $M$ is 23.}
      \label{fig:X2MHKappa}
  \end{center}
\end{figure}

The time-variation of  $\kappa_{X}$  for the same cutset shown in Fig.~\ref{fig:udsOnMCVar} -- \ref{fig:ccbarOffMCVar}  is shown in Fig.~\ref{fig:KappaMHVar}. 
The variation appears to be largely (though not entirely) statistical and systematic effects are not as obviously apparent as they are in  the individual MC efficiencies.
The  $\chi^{2}$ statistic
of $\kappa_{X}$  can be calculated, and for every cutset trialled it is above $M-1$, so  there is a small amount of variation which cannot be explained
by statistical uncertainties alone. For the cutset pictured in Fig.~\ref{fig:KappaMHVar}, the $ \chi^{2}$ statistic is  93.82, and the weighted mean value of 
$\kappa_{X}$ is
\begin{equation}
\hat{\kappa}_{X} = 0.98909   \pm  0.00019,
\end{equation}
where the uncertainty quoted is statistical only. The distribution of $\chi^{2}$ values for all cutsets is shown in Fig.~\ref{fig:X2MHKappa}. 
Across all cutsets, the mean value of $\chi^{2}$ is 86.53. The
weighted mean values of $\kappa_{X}$ are in the range $(0.98751$ -- $0.99238)$ with a typical statistical uncertainty of 0.00020 or less. 

\begin{figure}[hbt] 
  \begin{center}
    \includegraphics[width=4.0in]{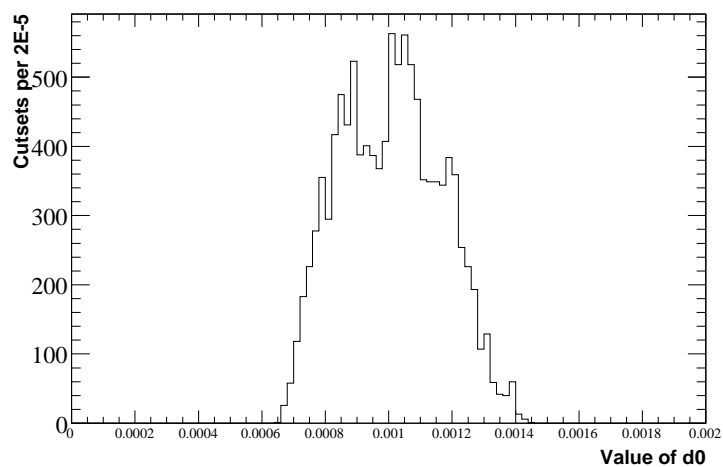}
    \caption[Distribution of $\kappa_{X}$ $d_0$ values.]{Distribution of $\kappa_{X}$ $d_0$ values for all trialled cutsets. }
      \label{fig:dMHKappa}
  \end{center}
\end{figure}

To estimate the  systematic uncertainty in each cutset's $\kappa_{X}$, again we  find $d_0$ satisfying
\begin{equation}\label{eqn:modchiMH}
\chi'^{2}(d_0) = M-1,
\end{equation} 
where 
\begin{equation}\label{eqn:modchiMH2}
\chi'^{2}(d)  =  \sum_{i=1}^{M} \frac{ ( \kappa_{X,i} - \hat{\kappa}_{X} )^2}{ (\Delta(\kappa_{X,i} ))^2 + d^2}.
\end{equation}
In this case, $M-1 = 22$. The  $d_0$ values found are between 0.00066  and 0.00144 and the distribution of these values is shown in Fig.~\ref{fig:dMHKappa}. 
The systematic uncertainty in each cutset is  added in quadrature with the 
statistical uncertainty to give the total uncertainty on each cutset's $\kappa_{X}$ value.

%% file: chapter7.tex
\chapter{Cutset Optimization}\label{chap:opt}

The main purpose of the research described in this thesis is to recommend an optimal set of cuts which count \B meson events with a minimal systematic uncertainty. This
uncertainty has contributions from the uncertainties in the counting of both  muon-pairs and hadronic events.  The number of \B meson events counted
with the  optimal cutsets should also agree with the original \B Counting code, which has been in place since 2000 and has a total uncertainty
of $\pm 1.1\%$. 

As was mentioned in Chapter \ref{chap:muonselection}, the \B Counting uncertainty for optimization purposes  found in this chapter is different to the final
estimate of systematic uncertainty in data (given in Chapter \ref{chap:systuncert}). In this chapter we find the \B Counting cutsets which give the lowest sensitivity to 
the time-variation of $uds$, $\ccbar$ and  mu-pair  Monte Carlo efficiencies. We also include the statistical uncertainty of the MC  
cross-sections (included in $\kappa$ uncertainties),

The method used to calculate the propagation of uncertainties is outlined in Appendix \ref{appendix:Uncert}.

\section{\B Counting Uncertainty}
The \B Counting formula of (\ref{eqn:nb}) is:
\begin{eqnarray}\label{eqn:nb7}
N^0_{B}      &  = & \frac{1}{\varepsilon_{B}}  (  N_{H} - N_{\mu}\cdot R_{o\!f\!\!f} \cdot\kappa ) \\
             &  = &   \frac{1}{\varepsilon_{B}}  (  N_{H} - \kappa_{\mu}\cdot\kappa_{X}\cdot R_{\mu}\cdot N'_{H} )\label{eqn:nb7b}
\end{eqnarray}
where  
\begin{equation}
R_{\mu} \equiv \frac{N_{\mu}}{N'_{\mu}},
\end{equation}
 $N'_{H}$ is the number of off-peak  hadronic events counted and $\kappa_{\mu}$ and $\kappa_{X}$ are defined in (\ref{eqn:Appkappadef}).
The subscripts ``$H$'' and ``$X$'' in an off-peak sample are interchangeable
since the off-peak hadronic data contains no \BB events from \FourS decays. 

The uncertainty in the number of \B meson events, $\Delta(N^0_{B})$   is  given by:
\begin{equation}\label{eqn:BBuncert}
\Delta(N^0_{B})^2 = \sum_i \left(\frac{\partial N^0_{B} }{\partial x_i} \Delta x_i\right)^2.
\end{equation}
Here, $x_i$ represents each quantity with associated uncertainty (such as $\varepsilon_{uds}$ or $\kappa_{\mu}$).  To estimate the uncertainty 
in the number of \B meson events in a sample of on-peak data of luminosity $\mathcal{L}$ with continuum subtracted by an off-peak sample
of luminosity $\mathcal{L}'$, we can rewrite (\ref{eqn:nb7}) as:
\begin{eqnarray}
N^0_{B} & = & \frac{1}{\varepsilon_{B}} \big[ \left( H  + H_O \right) -  \kappa_{\mu}\kappa_{X}R_{\mu} \left(H' + H'_O\right) \big]
\end{eqnarray}
where 
\begin{equation}
H \equiv \mathcal{L}(\varepsilon_{uds}\sigma_{uds} +\varepsilon_{\ccbar}\sigma_{\ccbar} +   \varepsilon_{B}\sigma_{B} )
\end{equation}
and
\begin{equation}
H' \equiv \mathcal{L'}(\varepsilon'_{uds}\sigma'_{uds} +\varepsilon'_{\ccbar}\sigma'_{\ccbar} ).
\end{equation}
$H_O$ and $H'_O$ (the subscript ``O'' represents ``other'' event-types) 
are the numbers of on- and off-peak  events which are not  \BB, uds or \ccbar   but pass the hadronic cuts. These terms are discussed in detail
in Chapter \ref{chap:contamination}, but are neglected during this optimization procedure. We can also write
\begin{equation}
R_{\mu} \equiv \frac{N_{\mu}}{N'_{\mu}} = \frac{\mathcal{L}\varepsilon_{\mu}\sigma_{\mu} + M_O}{\mathcal{L}'\varepsilon'_{\mu}\sigma'_{\mu} + M'_O}
\end{equation}
where $M_O$ ($M'_O$) is the number of on-peak (off-peak) events which are not $\epem \to \mumu$ but still pass the muon-pair cuts. We again neglect these
during the optimization.

In calculating the uncertainty in the number of \B mesons, we omit the uncertainties in efficiency of off-peak MC. The time-variations of 
on- and off-peak MC are correlated, and it is important not to double-count the uncertainty. Off-peak data (and hence also MC) is taken only for short 
periods of time  around longer periods of on-peak
 running. For this reason, off-peak running can be viewed as a `snapshot' of the non-\BB continuum at a particular time. 

It is clear the two are not independent, since $\kappa_{\mu}$ and $\kappa_{X}$
both have $\chi^2$ statistics much lower than any  MC type has alone. The time-variation visible in off-peak MC is to a large degree the same as that displayed
 in on-peak MC. For these reasons, the off-peak MC efficiencies are treated as constants in (\ref{eqn:BBuncert}).

Hence, we have:
\begin{eqnarray}\label{eqn:bigBUncert}
\Delta(N^0_{B})^2 & = & \left(\frac{\partial N^0_{B} }{\partial \varepsilon_{uds} } \Delta \varepsilon_{uds}\right)^2 + 
  \left(\frac{\partial N^0_{B} }{\partial \varepsilon_{\ccbar} } \Delta \varepsilon_{\ccbar}\right)^2 + \nonumber \\
 & & +  \left(\frac{\partial N^0_{B} }{\partial \varepsilon_{B} } \Delta \varepsilon_{B}\right)^2  + \left(\frac{\partial N^0_{B} }{\partial \varepsilon_{\mu} }
  \Delta \varepsilon_{\mu}\right)^2 + \nonumber \\
  & & + \left(\frac{\partial N^0_{B} }{\partial \kappa_{X} } \Delta \kappa_{X} \right)^2 +
  \left(\frac{\partial N^0_{B} }{\partial \kappa_{\mu} } \Delta \kappa_{\mu} \right)^2
\end{eqnarray}
where
\begin{equation}\label{eqn:totalerror}\left(\Delta \varepsilon_{uds} \right)^2 = \left(\Delta (\varepsilon_{uds} )_{\textrm{stat}} \right)^2 + (d_{0, uds})^2
\end{equation}
and similarly for $\varepsilon_{\ccbar}$, $\varepsilon_{\mu}$,  $\kappa_{X}$ and $\kappa_{\mu}$. Equation (\ref{eqn:totalerror}) gives the combined
statistical and systematic uncertainty found during the previous optimization procedure and has the same form as the denominator of the $\chi'^2$ statistic in
(\ref{eqn:modchi}). 

Once an optimal cutset is found, a value for $\Delta \varepsilon_{B}$ can be estimated by comparing \BB MC to data. During the optimization procedure, 
we estimate its value by setting:
\begin{equation}\label{eqn:effBError}
\Delta \varepsilon_{B}  = 0.2(1-\hat{\varepsilon}_B). 
\end{equation}
This is chosen to favour cutsets with higher \BB efficiency. For a cutset with $\hat{\varepsilon}_B$ of 0.96, (\ref{eqn:effBError}) gives
an uncertainty in $\hat{\varepsilon}_B$  of around 0.83\%. 

The differential quantities of (\ref{eqn:bigBUncert}) are:
\begin{eqnarray}\label{eqn:BUncertDiffs}
\frac{\partial N^0_{B} }{\partial \varepsilon_{uds} } & = & \frac{1}{\varepsilon_B}\mathcal{L}\sigma_{uds} \label{eqn:bigarray1} \\
\frac{\partial N^0_{B} }{\partial \varepsilon_{\ccbar} } & = & \frac{1}{\varepsilon_B}\mathcal{L}\sigma_{\ccbar} \\
\frac{\partial N^0_{B} }{\partial \varepsilon_{\mu} } & = & -\frac{1}{\varepsilon_B}\kappa_{\mu}\kappa_{X}
\frac{\mathcal{L}\sigma_{\mu}}{\mathcal{L}'\varepsilon'_{\mu}\sigma'_{\mu} } H'\\
\frac{\partial N^0_{B} }{\partial \kappa_{X}} & = & -\frac{1}{\varepsilon_B}\kappa_{\mu}R_{\mu}H' \\
\frac{\partial N^0_{B} }{\partial \kappa_{\mu}} & = & -\frac{1}{\varepsilon_B}\kappa_{X}R_{\mu}H'\label{eqn:bigarray2}
\end{eqnarray} 
and
\begin{eqnarray}
\frac{\partial N^0_{B} }{\partial \varepsilon_{B} } & = & - 
\frac{1}{\varepsilon_B^2}\big(\mathcal{L}\left(\varepsilon_{uds}\sigma_{uds} + \varepsilon_{\ccbar}\sigma_{\ccbar}\right) - \nonumber \\
 & &  \kappa R_{\mu}\mathcal{L}'\left(\varepsilon'_{uds}\sigma'_{uds} + \varepsilon'_{\ccbar}\sigma'_{\ccbar}\right)\big).
\end{eqnarray}
This last term is very small (and would be exactly zero if \B Counting were perfect) so is negligible in comparison to the other terms. 

As was mentioned above, the expression  (\ref{eqn:bigBUncert}) is useful for finding an optimal set of \B Counting cuts, but it will not be used
to estimate the systematic uncertainty of data.

\section{Optimization}
The optimal mu-pair and hadronic \B Counting cutsets are those  which minimize  $(\Delta N^0_B )/ N^0_{B}$ (the relative uncertainty in $N^0_B$) while meeting 
the selection goals
outlined in Section \ref{BCountMethod}. As a cross-check, the number of \B mesons counted in a sample by any pair of cutsets can be compared to that given
by the existing \B Counting code.

The total uncertainty of  (\ref{eqn:totalerror}) can be divided into two parts: one  dependent only on uncertainties related to muon-pair selection and the other only on
hadronic uncertainties. The muon-pair part
\begin{equation}
\Delta(N^0_{B})^2_{\mu} =  \left(\frac{\partial N^0_{B} }{\partial \varepsilon_{\mu} }
  \Delta \varepsilon_{\mu}\right)^2 +  \left(\frac{\partial N^0_{B} }{\partial \kappa_{\mu} } \Delta \kappa_{\mu} \right)^2 
\end{equation}
includes terms dependent on $\Delta\varepsilon_{\mu}$ and $\Delta\kappa_{\mu}$. The hadronic part $\Delta(N^0_{B})^2_{H}$  contains  all other terms. The two are not 
entirely independent --- for example the $\Delta\kappa_{\mu}$ term, (\ref{eqn:bigarray2}) depends specifically on the
 number of off-peak hadronic events counted by a chosen set of hadronic cuts.

The only quantities in (\ref{eqn:bigBUncert}) which are constant independent of cutset choice are $\mathcal{L}$, $\mathcal{L}'$ and the cross-sections. The variable
quantities of (\ref{eqn:bigarray1} -- \ref{eqn:bigarray2}) ---    the number of hadronic and   muon-pair data events, efficiencies, $\kappa_{X}$ and $\kappa_{\mu}$ ---
are specific to each cutset trialled during the procedure. 

The first stage of the optimization procedure calculates the number of \B mesons in a sample of on-peak data and the value of  $\Delta(N^0_{B})^2_{\mu}$  
for each muon-pair cutset. Continuum subtraction is provided by a corresponding off-peak data sample.  

To calculate these values, the number of hadronic events in the two samples must be input. Initially 
these are taken to be the number of isBCMultiHadron events in each sample, i.e. the number counted  by the existing \B Counting code. It is also
necessary to input the value of $\varepsilon_B$ for isBCMultiHadron (approximately 0.96) and a value of $\kappa_{X}$, which is initially taken to be the 
theoretical estimate of (\ref{eqn:kappacontest}).
The 61880 muon-pair cutsets are then sorted by  $\Delta(N^0_{B})_{\mu}/N^0_B$.

Assuming it meets the \B Counting selection goals, the `optimal' muon-pair cutset is the one with the lowest relative uncertainty in $N^0_B$.  
The numbers of  muon-pair  on- and off-peak data events counted  by this cutset, along with the value of $\kappa_{\mu}$  are then used as inputs to optimize the
hadronic selection. 

Following the same procedure, the values of $N^0_B$ and $\Delta(N^0_{B})^2_{H}$ are calculated for the  11232 hadronic cutsets, which are then sorted by 
$\Delta(N^0_{B})^2_{H}/N^0_B$.  The values of $\varepsilon_B$, $\kappa_{X}$ and the  number of on- and off-peak hadronic events counted by the optimal hadronic
 cutset are used as inputs to re-optimize
 the muon-pair cutsets.  This procedure continues until the optimal cutset choice stabilizes.

This procedure was performed on each major running period (Runs 1--6). There is typically less than 5\% variation in the value of  
$\Delta(N^0_{B})_{\mu}/N^0_B$  across the first five hundred muon-pair cutsets (this number is around 10\% for the same number of hadronic cutsets). 
With so many nearly-identical cutsets, the optimal choice for each Run varies and
statistical fluctuations can change each cutset's rank.  

Due mainly to these statistical fluctuations, no single cutset is  consistently the optimal choice, but some trends emerge. Muon-pair
selection cutsets tend to have a low values of $\Delta(N^0_{B})_{\mu}/N^0_B$ if:
\begin{itemize}
\item the masspair/eCM cut is high, typically in the range (0.756  -- 0.827),
\item the maxpCosTheta cut is high, typically above 0.735,
\item the acolincm cut is high, typically above 0.11.
\end{itemize}
No trend is obvious for the maxpEmcLab cut. For the hadronic selection, cutsets tend to have  low values of $\Delta(N^0_{B})_{H}/N^0_B$ if:
\begin{itemize}
\item the ETotal/eCM cut is high, typically above 0.378,
\item the p1Mag cut is low, typically less than 4 GeV/$c$. 
\end{itemize}
No trend is obvious for the R2 cut, and cutsets without a p1Mag cut do not generally perform near the top.

\section{Choice of Muon-Pair Selection Cutset}

As mentioned above, many cutsets have almost identical values of $\Delta(N^0_{B})_{\mu}/N^0_B$ or $\Delta(N^0_{B})_{H}/N^0_B$. The change to the new \B 
Counting particle lists (ChargedTracksBC and NeutralsBC) 
has increased stability (in comparison to isBCMuMu and isBCMultiHadron) even before any cuts were varied. Deciding on one cutset to recommend for
\B Counting is to some degree an arbitrary choice. 

The cutset we recommend to select muon-pairs is:
\begin{itemize}
\item BGFMuMu,
\item  masspair/eCM  $> 8.25/10.58$,
\item  acolincm $< 0.13$,
\item  nTracks $\geq 2$,  
\item  maxpCosTheta $< 0.745$, 
\item  either maxpEmcCandEnergy $> 0 $ GeV,  or at least one of the two highest-momentum tracks is identified as a muon in the IFR,
\item  maxpEmcLab $ < 1.1 $ GeV.
\end{itemize}
This can be compared to the isBCMuMu requirements listed in Table \ref{table:isBC}. Note that the meaning of nTracks is slightly changed, as it now refers to
the number of tracks on the ChargedTracksBC list. 

Some properties of this cutset and isBCMuMu for muon-pair MC are given in Table \ref{table:MuCutset}. 

\begin{figure}[htbp] 
  \begin{center}
    \includegraphics[width=4.3in]{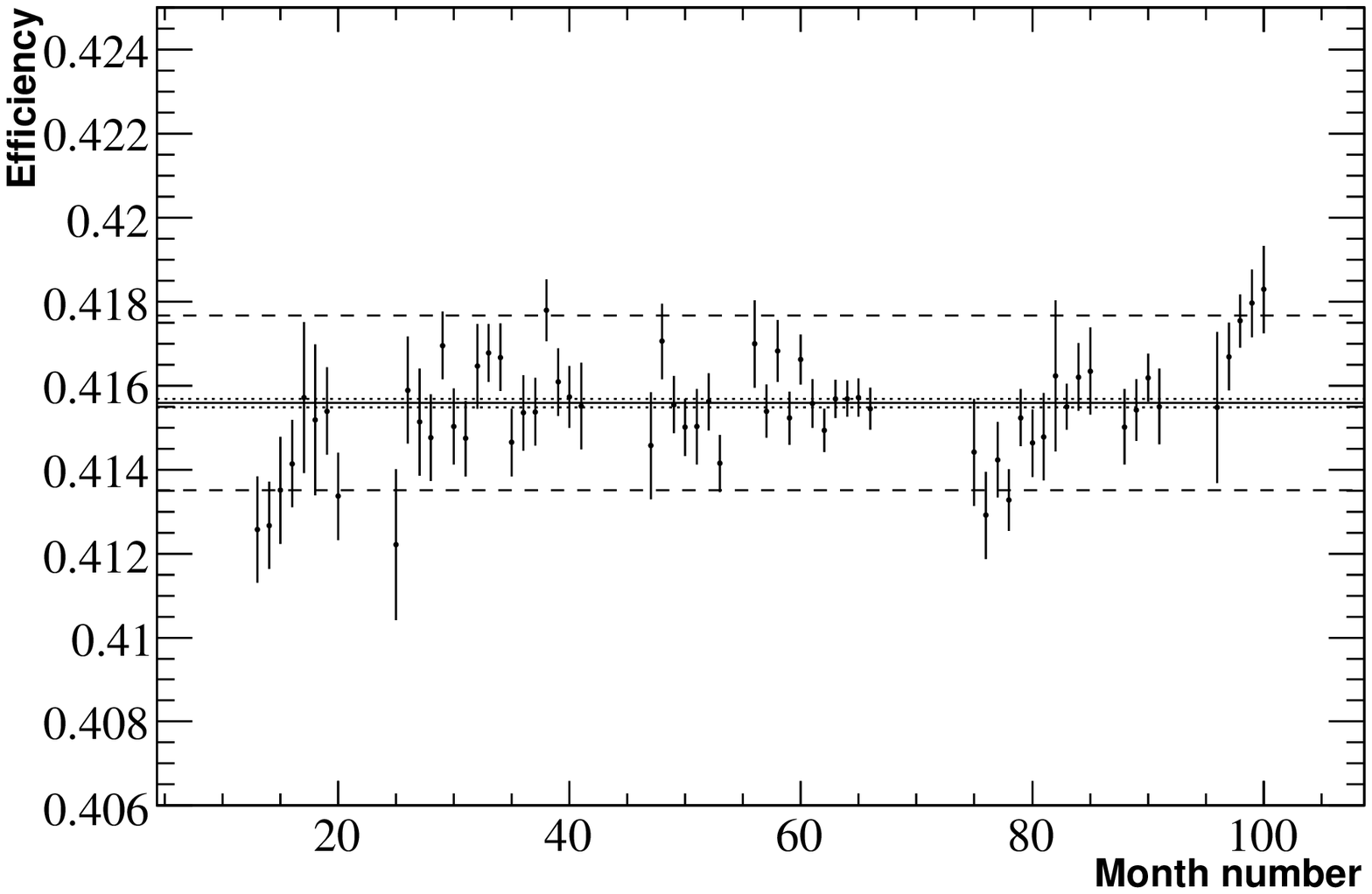}
    \caption[On-peak muon-pair MC efficiency variation for the new cuts.]{On-peak muon-pair MC efficiency variation for 
      the new proposed muon-selection cuts.  The dashed lines show $\pm 0.5\%$ from the mean.}
      \label{fig:MyMuCutsOnVar}
  \end{center}
\end{figure}

\begin{figure}[htbp] 
  \begin{center}
    \includegraphics[width=4.3in]{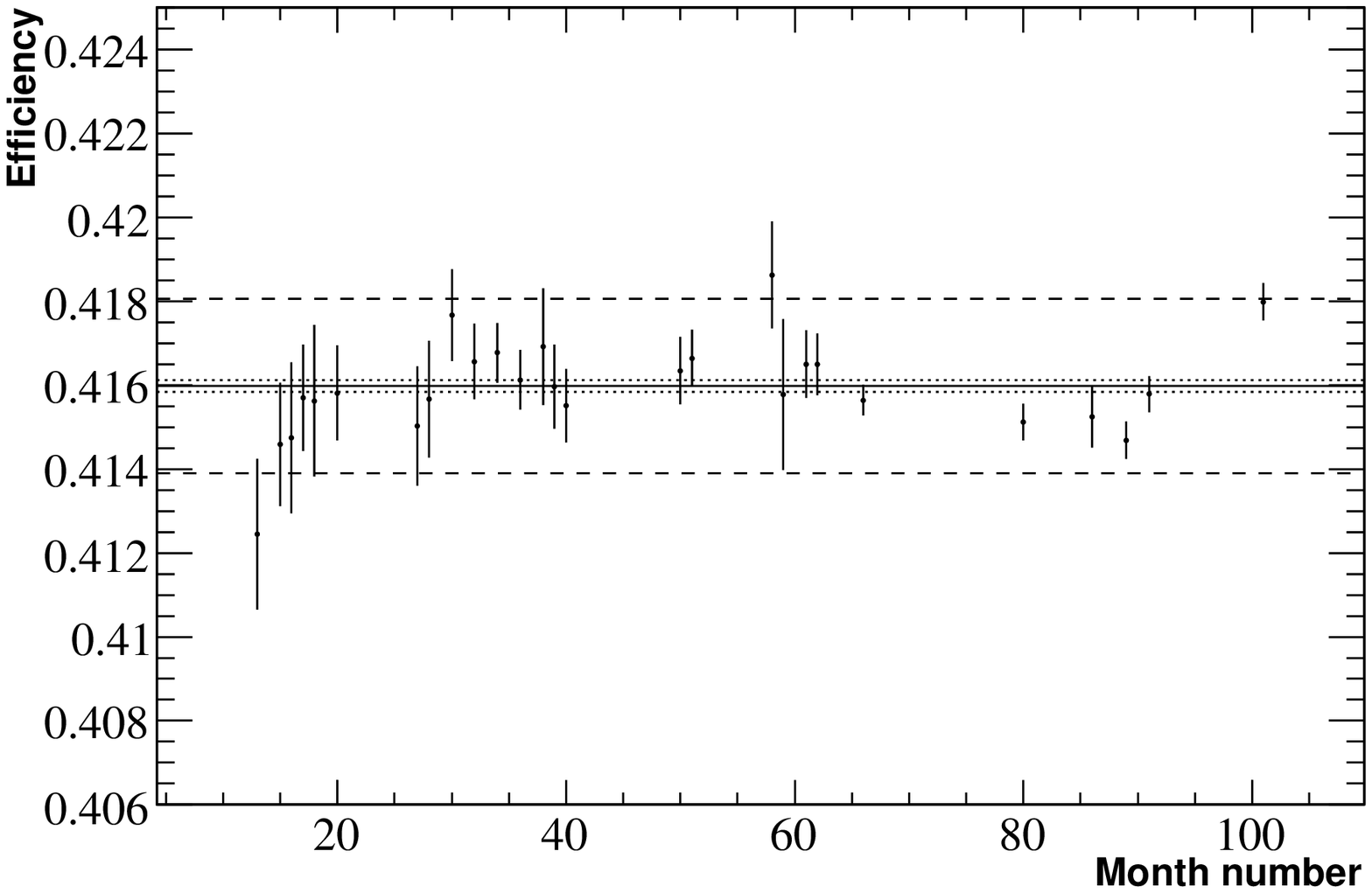}
    \caption[Off-peak muon-pair MC efficiency variation for the new cuts.]{Off-peak  muon-pair MC efficiency variation for 
      the new proposed muon-selection cuts.  The dashed lines show $\pm 0.5\%$ from the mean.}
      \label{fig:MyMuCutsOffVar}
  \end{center}
\end{figure}

\begin{figure}[htbp] 
  \begin{center}
    \includegraphics[width=4.3in]{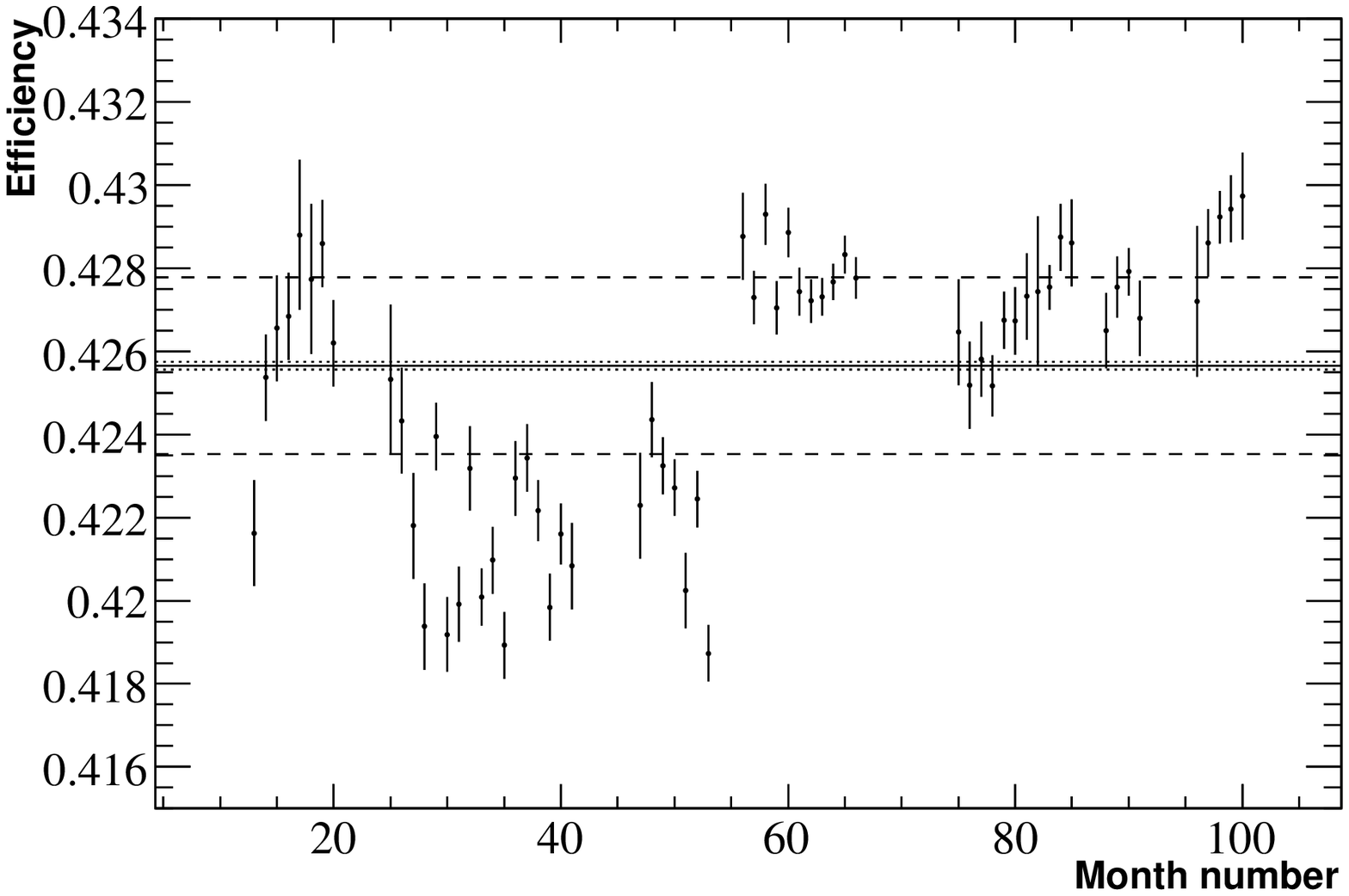}
    \caption[On-peak muon-pair MC efficiency variation for isBCMuMu.]{On-peak  muon-pair MC efficiency variation for 
isBCMuMu.  The dashed lines show $\pm 0.5\%$ from the mean.}
      \label{fig:isBCMuMuOnVar}
  \end{center}
\end{figure}

\begin{figure}[htbp] 
  \begin{center}
    \includegraphics[width=4.3in]{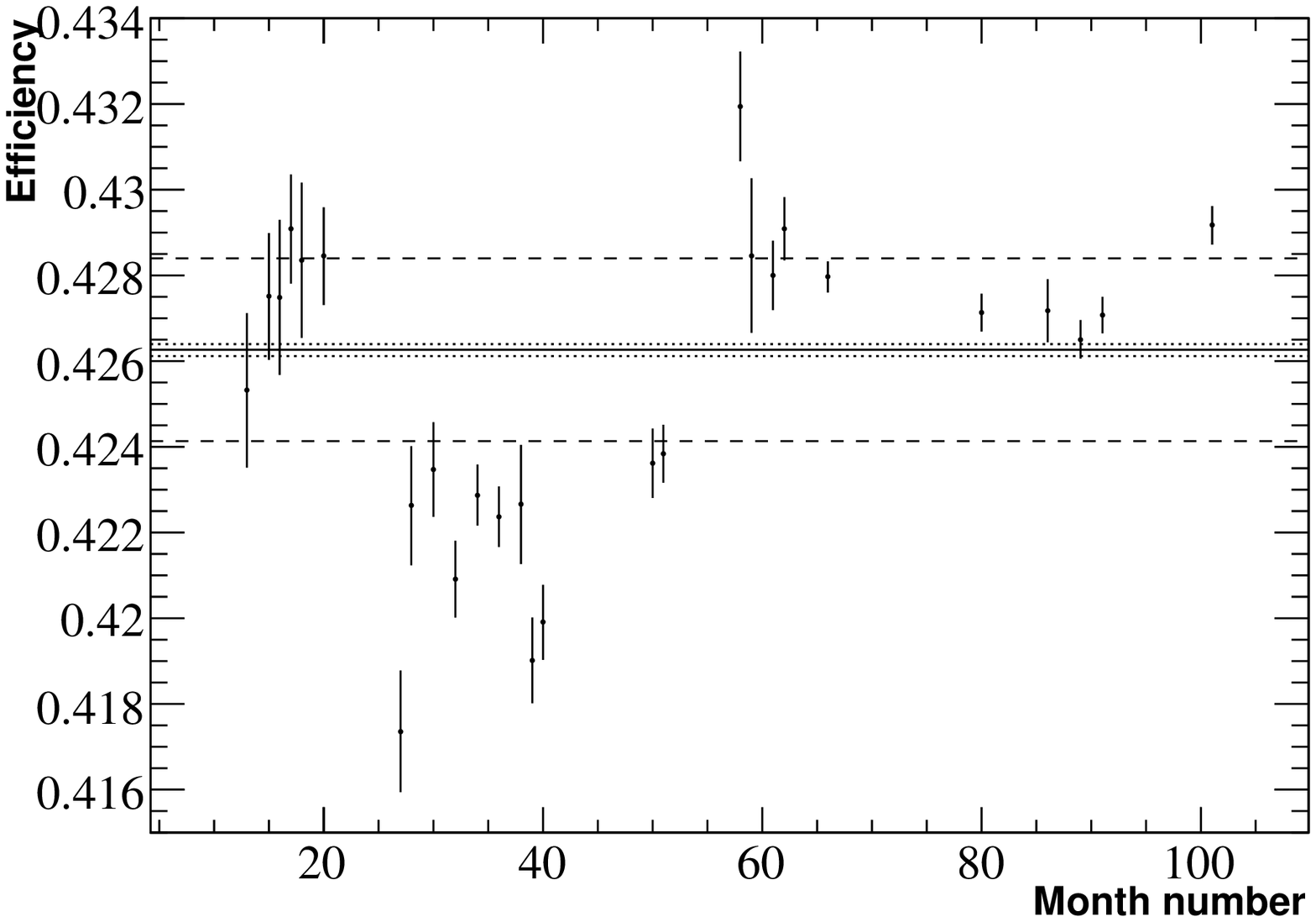}
    \caption[Off-peak muon-pair MC efficiency variation for isBCMuMu.]{Off-peak muon-pair MC efficiency variation for 
isBCMuMu. The dashed lines show $\pm 0.5\%$ from the mean. }
      \label{fig:isBCMuMuOffVar}
  \end{center}
\end{figure}

\begin{figure}[htbp] 
  \begin{center}
    \includegraphics[width=4.3in]{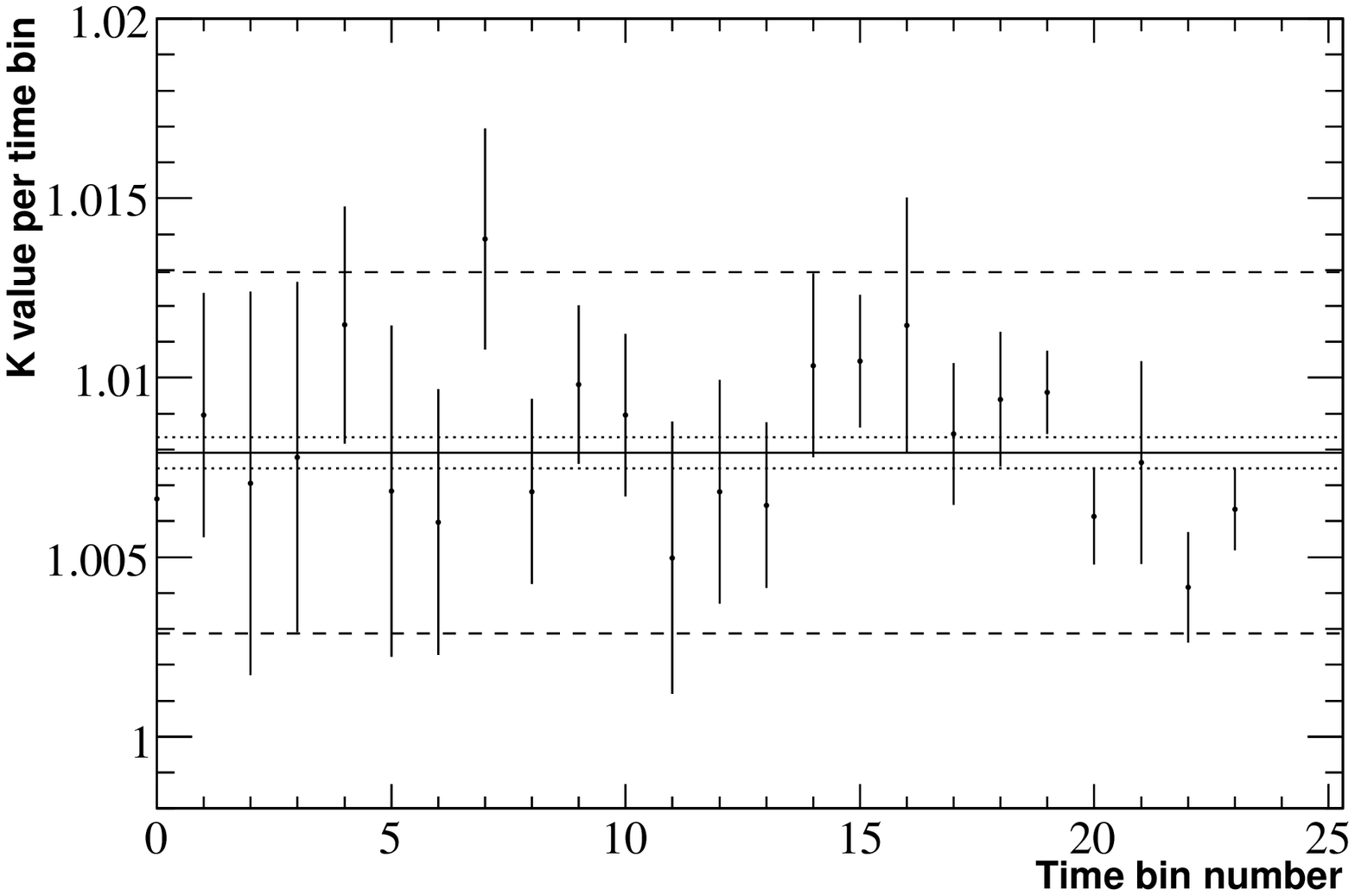}
    \caption[Time-variation of  $\kappa_{\mu}$  for the new cuts.]{Time-variation of  $\kappa_{\mu}$  for the new cutset. 
 The dashed lines show $\pm 0.5\%$ from the mean.}
      \label{fig:BestCutsKappaMu}
  \end{center}
\end{figure}

\begin{figure}[htbp] 
  \begin{center}
    \includegraphics[width=4.3in]{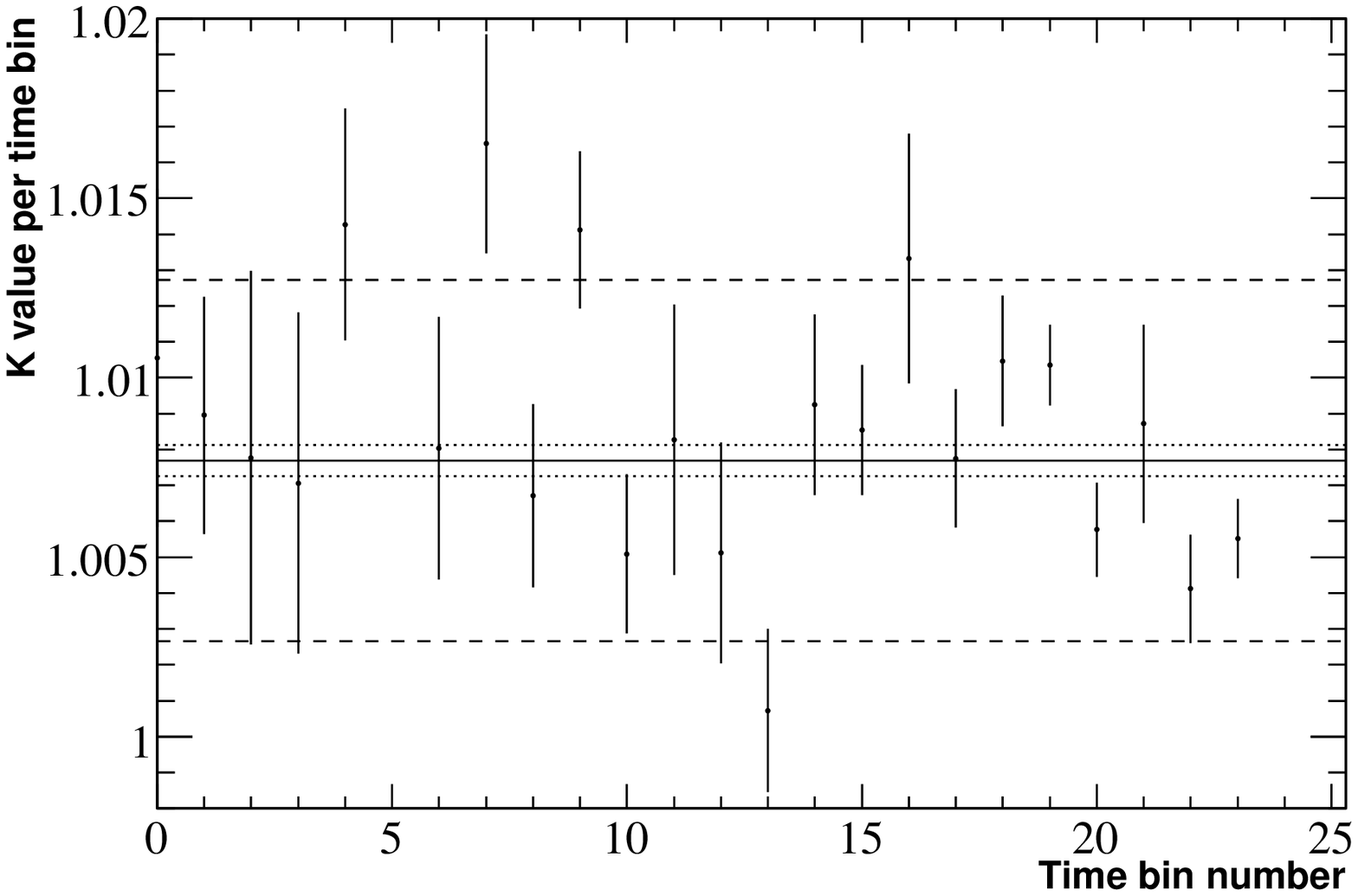}
    \caption[Time-variation of  $\kappa_{\mu}$  for isBCMuMu.]{Time-variation of  $\kappa_{\mu}$  for isBCMuMu. 
 The dashed lines show $\pm 0.5\%$ from the mean.}
      \label{fig:isBCMuMuKappa}
  \end{center}
\end{figure}

This new cutset clearly  displays less systematic variation 
than isBCMuMu. For example, the difference in $d^{\kappa}_0$ values shows an 
order of magnitude stability  improvement in  $\kappa_{\mu}$. The time-variation of on- and off-peak MC for these new cuts is shown in 
Fig.~\ref{fig:MyMuCutsOnVar} and \ref{fig:MyMuCutsOffVar}. The equivalent plot for the  isBCMuMu tag (with a vertical axis of the same scale) is shown in 
 Fig.~\ref{fig:isBCMuMuOnVar} and   \ref{fig:isBCMuMuOffVar}
and the  time variations 
of $\kappa_{\mu}$ are shown in Fig.~\ref{fig:BestCutsKappaMu} and \ref{fig:isBCMuMuKappa}.

We use one other method to estimate the systematic uncertainty on $\kappa_{\mu}$ --- we vary the mu-pair selection cuts on maxpCosTheta 
and measure how the change affects the quantity
\begin{equation}
\kappa_{\mu} \cdot \frac{N_{\mu}}{N'_{\mu}} = \kappa_{\mu} R_{\mu}.
\end{equation}
This quantity appears explicitly in the \B Counting formula (\ref{eqn:nb}). Any changes to it as the selection cuts vary show the combined effects 
of the altered cuts on both data and MC. This allows us to estimate the systematic uncertainty on $\kappa_{\mu}$ as the number of mu-pair events
counted should be almost  constant regardless of the arbitrary choice of mu-pair selection cuts.

Specifically we allow the cuts on maxpCosTheta to vary between 0.60 and 0.77 (in increments of 0.005) while the value of $\kappa_{\mu} R_{\mu}$ is recalculated for each
cut choice. The value of  $ \kappa_{\mu} R_{\mu}$ for the original cuts is 9.7126. The largest (positive or negative)
 variation from the
original value occurs when maxpCosTheta is equal to 0.635 --- this causes a decrease in $\kappa_{\mu}R_{\mu}$ of 0.047\%.  We attribute
this  percentage difference to an additional systematic uncertainty in  $\kappa_{\mu}$ as an estimate.

\begin{center}
\begin{table}[ht]  
  \caption[Properties of the proposed and existing  muon-pair selectors.]{
    \label{table:MuCutset}Properties of the recommended muon-pair cutset and isBCMuMu.}
  \centering
  \begin{tabular}{|c|l|c|c|}
    \hline
                &\ Cutset property &\  New Cutset &\ isBCMuMu \\
    \hline
    On-peak: &\  $\hat{\varepsilon}_{\mu}$     &\  0.41558      &\ 0.42566   \\
                &\   Stat. uncert. in $\hat{\varepsilon}_{\mu}$     &\ 0.00010 &\   0.00010 \\
                &\   $\chi^2$ statistic  of $\varepsilon_{\mu}$  &\ 118.11  &\ 1016.58         \\
                &\   Syst. uncert. in $\hat{\varepsilon}_{\mu}$ $(d_{0, \mu})$ &\ 0.00082  &\ 0.00311  \\
    \hline   
    Off-peak: &\   $\hat{\varepsilon}'_{\mu}$  &\ 0.41598   &\  0.42627  \\
               &\   Stat. uncert. in $\hat{\varepsilon}'_{\mu}$     &\  0.00014 &\ 0.00014 \\
                  &\   $\chi^2$ statistic  of $\varepsilon'_{\mu}$   &\ 51.71    &\ 402.25          \\
                &\   Syst. uncert. in $\hat{\varepsilon}'_{\mu}$ $(d'_{0, \mu})$ &\ 0.00060  &\ 0.00353  \\
    \hline
    $\kappa_{\mu}$: &\ $\hat{\kappa}_{\mu}$  &\  1.00791    &\ 1.00769    \\
               &\   Stat. uncert. in $\hat{\kappa}_{\mu}$     &\ 0.00044 &\  0.00043 \\
                  &\   $\chi^2$ statistic  of $\kappa_{\mu}$   &\ 23.82    &\ 62.11     \\
                &\   Syst. uncert. in $\hat{\kappa}_{\mu}$ $(d^{\kappa}_{0, \mu})$ &\  0.00032  &\   0.00306  \\
                &\   Syst. uncert. in $\hat{\kappa}_{\mu}$ ($\kappa_{\mu} R_{\mu}$) &\  0.00047  &\   N.A.  \\
    \hline
  \end{tabular}

\end{table}
\end{center}




\section{Choice of Hadronic Selection Cutset}
The cutset we recommend to select hadronic events is:
\begin{itemize}
\item BGFMultiHadron,
\item   R2All $\leq 0.65$, 
\item   nTracks $\geq 3$,
\item ETotal/eCM    $\geq 4.60/10.58$,
\item   PrimVtxdr $< 0.5$ cm,
\item $|$PrimVtxdz$| < 6.0$ cm,
\item p1Mag $< 3.0$ GeV/$c$ \ ($\geq 1$  DCH hit).
\end{itemize}

Some  properties of this cutset and isBCMultiHadron are shown in Table \ref{table:MHCutset}. The proposed cutset is more  stable in time than isBCMultiHadron, 
but the improvement from the existing tag is not as large as it was for the muon-pair selector. Nevertheless the new hadronic cutset has a lower 
$\chi^2$ statistic for all types of MC, and consequently the estimated systematic uncertainty ($d_0$) is lower in each case. The $\chi^2$ statistics of
each MC type decrease to between  26\% and 36\% of their isBCMultiHadron values.

\begin{figure}[htbp] 
  \begin{center}
    \includegraphics[width=4.3in]{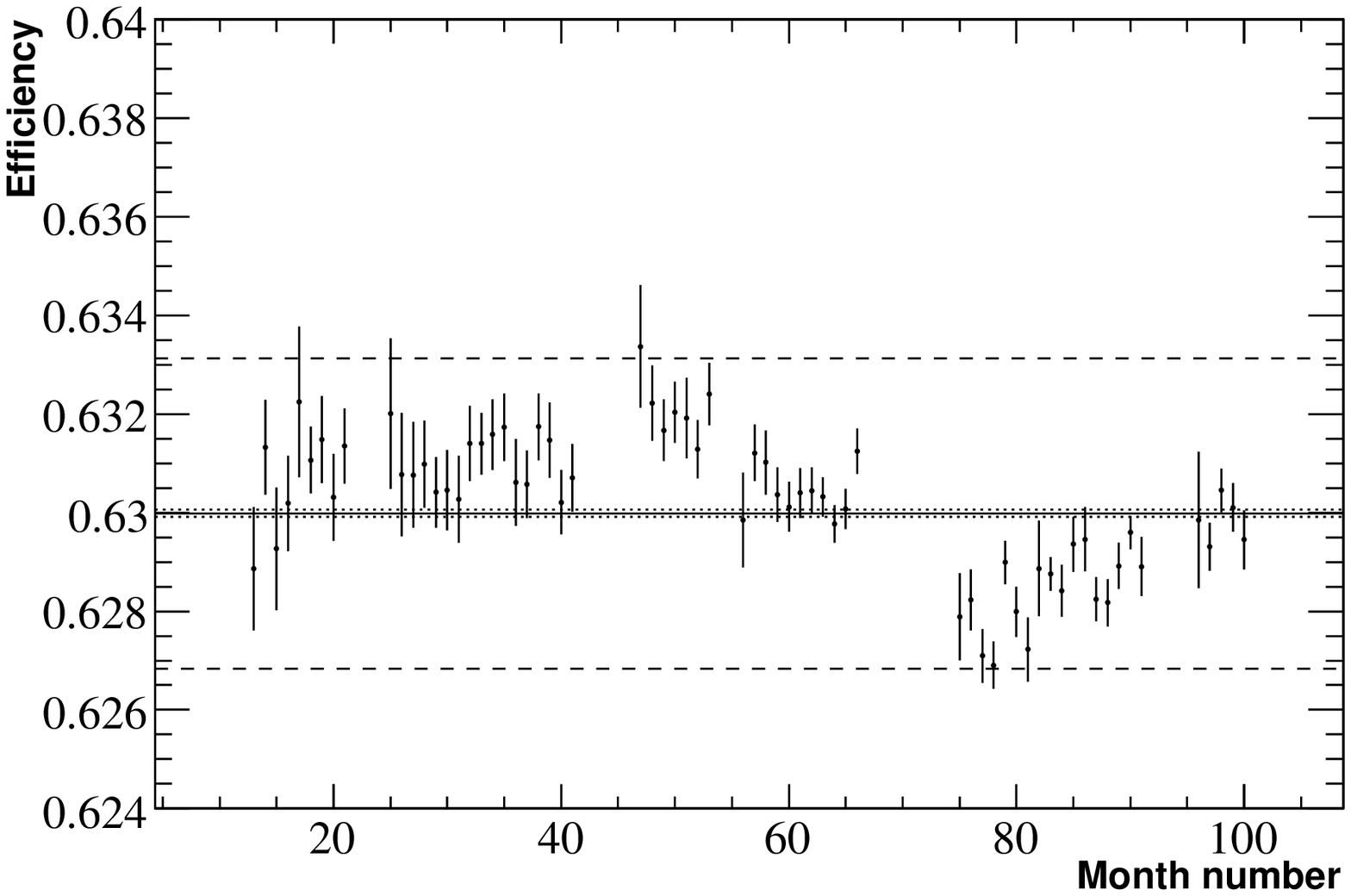}
    \caption[Time-variation of on-peak  $uds$ MC  for the new cuts.]{On-peak  $uds$ MC efficiency time-variation for 
      the new proposed hadronic cuts. The dashed lines show $\pm 0.5\%$ from the mean.}
      \label{fig:MyMHCutsudsOnVar}
  \end{center}
\end{figure}

\begin{figure}[htbp] 
  \begin{center}
    \includegraphics[width=4.3in]{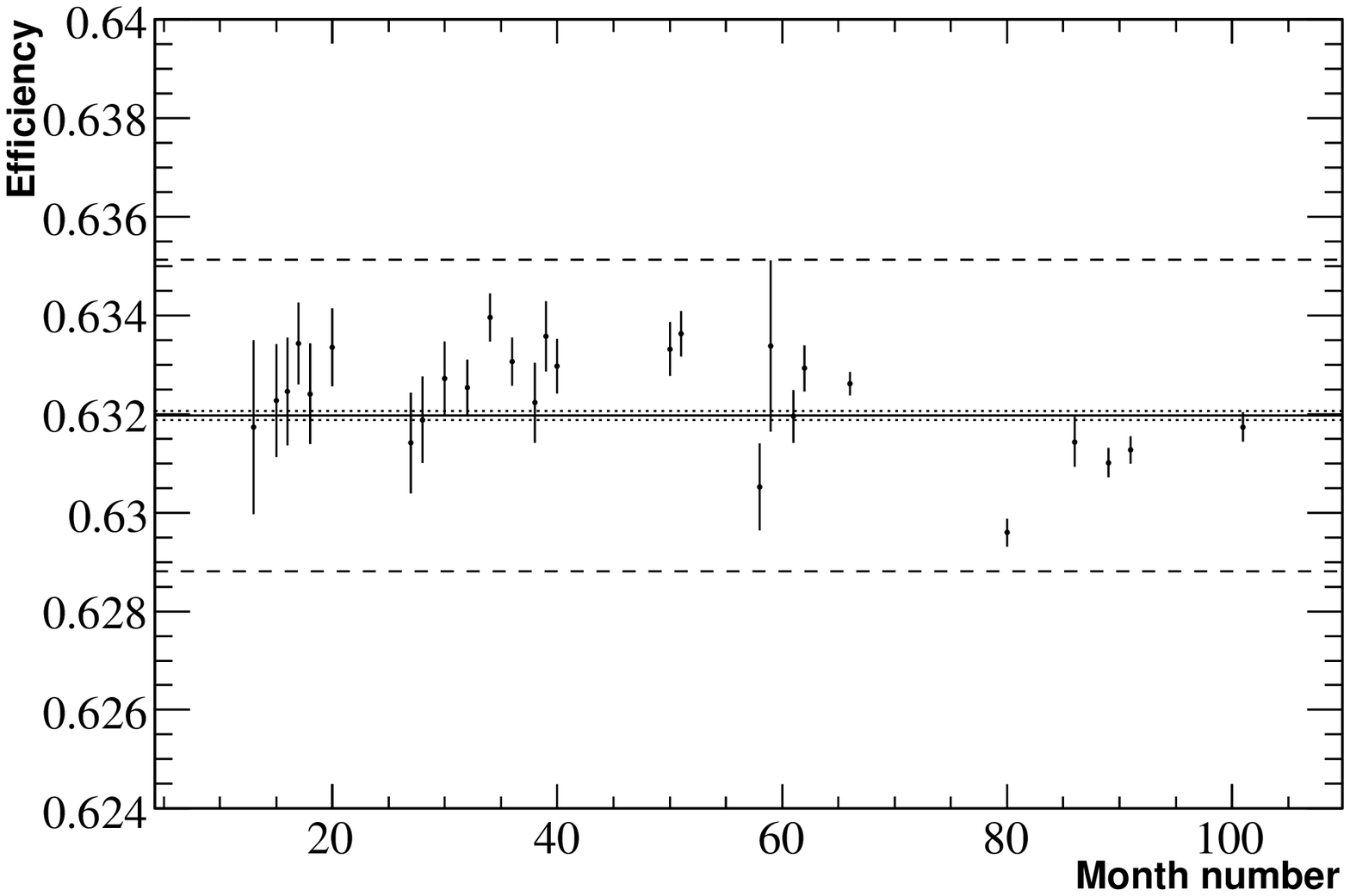}
    \caption[Time-variation of off-peak $uds$ MC  for the new cuts.]{Off-peak  $uds$ MC efficiency time-variation for 
      the new proposed hadronic cuts. The dashed lines show $\pm 0.5\%$ from the mean. }
      \label{fig:MyMHCutsudsOffVar}
  \end{center}
\end{figure}

\begin{figure}[htbp] 
  \begin{center}
    \includegraphics[width=4.3in]{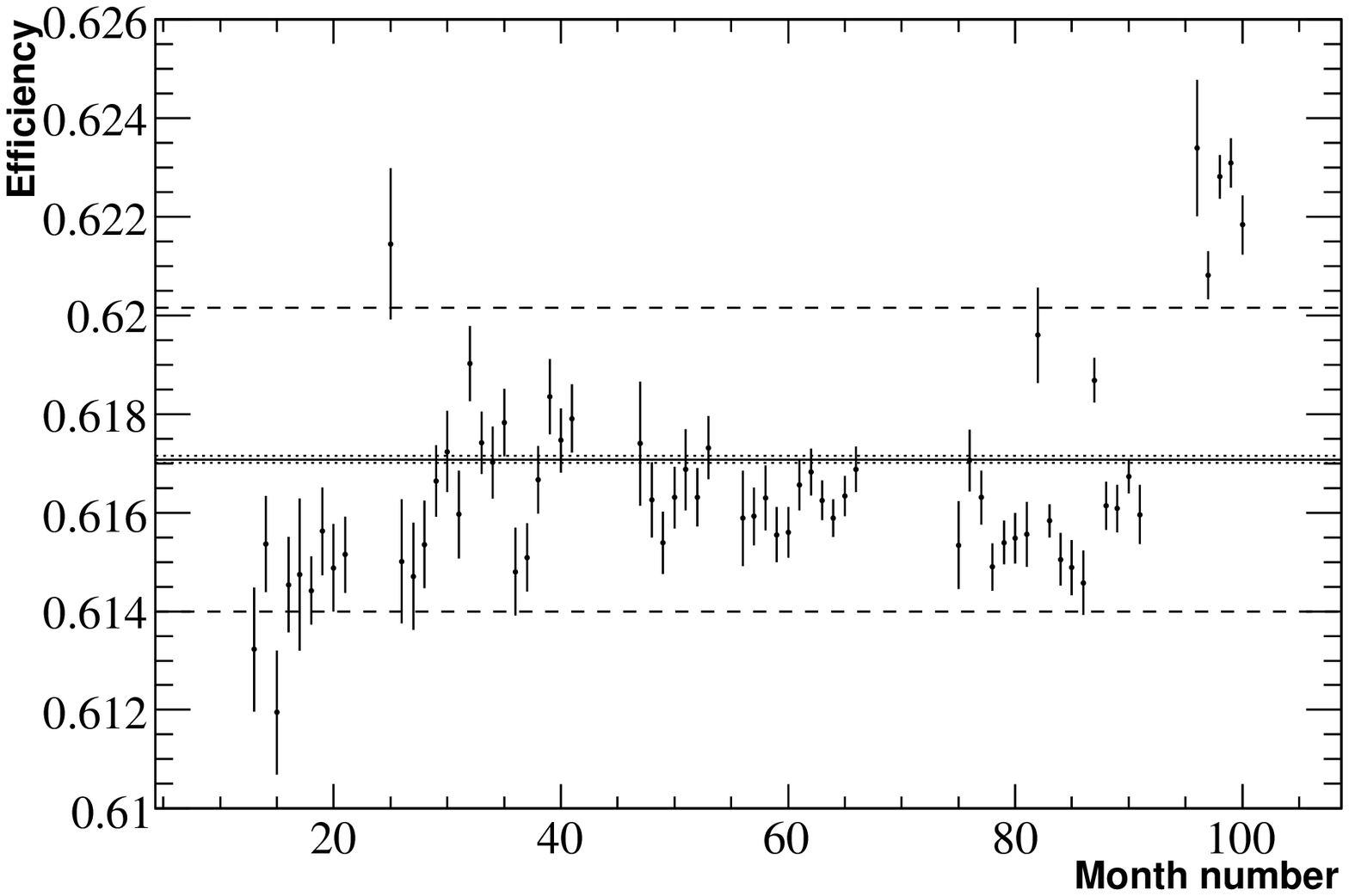}
    \caption[Time-variation of on-peak $uds$ MC   for isBCMultiHadron.]{On-peak  $uds$  MC efficiency time-variation for 
isBCMultiHadron. The dashed lines show $\pm 0.5\%$ from the mean.}
      \label{fig:isBCMHudsOnVar}
  \end{center}
\end{figure}

\begin{figure}[htbp] 
  \begin{center}
    \includegraphics[width=4.3in]{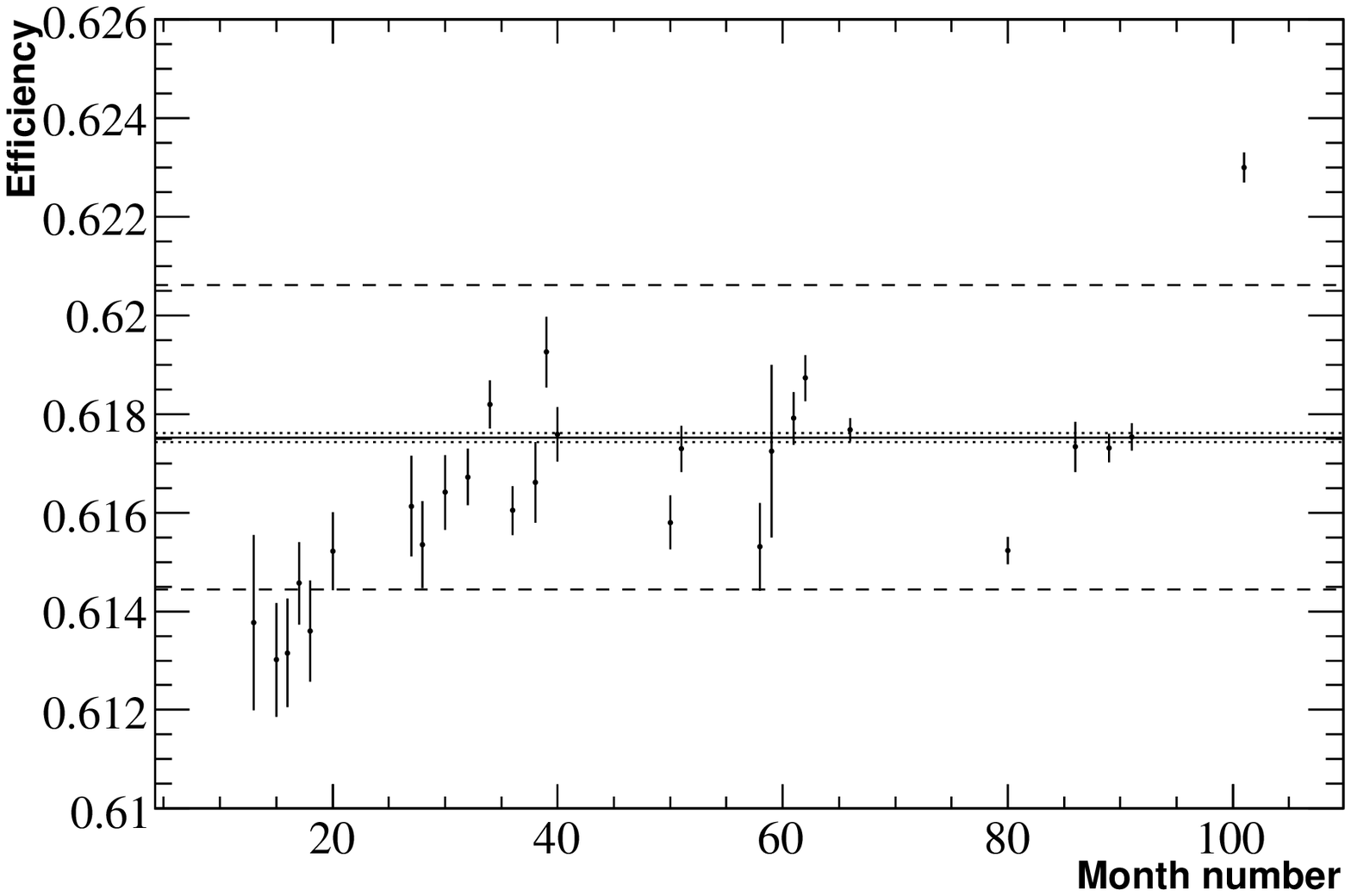}
    \caption[Time-variation of off-peak $uds$ MC   for isBCMultiHadron.]{Off-peak  $uds$  MC efficiency time-variation for 
isBCMultiHadron. The dashed lines show $\pm 0.5\%$ from the mean.}
      \label{fig:isBCMHudsOffVar}
  \end{center}
\end{figure}

\begin{figure}[htbp] 
  \begin{center}
    \includegraphics[width=4.3in]{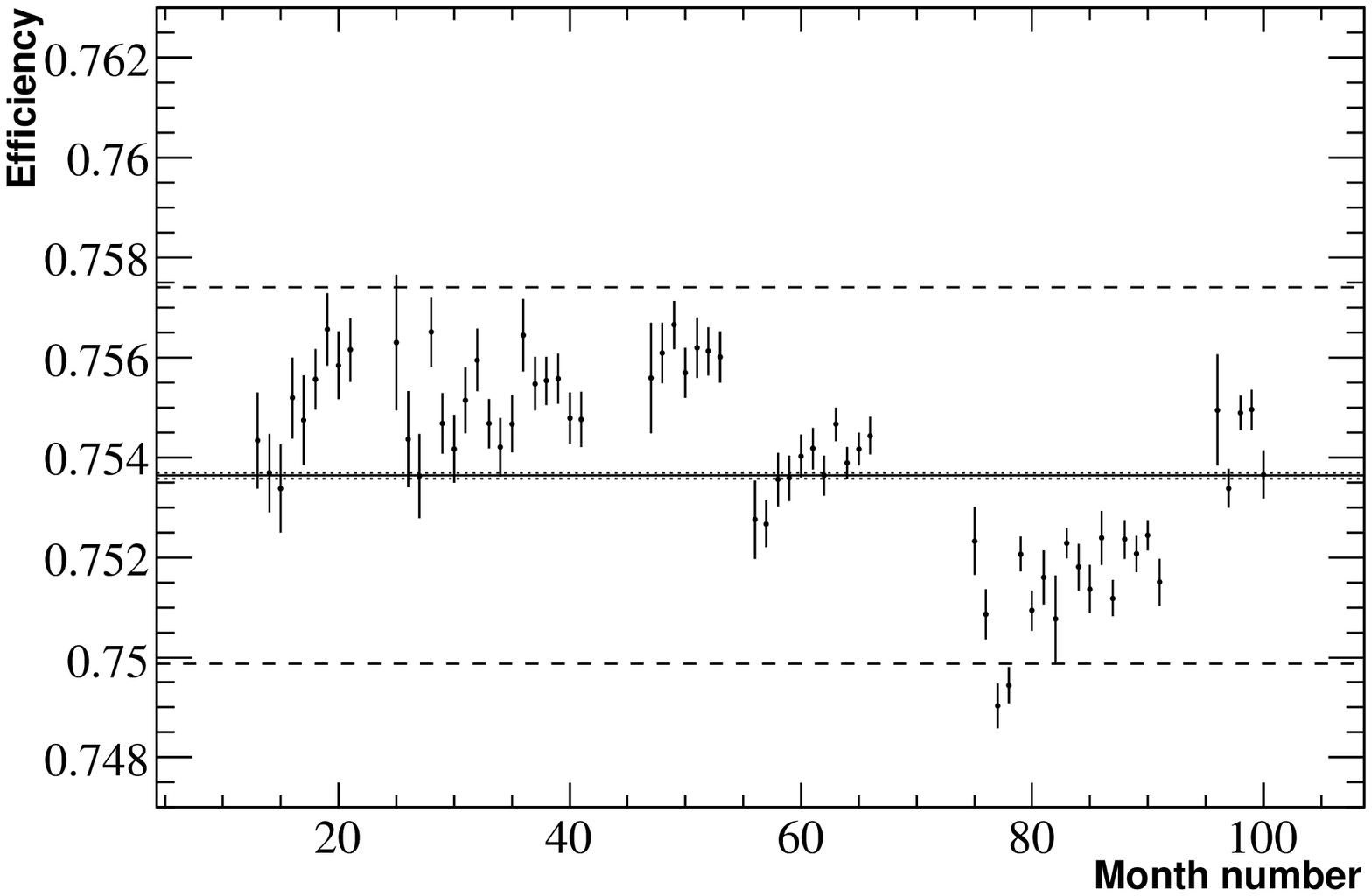}
    \caption[Time-variation of on-peak $\ccbar$ MC for the new cuts.]{On-peak $\ccbar$ MC efficiency time-variation for 
      the new proposed hadronic cuts. The dashed lines show $\pm 0.5\%$ from the mean.}
      \label{fig:MyMHCutsccbarOnVar}
  \end{center}
\end{figure}

\begin{figure}[htbp] 
  \begin{center}
    \includegraphics[width=4.3in]{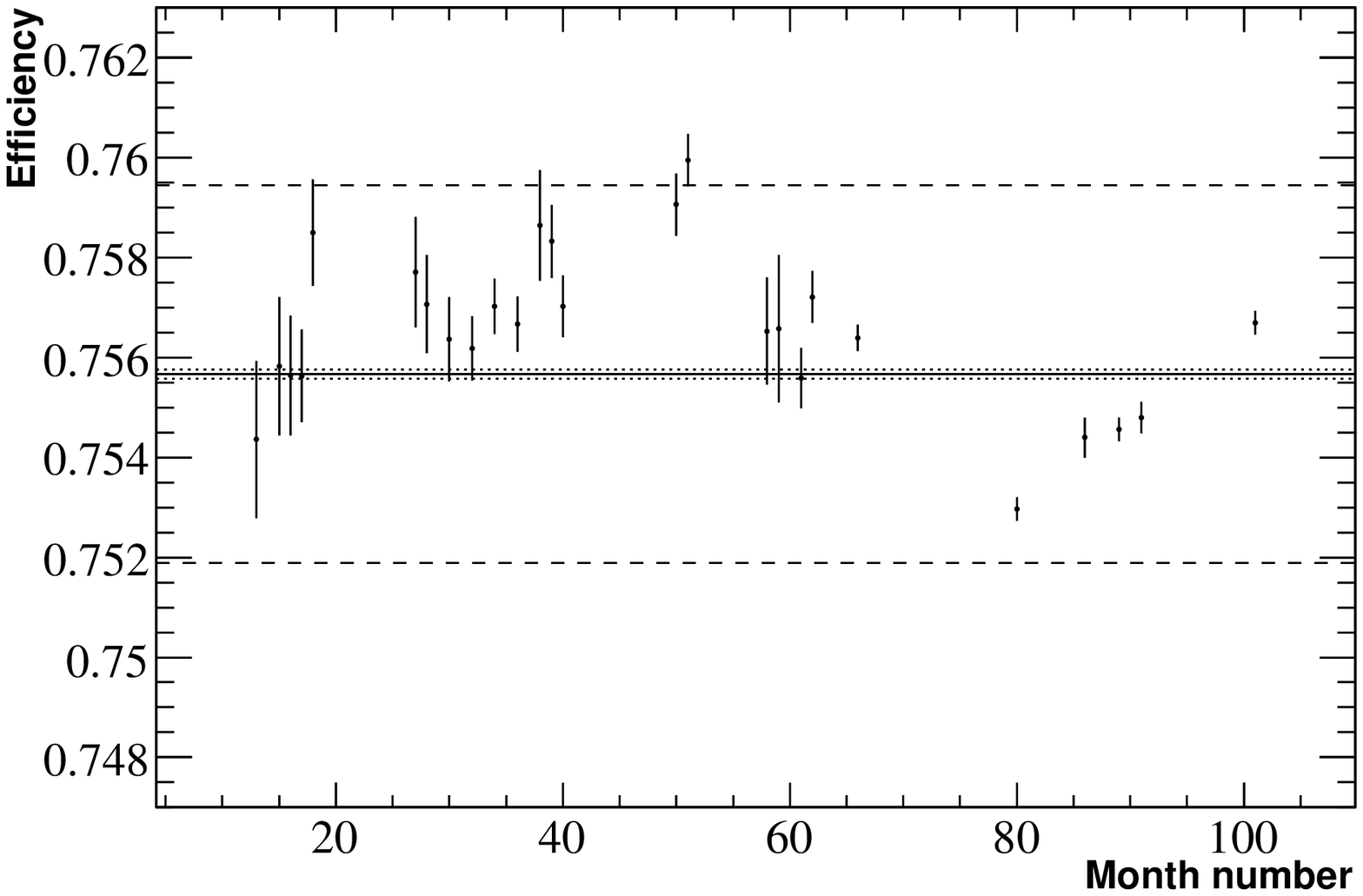}
    \caption[Time-variation of off-peak $\ccbar$ MC  for the new cuts.]{Off-peak  $\ccbar$ MC efficiency time-variation for 
      the new proposed hadronic cuts. The dashed lines show $\pm 0.5\%$ from the mean. }
      \label{fig:MyMHCutsccbarOffVar}
  \end{center}
\end{figure}

\begin{figure}[htbp] 
  \begin{center}
    \includegraphics[width=4.3in]{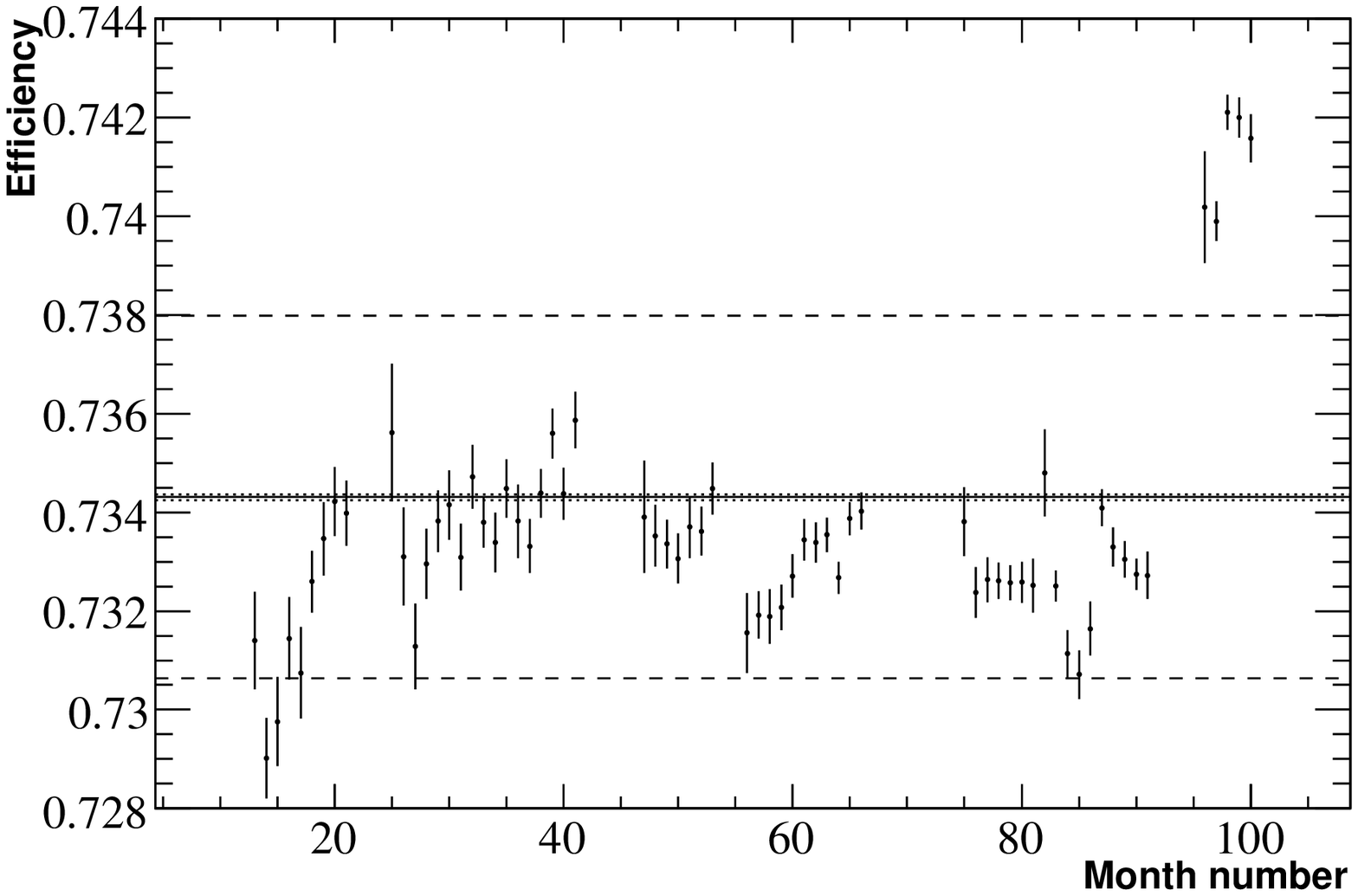}
    \caption[Time-variation of on-peak $\ccbar$ MC for isBCMultiHadron.]{On-peak  $\ccbar$  MC efficiency time-variation for 
isBCMultiHadron. The dashed lines show $\pm 0.5\%$ from the mean.}
      \label{fig:isBCMHccbarOnVar}
  \end{center}
\end{figure}

\begin{figure}[htbp] 
  \begin{center}
    \includegraphics[width=4.3in]{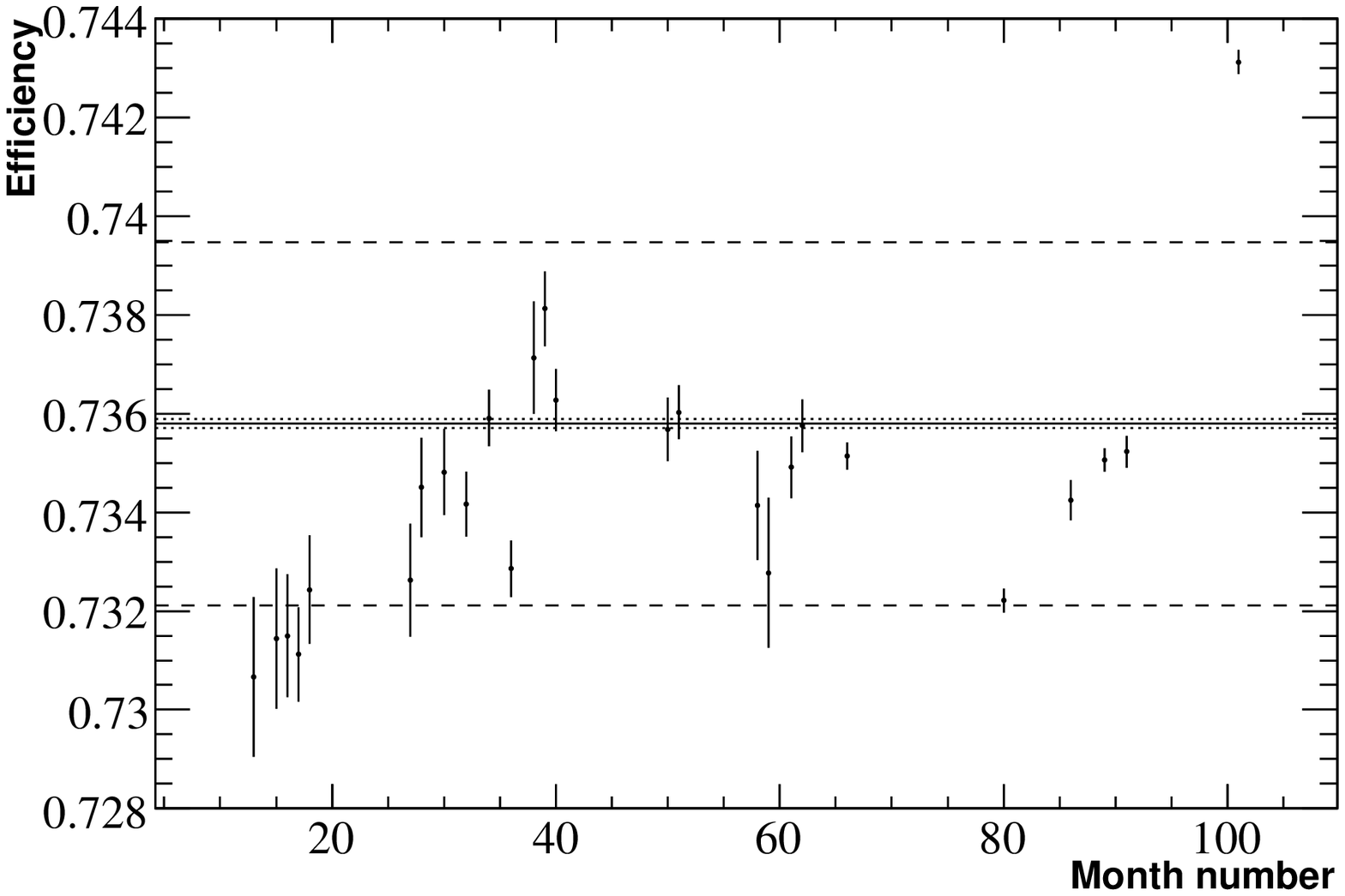}
    \caption[Time-variation of off-peak $\ccbar$ MC for isBCMultiHadron.]{Off-peak $\ccbar$  MC efficiency time-variation for 
isBCMultiHadron. The dashed lines show $\pm 0.5\%$ from the mean.}
      \label{fig:isBCMHccbarOffVar}
  \end{center}
\end{figure}

\begin{figure}[htbp] 
  \begin{center}
    \includegraphics[width=4.3in]{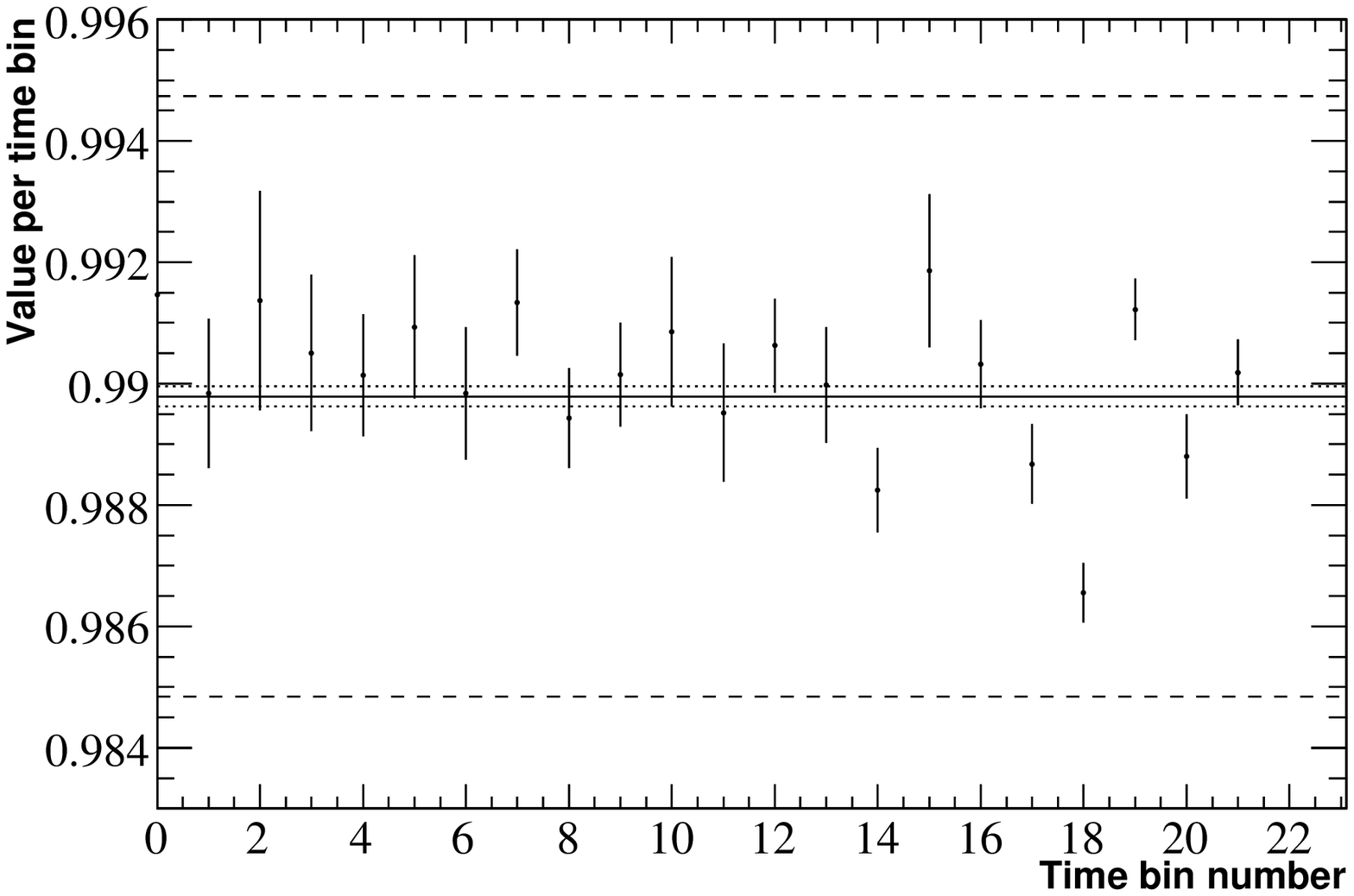}
    \caption[Time-variation of  $\kappa_{X}$ for the new cuts.]{Time-variation of  $\kappa_{X}$  for the new cutset. 
The dashed lines show $\pm 0.5\%$ from the mean.}
      \label{fig:MyMHCutsKappaMH}
  \end{center}
\end{figure}

\begin{figure}[htbp] 
  \begin{center}
    \includegraphics[width=4.3in]{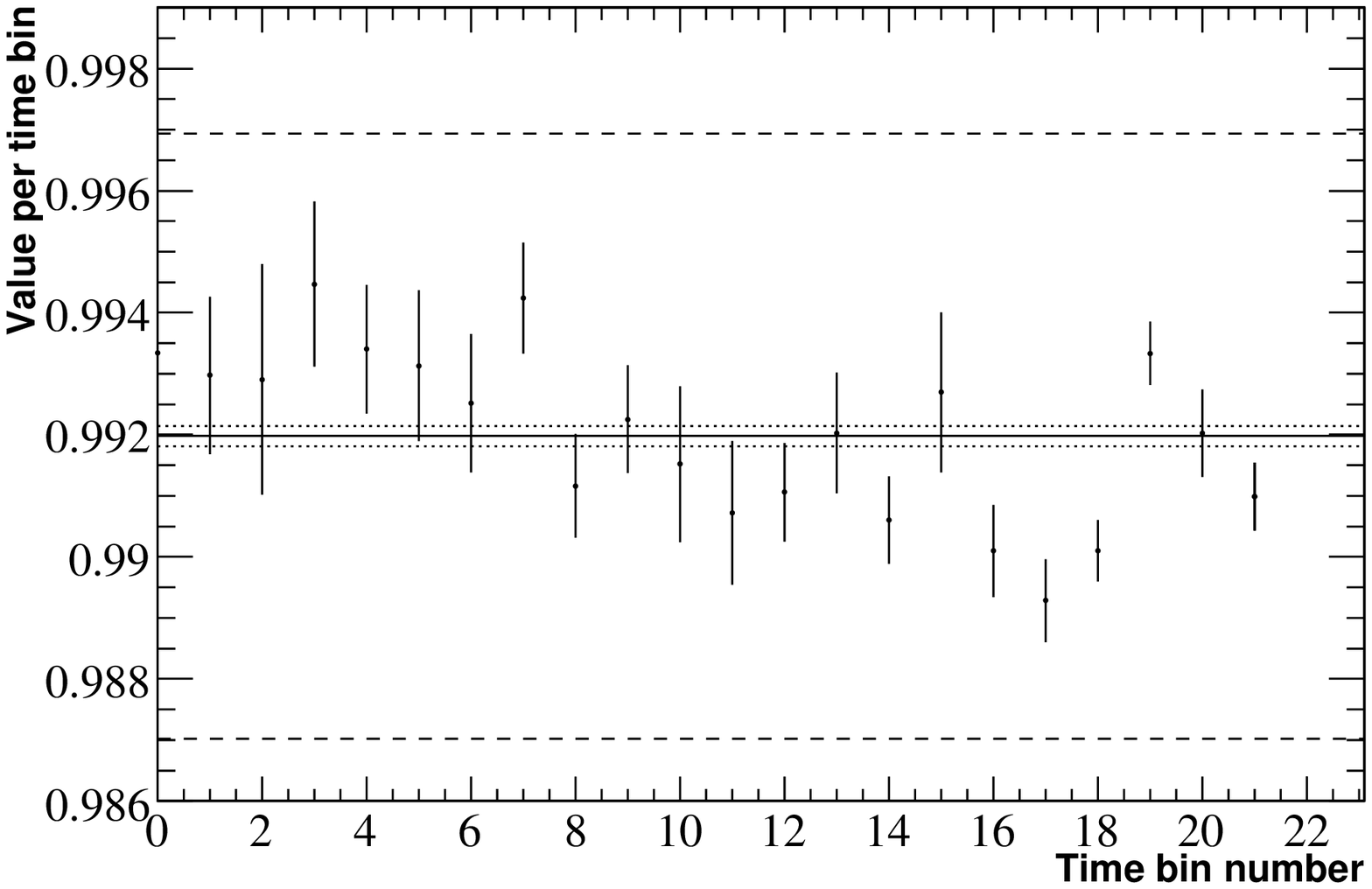}
    \caption[Time-variation of  $\kappa_{X}$ for isBCMultiHadron.]{Time-variation of  $\kappa_{X}$  for  isBCMultiHadron. 
The dashed lines show $\pm 0.5\%$ from the mean.}
      \label{fig:isBCMHKappaMH}
  \end{center}
\end{figure}

The time-variation of  $\kappa_{X}$ and the efficiency of each type of continuum MC  for the proposed cutset and isBCMultiHadron  are shown
in Fig.~\ref{fig:MyMHCutsudsOnVar} -- \ref{fig:isBCMHKappaMH}. The time-variation of $uds$ MC efficiency for the new cuts  and isBCMultiHadron
are shown in Fig.~\ref{fig:MyMHCutsudsOnVar} --  \ref{fig:isBCMHudsOffVar}. The equivalent plots for   \ccbar MC efficiency (with vertical axes of the 
same scale) are shown in 
Fig.~\ref{fig:MyMHCutsccbarOnVar} -- \ref{fig:isBCMHccbarOffVar}. Finally,  $\kappa_{X}$ for the new cutset and isBCMultiHadron is
shown in Fig.~\ref{fig:MyMHCutsKappaMH} -- \ref{fig:isBCMHKappaMH}.

The weighted mean \BB efficiency ($\hat{\varepsilon}_B$) for this cutset is
$0.94046 \pm 0.00003$ (where the uncertainty is statistical only), which is slightly lower than the isBCMultiHadron value 
of 0.9613. The main reasons for this decrease are the addition of a 
p1Mag cut and the slightly higher cut on ETotal. 

\subsection{ISR Production and Two-Photon Events}
We have not yet considered the effect on $\kappa_{X}$  from events which are difficult to simulate with Monte Carlo; in particular
Initial State Radiation (ISR) and two-photon events. Our limited knowledge of these events introduces an additional systematic uncertainty
to the value of $\kappa_{X}$.

ISR events are those in which one of the incoming beam particles
emits a photon before interacting.  The theoretical ISR cross-section used in this analysis is based on that described by Benayoun \emph{et al.} \cite{Benayoun}. 
For ISR production of any state (labelled ``V''), the
ISR cross-section is taken to be:

\begin{equation}
\sigma_{ISR}^V \approx \frac{12\pi^2 \Gamma_{ee}}{m_V\cdot s}\cdot W(s, 1 - \frac{m^2_V}{s}),
 \end{equation}
where $W$ is a function defined in \cite{Benayoun}, and $m_V$ and $\Gamma_{ee}$ are the resonance mass and partial width of $V\to\epem$ respectively. The total
cross-section is estimated to be the sum of each possible $J^{PC} = 1^{--}$ resonance. The dominant terms in this sum are to the $\rho^0$ (by an 
order-of-magnitude), \jpsi and \ThreeS resonances.

Two-photon events are those where the two colliding particles emit virtual photons which in turn interact
to produce some system of particles. A review of two-photon particle production can be read in \cite{Budnev:1974de}. For high energies (such as those provided
by PEP-II), the cross-section for two-photon production of a lepton pair is given approximately by:
\begin{equation}
\sigma_{\epem \to \epem \ellell} = \frac{28\alpha^4}{27\pi m_l^2}\left(\ln \frac{s}{m^2_e}\right)^2 \ln\frac{2}{m^2_l},
\end{equation}
where $l$ can be either $e$ or $\mu$ and $\sqrt{s}$ is the CM energy. 

The (high-energy limit) cross-section for hadron production ($h$) is given approximately by:
\begin{equation}
\sigma_{\epem \to \epem h} = \frac{\alpha^4}{18\pi^2 m_{\pi}^2}\ln\frac{s m^2_{\rho}}{m^2_e m^2_{\pi}}
\ln\frac{s m^6_{\rho}}{m^6_e m^2_{\pi}}\left(\ln \frac{s}{m^2_\pi}\right)^2,
\end{equation}
where $m_{\rho}$ and $m_{\pi}$ are the masses of the  $\rho$ and $\pi$ mesons. 

Since these two processes are not known especially well, we attempt to quantify  their effect on systematic uncertainty. We do this by using
\ccbar, $uds$, \tautau and high-angle Bhabha MC to estimate the effective cross-section for other (i.e. not of these four types) events in off-peak data of passing
the hadronic selection, and the corresponding systematic uncertainty on $\kappa$. 

The hadronic part of $\kappa$  takes the form:
\begin{eqnarray}
\kappa_X & =  & \frac{\varepsilon_X \sigma_X}{\varepsilon'_X \sigma'_X }  \\
     & = & \frac{\sum^{known}  \varepsilon_i \sigma_i +\sum^{unknown} \varepsilon_j \sigma_j}{\sum^{known}  \varepsilon'_i \sigma'_i +\sum^{unknown} \varepsilon'_j \sigma'_j}.
 \end{eqnarray}

Here, by \emph{known} we mean events for which \babar\ has a reliable MC model and well-known cross-sections. The effects of the \emph{unknown} types can  
be estimated by comparing off-peak data and the known MC. We assume that the unknown component is comprised solely of two-photon, ISR and low-angle Bhabha events. 

\begin{figure}[htbp] 
  \begin{center}
    \includegraphics[width=4.5in]{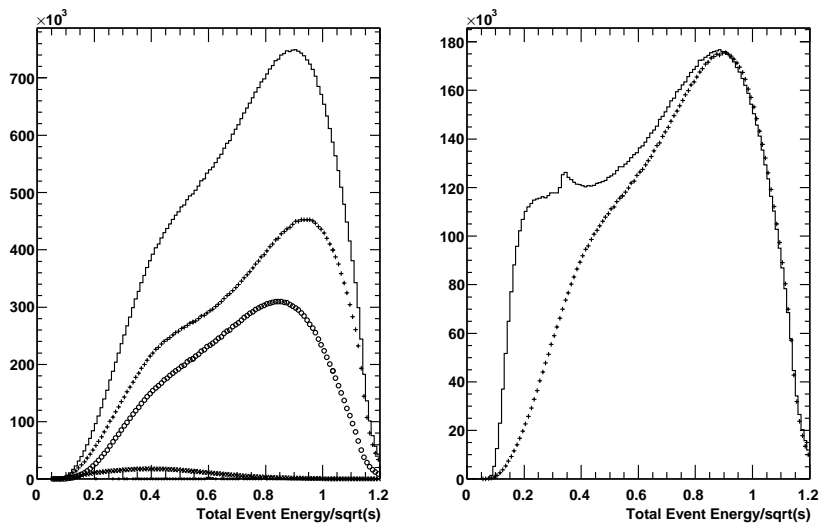}
    \caption[Comparison of ETotal for known MC types to off-peak data.]
	    {Comparison of ETotal$/\sqrt{s}$ for known MC types to off-peak data for events passing all other hadronic selection cuts: (left) in decreasing order
distributions of sum of all MC types, $uds$, \ccbar, \tautau, low-angle Bhabha MC; (right) the sum (points) is shown overlaid with data  -- the plots are normalized to 
have the same value at 0.9. }
            \label{fig:Cont1}
  \end{center}
\end{figure}

We compare the known MC types to off-peak data. The cross-sections for each
MC type are used to combine them in the correct proportions to model data as accurately as possible. Normalization is achieved by forcing the MC to have
the same value as data at 0.9 on the plot of total event energy. The MC distributions of total energy are shown in the left plot of Fig.~\ref{fig:Cont1} and the sum
overlaid with data on the right.

The excess in data over the cut range (above 0.4348 on the ETotal$/\sqrt(s)$ plot) is found to be equivalent to an effective cross-section of approximately 
0.10 nb. The value of $\kappa_X$ for the optimal cuts  for only the known types of MC has the value  $0.98979 \pm 0.00016$ (where the uncertainty
is statistical only). 
To estimate $\kappa_X$ including two-photon and ISR events we estimate the effective on-peak cross-sections by scaling by the theoretical cross-sections.
These considerations  increase the value of $\kappa_X$ and introduce an additional systematic uncertainty on its value. We combine this new
uncertainty in quadrature with that due to MC time-variation. 

Note that we do not use the (fairly old) theoretical models to predict the absolute cross-section, but use the ratio of theoretical on- and off-peak
cross-sections for the processes to estimate the total effective on-peak cross-section.

The proportions of two-photon and ISR events in the data are unknown, so for a range of $\alpha$ values from 0.01 to 0.99 we let
\begin{equation}
\sum^{unknown} \varepsilon'_j \sigma'_j = \alpha\cdot \varepsilon'_{ISR} \sigma'_{ISR} + (1-\alpha) \cdot \varepsilon'_{2\gamma} \sigma'_{2\gamma}.
 \end{equation}
and use the ratios of $\sigma_{ISR}/\sigma'_{ISR}$ and $\sigma_{2\gamma}/\sigma'_{2\gamma}$ from theory to estimate the on-peak unknown 
effective cross-section (assuming the on- and off-peak efficiencies are identical).  

The value of $\kappa_X$ including the unknown component can be estimated. We take its value to be the $\alpha=0.5$ value, and equate the difference 
as $\alpha$ varies to be systematic uncertainty. Performing the identical analysis for the theoretical Bhabha cross-section does not change the solution ---
for 100\% Bhabhas or a 1:1:1 mixture of two-photon, ISR and Bhabha events, the value of $\kappa_X$ stays within
the range found using two-photon and ISR events alone.

\begin{figure}[htbp] 
  \begin{center}
    \includegraphics[width=4.0in]{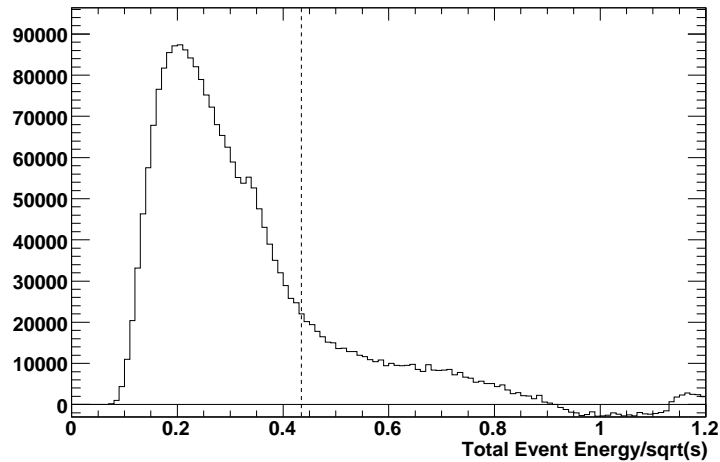}
    \caption[Difference between known MC types and off-peak data.]
	    {Difference  in ETotal between known MC types and off-peak data. The cut at 0.4348 is marked with a dashed line. }
            \label{fig:Cont_diff}
  \end{center}
\end{figure}

Fig.~\ref{fig:Cont_diff} shows the ETotal/eCM distribution of the ``unknown'' events (i.e. the difference between the MC and off-peak data distributions).

The result is: $\kappa_X =  0.9901 \pm 0.0002$. Note that the addition of two-photon and ISR
events give a central value of $\kappa_X$ higher than that calculated for the known types only (although they agree within uncertainties). 
We assume this uncertainty is independent of that calculated in Section~\ref{sec:kappacont}.

Combining the two uncertainties (from the MC time-variation and this two-photon, ISR and low-angle Bhabha analysis) in quadrature, the value for the optimal cutset is:
 $\kappa_X =  0.9901 \pm 0.0010$.

\begin{center}
\begin{table}[ht] 
   \caption[Properties of the proposed and existing  hadronic selectors.]{
    \label{table:MHCutset}Properties of the recommended hadronic cutset and isBCMultiHadron.}
  \centering
  \begin{tabular}{|c|l|c|c|}
    \hline
                &\ Cutset property &\  New Cutset &\ isBC- \\
                    &\ &\  &\ MultiHadron \\
    \hline
    On-peak: &\  $\hat{\varepsilon}_{uds}$     &\    0.62998        &\    0.61708   \\
                &\   Stat. uncert. in $\hat{\varepsilon}_{uds}$    &\ 0.00007   &\ 0.00007  \\
                &\   $\chi^2$ statistic  of $\varepsilon_{uds}$  &\  344.93        &\ 1081.82        \\
                &\   Syst. uncert. in $\hat{\varepsilon}_{uds}$ $(d_{0, uds})$ &\     0.00121    &\  0.00227    \\
                &\  $\hat{\varepsilon}_{\ccbar}$     &\  0.75364       &\   0.73431     \\
                &\   Stat. uncert. in $\hat{\varepsilon}_{\ccbar}$    &\    0.00006  &\ 0.00006  \\
                &\   $\chi^2$ statistic  of $\varepsilon_{\ccbar}$  &\ 939.95        &\ 2669.32        \\
                &\   Syst. uncert. in $\hat{\varepsilon}_{\ccbar}$ $(d_{0, \ccbar})$ &\ 0.00175  &\  0.00278  \\
    \hline   
    Off-peak: &\   $\hat{\varepsilon}'_{uds}$  &\   0.63200       &\ 0.61752    \\
               &\   Stat. uncert. in $\hat{\varepsilon}'_{uds}$    &\    0.00009    &\ 0.00009  \\
                  &\   $\chi^2$ statistic  of $\varepsilon'_{uds}$   &\ 164.46         &\ 527.27   \\
                &\   Syst. uncert. in $\hat{\varepsilon}'_{uds}$ $(d'_{0, uds})$   &\  0.00094   &\ 0.00216     \\
               &\   $\hat{\varepsilon}'_{\ccbar}$  &\   0.75568      &\  0.73580       \\
               &\   Stat. uncert. in $\hat{\varepsilon}'_{\ccbar}$     &\ 0.00009  &\  0.00009  \\
                  &\   $\chi^2$ statistic  of $\varepsilon'_{\ccbar}$   &\  349.56          &\ 1304.13           \\
                &\   Syst. uncert. in $\hat{\varepsilon}'_{\ccbar}$ $(d'_{0, \ccbar})$ &\   0.00167   &\  0.00274    \\
    \hline
    $\kappa_{X}$: &\ $\hat{\kappa}_{X}$  (\emph{known} MC only) &\  0.98979  &\  0.99198     \\
         &\   Stat. uncert. in $\hat{\kappa}_{X}$     &\  0.00016   &\ 0.00017  \\
                  &\   $\chi^2$ statistic  of $\kappa_{X}$   &\ 80.19         &\ 101.19      \\
                &\   Syst. uncert. in $\hat{\kappa}_{X}$ $(d^{\kappa}_{0, cont})$ &\  0.00095  &\   0.00122   \\
                &\   Syst. uncert. in $\hat{\kappa}_{X}$ (ISR/2$\gamma$) &\  0.0002  &\   0.0002   \\
    \hline
   &\ $\hat{\kappa}_{X}$  (incl. ISR/2$\gamma$) &\  0.9901  &\  0.9918    \\
                &\  Combined uncert. in $\hat{\kappa}_{X}$  &\  0.0010  &\   0.0012   \\
    \hline
  \end{tabular}

\end{table}
\end{center}




\section{\B Counting Results}
The numbers of \B meson events counted in data by this pair of new proposed cutsets for each major running period are given in 
Table \ref{table:BCountSummary}. The two
uncertainty terms, $\Delta(N^0_{B})_{\mu}/N^0_B$ and $\Delta(N^0_{B})_{H}/N^0_B$  are added in quadrature to give the total uncertainty
in the number of \B mesons. For comparison, the number counted by the existing code (with a total uncertainty of $1.1\%$ in each case) is also provided.

For every Run, the number of \B meson events counted by the new cutsets agrees with the existing value within uncertainty. The total uncertainty ranges
between $0.615\%$ and  $0.634\%$. Note that these values are calculated only to find  the optimal pair of cutsets. The final estimate of systematic
uncertainty in data using the optimal cutset is discussed in Chapter \ref{chap:systuncert}.

    \begin{table}   
      \caption[Uncertainty and number of \B meson events in Runs 1--6.]{
	\label{table:BCountSummary}Uncertainty terms and number of \B meson events with overall uncertainty found during optimization procedure  in Runs 1--6.}
        \centering
      \begin{tabular}{|l||c|c|}
	\hline
        &\  \bf{Run 1} &\     \bf{Run 2}  \\
       	\hline
	$\Delta(N^0_{B})_{\mu}/N^0_B$  &\ 0.461\%  &\  0.456\% \\
	$\Delta(N^0_{B})_{H}/N^0_B$  &\ 0.416\%  &\  0.412\% \\
	$N^0_B$ (new cutsets)  &\  $(22.39 \pm 0.14) \times 10^6$   &\  $(67.61 \pm 0.42) \times 10^6$   \\
	\hline
	$N^0_B$ (existing code)   &\ $(22.40 \pm 0.26) \times 10^6$ &\ $(67.39 \pm 0.74) \times 10^6$    \\
	\hline
	\hline
      &\  \bf{Run 3} &\     \bf{Run 4}  \\
       	\hline
	$\Delta(N^0_{B})_{\mu}/N^0_B$  &\ 0.463\%  &\  0.460\% \\
	$\Delta(N^0_{B})_{H}/N^0_B$  &\ 0.417\%  &\  0.414\% \\
	$N^0_B$ (new cutsets)  &\  $(35.75 \pm 0.22) \times 10^6$   &\  $(110.70 \pm 0.69) \times 10^6$   \\
	\hline
	$N^0_B$ (existing code)   &\ $(35.57 \pm 0.39) \times 10^6$ &\ $(110.45 \pm 1.21) \times 10^6$    \\
	\hline
	\hline
      &\  \bf{Run 5} &\     \bf{Run 6}  \\
       	\hline
	$\Delta(N^0_{B})_{\mu}/N^0_B$  &\ 0.462\%  &\  0.471\% \\
	$\Delta(N^0_{B})_{H}/N^0_B$  &\ 0.416\%  &\  0.424\% \\
	$N^0_B$ (new cutsets)  &\  $(146.19 \pm 0.91) \times 10^6$   &\  $(82.19 \pm 0.52) \times 10^6$   \\
	\hline
	$N^0_B$ (existing code)   &\ $(147.19 \pm 1.62) \times 10^6$ &\ $(82.04 \pm 0.90) \times 10^6$    \\
	\hline
	\end{tabular}

    \end{table}

%% file: chapter8.tex
\chapter{Sources of Contamination}\label{chap:contamination}

As mentioned in previous Chapters, there are several possible contamination effects which
are  considered negligible for \B Counting. This includes $\epem \to \tautau$ decays incorrectly passing hadronic selection and  Bhabha events
incorrectly passing muon-pair selection among others. In this Chapter we quantify these effects.

\section{Bhabha Events}
The process with by far the highest cross-section in PEP-II is Bhabha scattering.  Most processes, like $\epem \to \mumu$ or $\epem \to \qqbar$ have cross-sections
proportional to $1/E_{CM}^2$, where $E_{CM}$ is the event's CM energy.  The cross-section for Bhabha scattering is more complicated, and at PEP-II's
on-peak CM energy of 10.58 GeV  is around 40 times larger than most other processes. The differential cross-section   takes the form:
\begin{equation}
\frac{\mathrm{d}\sigma}{\mathrm{d}\cos{\theta}} = \frac{\pi \alpha^2}{s}\left[u^2\left(\frac{1}{s} + \frac{1}{t}\right)^2  + \left(\frac{t}{s}\right)^2  +\bigg(\frac{s}{t}\bigg)^2 \right]
\end{equation}
where $s$, $t$ and $u$ are the Mandelstam variables, which relate the four-momenta of the incoming and outgoing particles  \cite{Peskin:1995}. 

The majority of Bhabha events scatter at low angles (i.e. continue down the beam-pipe),
but the cross-section for large-angle Bhabha scattering (i.e. within the angular limits of ChargedTracksBC) is still substantial. Based on samples of one million generated
Monte Carlo Bhabha scattering events,  the values for on- and off-peak running are  $(13.8910 \pm 0.0040)$ nb and $(14.0918 \pm 0.0041)$ nb respectively. 
These are more than ten  times the \BB cross-section of 1.1 nb. 

Ideally, the \B Counting muon-pair and hadronic selectors should reject all Bhabha events, but due to many imperfections a small number incorrectly  pass.  
To estimate the size of this effect, we apply the selectors to 7.725 million high-angle Bhabha MC events. The numbers of Bhabha MC events passing the
selectors are given in Table \ref{tab:BhabhaFakes}. For the new cuts, the Bhabha hadronic fake-rate is approximately 0.006\%, and the Bhabha  muon-pair fake-rate
is approximately 0.002\%.

\begin{table}[ht]  
  \caption[Fake-rates for Bhabha MC events.]{
    \label{tab:BhabhaFakes}Fake-rates for Bhabha MC events passing hadronic and mu-pair selectors. The sample contained 7.725 million large-angle on-peak Bhabha MC events.}
  \centering
  \begin{tabular}{|l|c|c|}
    \hline
    &\ Number passing selector &\ Percent of total \\
    \hline
 New hadronic selector &\  449 &\ 0.0058\% \\
 isBCMultiHadron &\ 1559 &\ 0.0202\% \\
\hline
 New mu-pair selector &\ 149 &\ 0.0019\% \\
 isBCMuMu  &\ 22  &\ 0.0003\% \\
    \hline
  \end{tabular}

\end{table}

We can estimate the ratio of Bhabha events to \B meson events passing the hadronic cuts in an on-peak sample of luminosity $\mathcal{L}$:
\begin{eqnarray}
\frac{N_{Bha} \ \textrm{(pass had. cuts)}}{N_B \ \textrm{(pass had. cuts)}} & = &\frac{\varepsilon^{had.}_{Bha} \sigma_{Bha} \mathcal{L}}{\varepsilon^{had.}_B \sigma_B \mathcal{L}}\\
\frac{N_{Bha} \ \textrm{(pass had. cuts)}}{N_B \ \textrm{(pass had. cuts)}}   & \approx & \frac{0.00006 \times 13.9 \ \textrm{nb}}{0.941 \times 1.1 \  \textrm{nb}} \\
   & \approx & 0.08\%.
\end{eqnarray}
Thus, even though the cross-section for high-angle  Bhabha events is more than ten times the other continuum  cross-sections, the Bhabha hadronic fake-rate is very much
smaller than  the selector's systematic uncertainty. The equivalent number for the muon selector (of fake Bhabhas to genuine muon-pairs) 
is  factor of three smaller. 

The final possible source of Bhabha contamination relates to the energy-dependence of the Bhabha cross-section. One of the most important assumptions for \B Counting 
is that  
background events scale in the same way with CM energy as muon-pair events. This assumption enables
backgrounds to be  subtracted by using the ratio of on-peak to  off-peak muon-pair events.  For Bhabha events however,  this assumption is not true. 
 The ratio of on-peak to off-peak muon-pair MC events from Section \ref{sec:MuonMC} is:
\begin{equation}
\frac{\sigma_{\mu}}{\sigma'_{\mu}} = \frac{1.11853 \ \textrm{nb}}{1.12647 \ \textrm{nb}} = 0.99295 \quad \ \textrm{(5 sig. fig.)}
\end{equation}
and for Bhabhas, this ratio is
\begin{equation}
\frac{\sigma_{Bha}}{\sigma'_{Bha}} =  \frac{13.8910\  \textrm{nb} }{14.0918 \  \textrm{nb}} = 0.98575 \quad \ \textrm{(5 sig. fig.)}.
\end{equation}

This difference means that when subtracting continuum events,  the number of on-peak Bhabha events passing the hadronic cuts will be over-estimated by a factor  
of approximately 0.7\%.  The total effect is 0.7\% of the 0.006\% of Bhabha events passing hadronic selection. 
Since this number is negligibly small it can  be ignored safely.

\section{Continuum Fake-Rates}

There is a  small probability of hadronic continuum events passing  muon-pair selection. These events scale with energy in the same way as 
muon-pairs. When counting \B meson events, the ratio of muon-pairs ($R_{\mu}$) is used, so if the percentage of on- and off-peak fake muons from 
hadronic  continuum events are the same, any hadronic continuum fakes will have no effect on \B Counting.

Table \ref{tab:ContinuumFakes} shows the percentage of hadronic continuum MC events which pass the new muon-pair selector and isBCMuMu. Clearly $uds$ events
have a higher fake-rate, but the overall effect is too small (in both the number of fakes and on-/off-peak difference) to have any consequences for \B Counting.
\begin{table}[ht]  
  \caption[Fake-rates for hadronic continuum MC events.]{
    \label{tab:ContinuumFakes}Fake-rates for hadronic continuum MC events passing the mu-pair selector. The sample contained 5.20 million on-peak \ccbar, 
    4.20 million off-peak \ccbar, 6.40 million on-peak $uds$ and 4.20 million off-peak $uds$ MC events.}
  \centering
  \begin{tabular}{|c|c|c|}
    \hline
    &\ \% passing new muon-pair selector &\ \% passing isBCMuMu\\
    \hline
 \ccbar (on-peak) &\  0.00014\% &\ 0.00036\% \\
 \ccbar (off-peak) &\ 0.00002\%  &\ 0.00021\%\\
\hline
 $uds$ (on-peak) &\ 0.009\% &\ 0.024\% \\
 $uds$ (off-peak)  &\ 0.010\%  &\ 0.026\% \\
    \hline
  \end{tabular}

\end{table}

There is also a small chance that muon-pair events can pass the hadronic selector. The probabilities for this are shown in Table \ref{tab:MuHadronicFakes}. The
effect is negligibly small.

\begin{table}[ht] 
   \caption[Fake-rates for muon-pair MC events.]{
    \label{tab:MuHadronicFakes}Fake-rates for muon-pair MC events passing the hadronic selector. The sample contained 4.05 million on-peak 
and 4.08 million off-peak \mumu MC events.}
  \centering
  \begin{tabular}{|c|c|c|}
    \hline
    &\ \% passing new  &\ \% passing isBCMulti-\\
    &\ hadronic selector &\ Hadron \\
    \hline
 \mumu (on-peak) &\  0.0078\% &\ 0.0359\% \\
 \mumu (off-peak) &\ 0.0082\%  &\ 0.0377\%\\

    \hline
  \end{tabular}

\end{table}

\section{Tau Production}

Another process with a cross-section close to 1 nb is $\epem \to \tautau$. The $\tau$ leptons are short-lived  and decay 
within the beam pipe. These events have only a small probability of passing the muon-pair selector, but the probability of passing the hadronic selector
is much larger. 

Similarly to mu-pair events, tau-pair events are simulated with the KK2F MC generator, GEANT 4 and background addition. The production cross-sections 
for the process  $\epem \to \tautau$ based on 10 million
generated on- and off-peak events are $(0.918797 \pm 0.000066)$ nb and $(0.925180 \pm 0.000066)$ nb respectively. 

The hadronic and mu-pair selector efficiencies
are shown in Table \ref{tab:TauFakes}. The hadronic efficiency for $\tau$ pairs is approximately 4\% of the $uds$ and \ccbar hadronic efficiencies.

\begin{table}[ht]  
  \caption[Fake-rates for \tautau MC events.]{
    \label{tab:TauFakes}Fake-rates for \tautau  MC events passing hadronic  and  mu-pair selectors. The sample contained 4.55 million on-peak and
    4.26 million off-peak $\epem \to \tautau$ MC events.}
  \centering
  \begin{tabular}{|c|c|c|}
    \hline
    &\ \% passing new  &\ \% passing isBCMulti-\\
    &\ hadronic selector &\ Hadron \\
    \hline
 \tautau (on-peak) &\  2.67\% &\ 1.37\% \\
 \tautau (off-peak) &\ 2.68\%  &\ 1.37\%\\
\hline
\hline
    &\ \% passing new  &\ \% passing isBCMuMu\\
    &\ muon-pair selector &\  \\
    \hline
 \tautau (on-peak) &\ 0.052\% &\ 0.126\% \\
 \tautau (off-peak)  &\ 0.051\%  &\ 0.117\% \\
    \hline
  \end{tabular}

\end{table}

The rate at which $\tau$ leptons pass the hadronic cuts is a factor of two higher than for isBCMultiHadron, but this largely does not affect \B Counting. 
The increase in efficiency is mainly due to the increased R2 cut (which changed from 0.5 to 0.65). The cross-section for $\tau$ production changes with 
energy identically to muon-pairs, and since the on-peak and off-peak efficiencies are almost
identical, the overall effect on \B Counting is small.

As for other MC types, we can calculate the $\chi^2$ statistic for the time-variation of \tautau MC. Again, we find 
the  $d_0$ value of (\ref{eqn:modchi})  such that
\begin{equation}\label{eqn:modchi2c}
\chi'^{2}(d_0) = M-1.
\end{equation}
This value provides an estimate of the systematic uncertainty in the $\tau$ lepton efficiency.

\begin{figure}[htbp] 
  \begin{center}
    \includegraphics[width=4.3in]{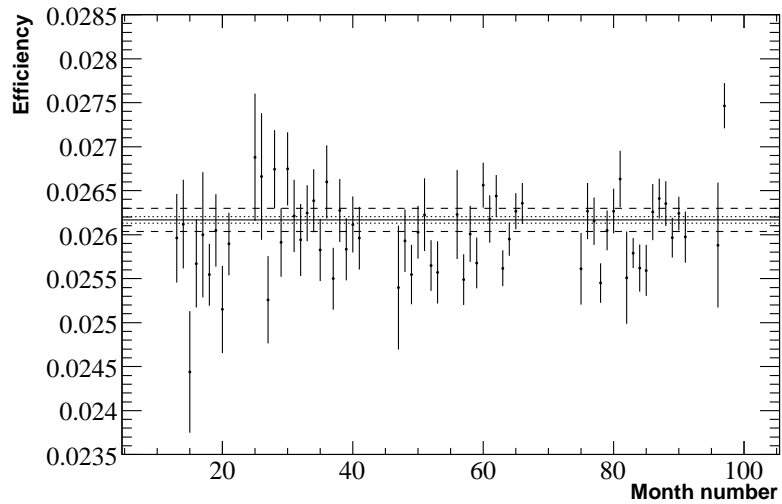}
    \caption[On-peak tau lepton MC efficiency variation for the new cuts.]{On-peak  tau lepton MC efficiency variation for 
      the new proposed hadronic cuts.  The dashed lines show $\pm 0.5\%$ from the mean.}
      \label{fig:MyCutsTauOnVar}
  \end{center}
\end{figure}

\begin{figure}[htbp] 
  \begin{center}
    \includegraphics[width=4.3in]{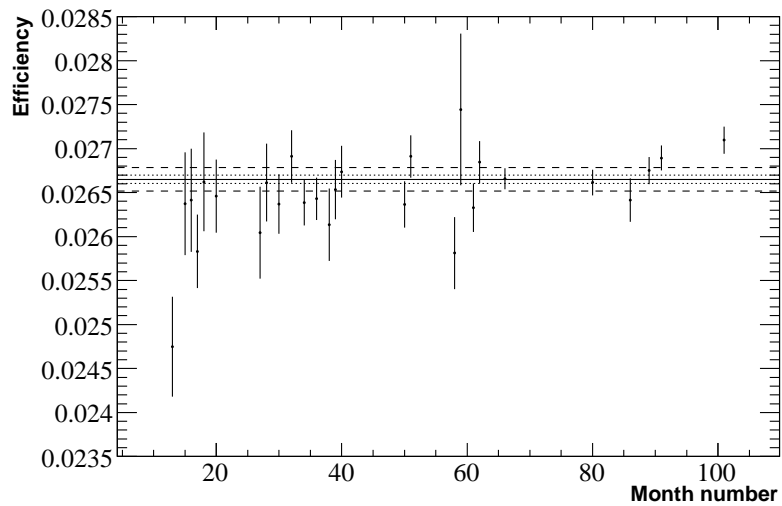}
    \caption[Off-peak tau lepton MC efficiency variation for the new cuts.]{Off-peak tau lepton MC efficiency variation for 
      the new proposed hadronic cuts. The dashed lines show $\pm 0.5\%$ from the mean. }
      \label{fig:MyCutsTauOffVar}
  \end{center}
\end{figure}

The time-variation of on- and off-peak \tautau MC is shown are Fig.~\ref{fig:MyCutsTauOnVar} and  \ref{fig:MyCutsTauOffVar}.
 The overall uncertainties (combining systematic and statistical uncertainties
in quadrature) are $\pm1.5\%$ for on-peak and  $\pm0.99\%$ for off-peak $\tau$ lepton MC. When combined with the efficiency for passing the hadronic cuts of around 
$2.7\%$, the contribution to \B Counting uncertainty  from variations in $\tau$ lepton efficiency is an order of magnitude lower than the $uds$ and \ccbar contribution.


%% file: chapter9.tex
\chapter{Systematic Uncertainty}\label{chap:systuncert}

In this chapter we estimate the total systematic uncertainty on \B Counting. We consider the uncertainties on \BB efficiency, $\kappa_X$ and 
$\kappa_{\mu}$ and estimate the combined uncertainty.

\section{BGFMultiHadron and Tracking Efficiency}
One contributor to  systematic uncertainty in $\varepsilon_B$ is the efficiency of the BGFMultiHadron tag. To examine this, we studied  samples of random
on- and off-peak events which pass the Level Three trigger (the stage at which BGF tags are assigned). This dataset has the 
name \emph{DigiFL3Open} and consists
of 0.5\% of all Level Three events.   The samples contained   17.57 million on-peak and 1.67 million off-peak data events  from Run 4.

BGFMultiHadron requires at least three charged tracks and R2 of  less than 0.98 in the CM frame. For events with a low number of charged tracks,
the quantities BGFMultiHadron and nTracks are closely correlated. Events which fail hadronic selection because nTracks is too low 
(less than three) also usually fail because they do not have the BGFMultiHadron tag. In the on-peak sample, 1.89 million 
events passed all hadronic cuts except those on BGFMultiHadron and nTracks. The combination of
BGFMultiHadron and nTracks cuts rejected a further 0.19 million events; of these, 86\% failed both cuts. This number was
87\% in the off-peak data. 

\begin{figure}[tbp] 
  \begin{center}
    \includegraphics[width=4.1in]{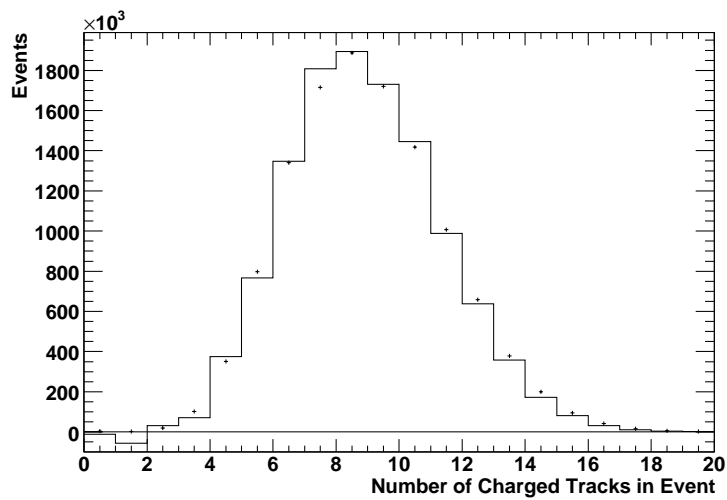}
    \caption[Distribution of nTracks for \BB data and MC.]
    {Distribution of nTracks for \BB data (solid) and MC (points). All hadronic cuts except those on 
BGFMultiHadron and nTracks are applied. The two are normalized to have the same area from 3--20.}\label{fig:nTracks_noBGF}
  \end{center}
\end{figure}

To find the component of systematic uncertainty due to the combination of the  BGFMultiHadron and nTracks cuts in hadronic selection, we  
compare the low-end tail of the nTracks distribution in \BB  data and MC for events passing all hadronic cuts except those two. The distribution
for data (luminosity-scaled off-peak  subtracted from on-peak) overlaid with \BB MC is shown in Fig.~\ref{fig:nTracks_noBGF}. The \BB MC sample
contained 12.43 million events.

After scaling, the off-peak data has more zero- and one-track events than the on-peak data.  This causes the equivalent data bins
in  Fig.~\ref{fig:nTracks_noBGF} to be negative. This is due in a large part to contamination by Bhabha events. The proportion of on-peak
(off-peak) data events in this plot with exactly one charged track, where that track is identified as an electron is 83.7\% (83.8\%). This compares
to 12.1\% for \BB MC. Similarly the proportion of two-track events where both are identified as electrons is significantly higher in data 
(26.2\% and 28.3\% in on-peak and off-peak respectively) than in MC (3.6\%).

\begin{figure}[htb] 
  \begin{center}
    \includegraphics[width=4.1in]{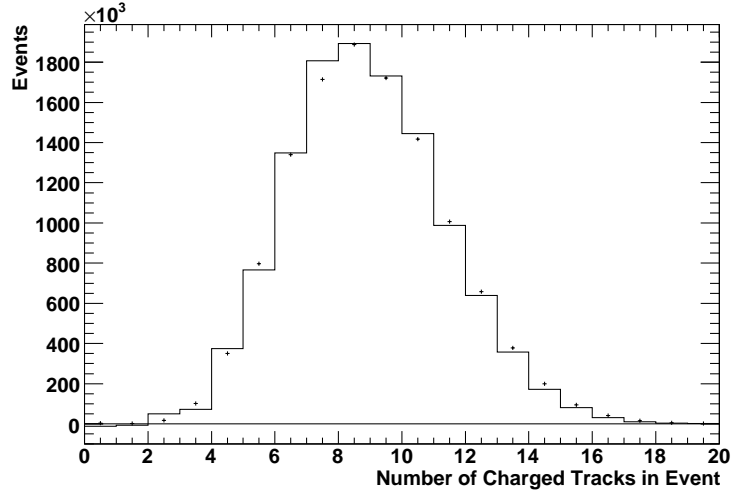}
    \caption[Distribution of nTracks without `obvious' Bhabhas.]
    {Distribution of nTracks for \BB data (solid) and MC (points) with `obvious' Bhabha events removed. All hadronic cuts except those on 
BGFMultiHadron and nTracks are applied. In addition events are not plotted if the number of electrons is non-zero and equal to the number
of charged tracks. The two are normalized to have the same area from 3--20.
}\label{fig:nTracks_noBGF_noBhabhas}
  \end{center}
\end{figure}

Events with one or two charged tracks where the  number of charged tracks is equal to the number of electrons in the event are likely to be Bhabhas.  When tracks 
of this type are removed, the agreement between data and MC improves, especially in the bin of data events with exactly one track. The revised distribution
is shown in Fig.~\ref{fig:nTracks_noBGF_noBhabhas}. The MC distribution is largely unchanged.

To estimate the contribution to systematic uncertainty by these two cuts we can compare
the proportion of events in  Fig.~\ref{fig:nTracks_noBGF_noBhabhas} with nTracks less than three  for  \BB data and MC. For data
we sum the absolute value of the bin contents (since they are negative for the first two bins) and find 0.56\% of all events in the plot
are in the first three bins. For \BB MC, this number is 0.20\%. The difference (0.36\%) is attributed to systematic uncertainty in $\varepsilon_B$ from
the BGFMultiHadron and nTracks cuts.

\section{Comparison of Data and MC }

Next we estimate the systematic uncertainty in $\varepsilon_B$  for the optimal cutset choice due to the difference between \BB MC and data. 
We assumed in Chapter \ref{chapt:BCount}  that the proportions of
charged and neutral \BB decays of the \FourS are the same, i.e. that
\begin{equation}\label{eqn:Rpm}
R^{+\!/0}  =  \frac{\BR(\FourS \to \BpBm)}{\BR(\FourS \to \BzBzb)} =1.
\end{equation}
Since the decays of charged \B mesons are slightly different to those of neutral ones, a source of uncertainty arises if this assumption is false. 

Two of the most precise measurements of $f_{00}\equiv \BR(\FourS \to \BzBzb)$ and $f_{+-}\equiv \BR(\FourS \to \BpBm)$ are described in \cite{Aubert:2005bq}.
That analysis uses a partial reconstruction of the decay $\Bzb \to D^{*+} \ell^{-} \nu$ to obtain:
\begin{equation}
f_{00} = 0.487 \pm 0.010 \textrm{(stat)} \pm 0.008 \textrm{(syst)}
\end{equation}
and infer:
\begin{equation}
f_{+-} = 0.490 \pm 0.023.
\end{equation}
Since  $f_{00}$ and $f_{+-}$ agree within experimental uncertainty, in this thesis we assume  that they are equal and that the effects of any variation from (\ref{eqn:Rpm})
are negligibly small.

To finally estimate  $\Delta\varepsilon_B$, we compare the cut-quantity distributions
of  \BB data  and MC. We shift the MC distributions by small increments (the nature of which depend on the original distributions) 
to determine whether any variation from the original value improves the 
agreement between data and MC. To provide maximum statistics, we examine all Runs simultaneously (approximately 464 million \BB events). 

If a better agreement is found, the \BB efficiency of the shifted MC is calculated and the difference between this and the original value
is attributed to systematic uncertainty. To measure goodness-of-fit between data and MC histograms of $n$ bins, we use again use a $\chi^2$ statistic of the form:
\begin{equation}
\chi^2 =  \sum_{i=1}^{n} \frac{ \left( N^{data}_i - N^{M\!C}_i \right)^2 }{ \left( \sigma^{data}_i \right)^2 +  \left( \sigma^{M\!C}_i \right)^2   }
\end{equation}
where $N^{data}_i$ ($N^{M\!C}_i$) is the number of data (MC) events in bin $i$, and for estimation purposes we make the approximations
\begin{equation}
 \sigma^{data}_i = \frac{1}{\sqrt{N^{data}_i}}    \quad \textrm{and} \quad  \sigma^{M\!C}_i = \frac{1}{\sqrt{N^{M\!C}_i}}.
\end{equation}
 
The increments made to each relevant quantity are listed below. Here,  $\delta$ is an integer, varied between 0 and 
200.
\begin{itemize}

\item   ETotal is shifted by an increment, up to 0.2 GeV:  \begin{itemize}\item ETotal/eCM  $\to$  (ETotal + $0.001\delta$)/eCM.\end{itemize}
\item   R2 is multiplied by a constant between 0.96 and 1.04: \begin{itemize}\item R2  $\to $  $(0.96 + 0.0004\delta)\times$R2. \end{itemize}   
\item   To simulate the effect of tracks being lost due to tracking inefficiencies (when a genuine track is not correctly reconstructed as such), the tracking
efficiency of each track  $(p_{track})$ is varied between 90\% and 100\%. Each track is given
  a random number between 0 and 1, and it is kept if the value is less than  $(0.9 + 0.0005\delta)$.  Any rejected track reduces the value
 of nTracks for the event:\begin{itemize} \item $p_{track}    =  (0.9 + 0.0005\delta)$. \end{itemize}
\item The value of p1Mag is shifted by an increment, positive if p1Mag is below the peak of the distribution and negative if it is above the peak. This has
the effect of narrowing the distribution: \begin{itemize} \item p1Mag $\to (1.0 + 0.0005\delta)\times$p1Mag if p1Mag$\ <0.97$ GeV/$c$,
\item  p1Mag $\to (1.0 - 0.0005\delta)\times$p1Mag if p1Mag$\ >0.97$ GeV/$c$.\end{itemize}
\item   PrimVtxdr is multiplied by a constant between 0.8 and 1.0: \begin{itemize} \item PrimVtxdr $\to$   $(1.0 - 0.001\delta)\times$PrimVtxdr. \end{itemize}  
\item    PrimVtxdz  is multiplied by a constant between 0.6 and 1.0: \begin{itemize} \item PrimVtxdz  $\to$  $(1.0 - 0.002\delta)\times$PrimVtxdz. \end{itemize}

\end{itemize}

In each case, once the value of $\delta$ resulting in the lowest $\chi^2$ is found ($\delta_{opt.}$), the \BB MC efficiency is 
recalculated for the new quantity while all other cuts
remain unchanged ($\hat{\varepsilon}_B^{1}$). For some quantities like PrimVtxdz, even though the initial MC and \BB distributions do not agree perfectly,
the revised efficiency  is almost identical to the original. This is because the cut value is so far from the distribution's peak that only a small number
of events in the tail move from failing to passing the cut.  The results are shown in Table
\ref{table:BBEffResults}. Before any increments are made, the original value of $\hat{\varepsilon}_B$ is 
\begin{equation}
\hat{\varepsilon}_B = 0.940456    \pm   0.000033
\end{equation}
where the uncertainty stated is statistical only.

\begin{center}
\begin{table}[ht] 
   \caption[Revised efficiencies for hadronic cut quantities.]{
    \label{table:BBEffResults}Revised efficiencies for hadronic cut quantities after shifting \BB MC. The statistical uncertainty on each $\varepsilon_B^{1}$ 
value is $\pm  0.000033$ (or  $\pm  0.003\%$)}
  \centering
  \begin{tabular}{|c|c|c|c|}
    \hline
    Quantity &\ $\delta_{opt.}$  &\ $\hat{\varepsilon}_B^{1}$  &\  $\frac{  \hat{\varepsilon}_B^1 - \hat{\varepsilon}_B}{\hat{\varepsilon}_B}$ \\
    \hline
    ETotal &\ 87 &\ 0.944242  &\ $+0.403 \%$   \\
    R2  &\ 110 &\ 0.940378   &\ $-0.008\%$ \\
    $p_{track}$  &\ 191 &\ 0.940308 &\ $-0.016\%$ \\
    p1Mag  &\ 8 &\   0.940408  &\ $-0.005\%$   \\
    PrimVtxdr  &\ 105 &\  0.940644 &\ $+0.020\%$  \\
    PrimVtxdz  &\ 59 &\  0.940408 &\ $-0.005\%$   \\
    \hline 
  \end{tabular}

\end{table}
\end{center}

\begin{figure}[tbp] 
  \begin{center}
    \includegraphics[width=3.5in]{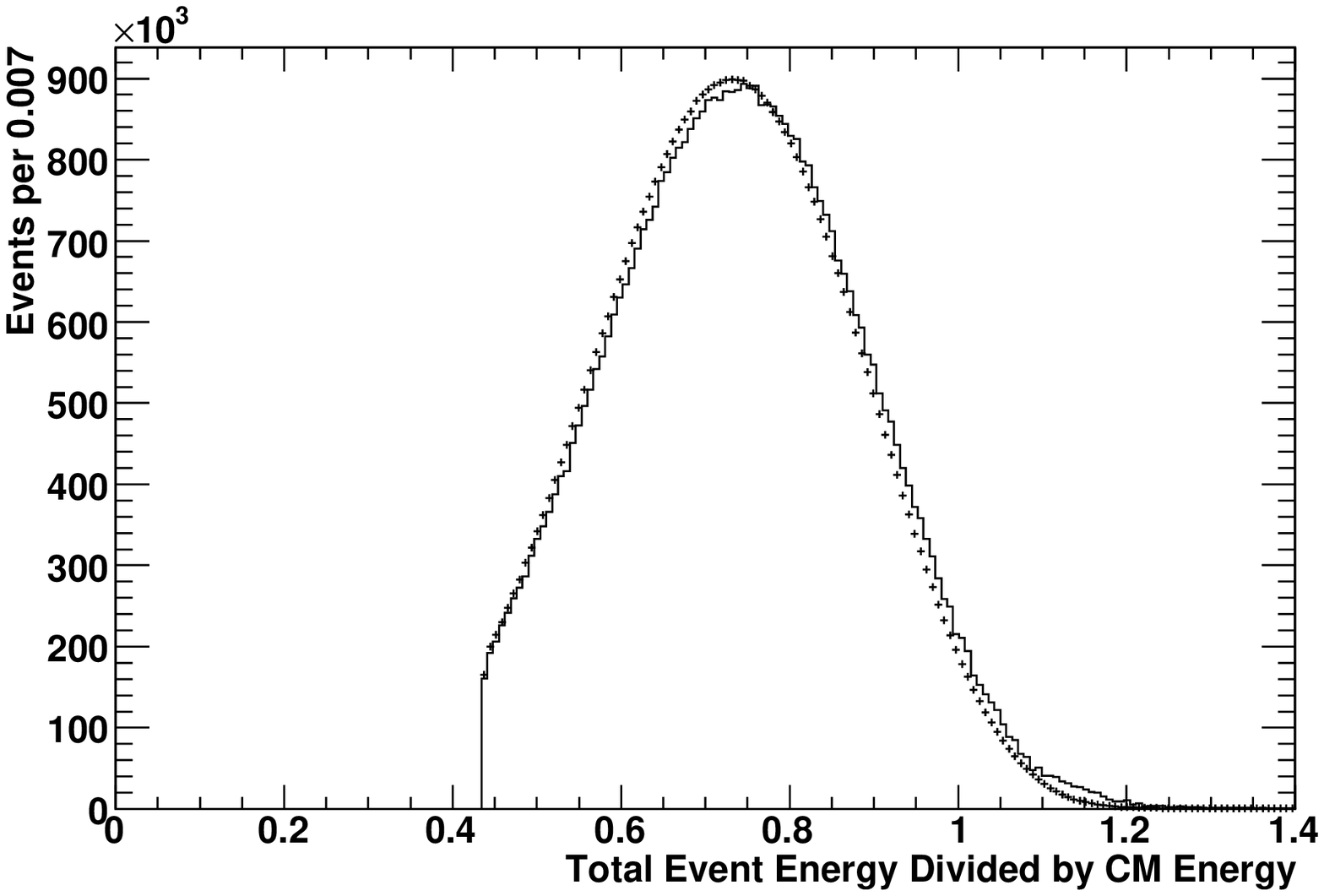}
    \caption[ETotal/eCM for \BB data overlaid with original MC.]{ETotal/eCM for \BB data overlaid with \BB Monte Carlo for the original MC ETotal cut. 
The solid histogram  represents data (off-peak subtracted from on-peak) and the dotted (+) histogram 
is from MC. The histograms are normalised to have the same area from 0.45--1.0.}\label{fig:OrigETotal}
  \end{center}
\end{figure}

\begin{figure}[tbp] 
  \begin{center}
    \includegraphics[width=3.5in]{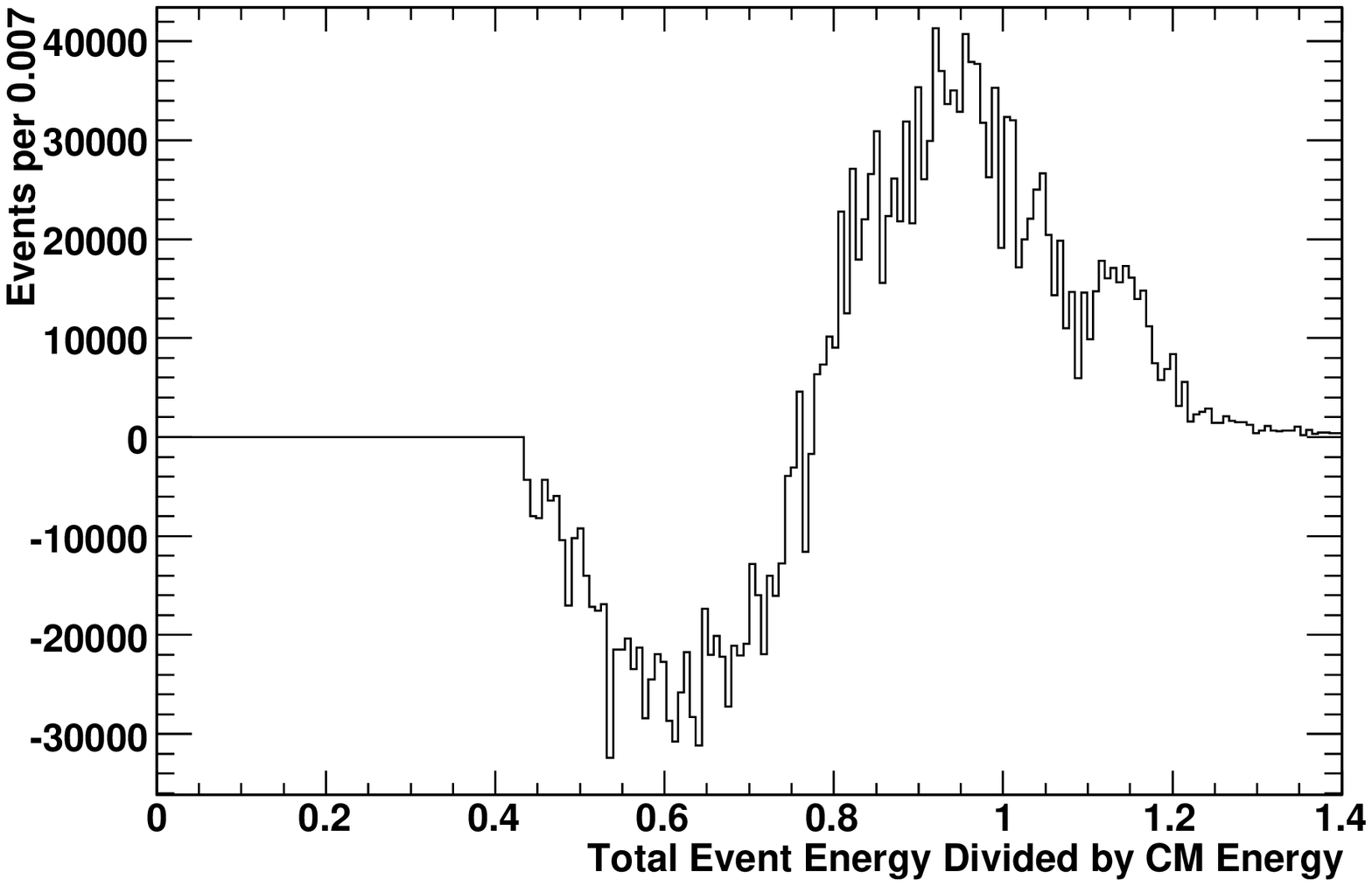}
    \caption[Difference between data and MC of Fig.~\ref{fig:OrigETotal}.]{The difference between the data and MC histograms of Fig.~\ref{fig:OrigETotal}. 
}\label{fig:OrigETotalDiff}
  \end{center}
\end{figure}

\begin{figure}[tbp] 
  \begin{center}
    \includegraphics[width=3.5in]{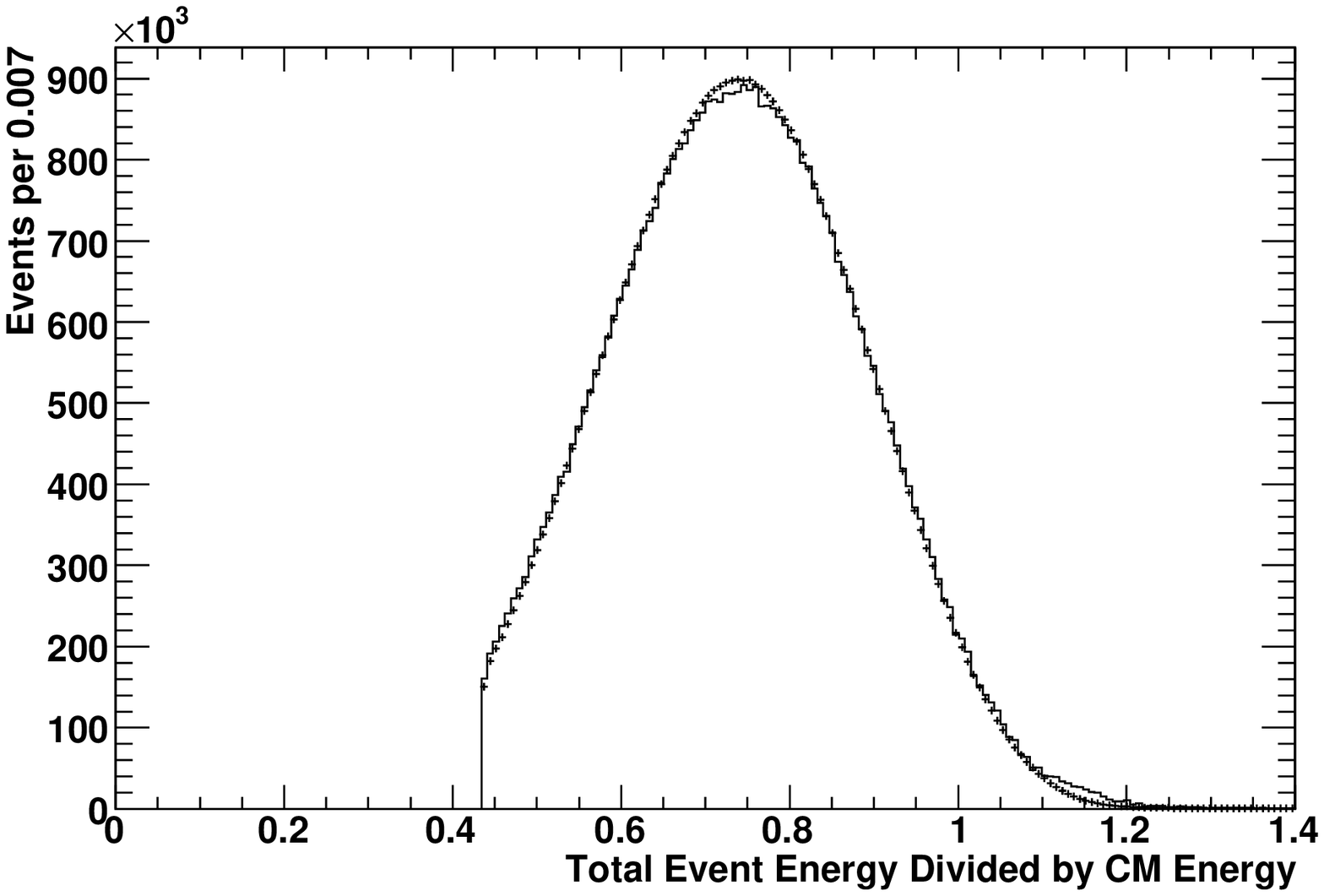}
    \caption[ETotal/eCM for \BB data overlaid with revised MC.]{ETotal/eCM for \BB data overlaid with \BB Monte Carlo for the  revised MC ETotal cut. 
The solid histogram  represents data (off-peak subtracted from on-peak) and the dotted (+) histogram 
is from MC. The histograms are normalised to have the same area from 0.45--1.0.}\label{fig:OptETotal}
  \end{center}
\end{figure}

\begin{figure}[tbp] 
  \begin{center}
    \includegraphics[width=3.5in]{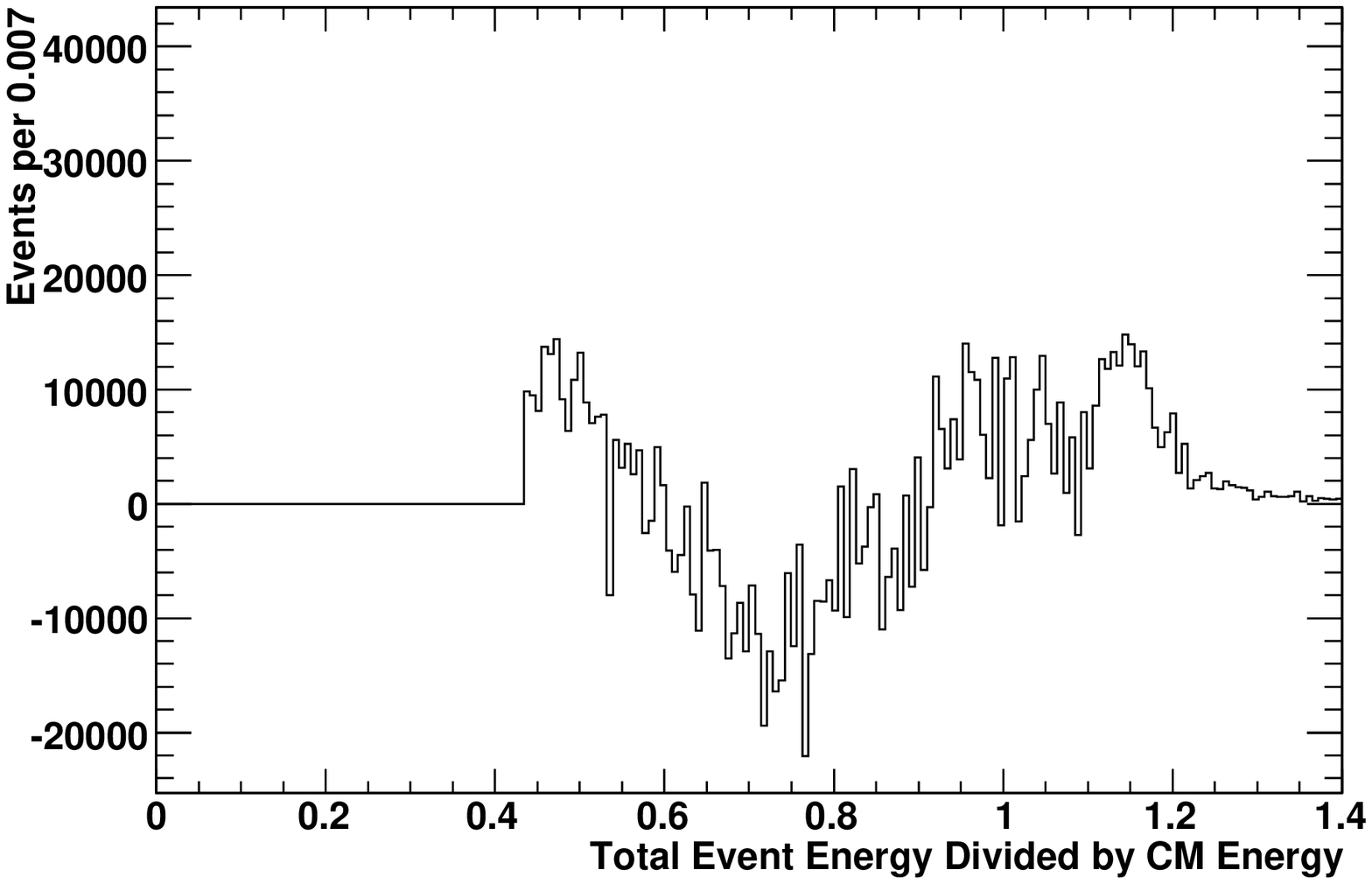}
    \caption[Difference between data and MC of Fig.~\ref{fig:OptETotal}.]{The difference between the data and MC histograms of Fig.~\ref{fig:OptETotal}. 
}\label{fig:OptETotalDiff}
  \end{center}
\end{figure}

The quantity with the largest change in efficiency is ETotal. The value of $\delta_{opt.}$ in this case  implies that adding 87 MeV to each MC event improves
the fit  to data the most.  The efficiency increases by 0.403\%  because  events  up to  87 MeV below the cut  now pass selection. The original and revised
distributions are shown in Fig.~\ref{fig:OrigETotal} and \ref{fig:OptETotal}, and the differences between data and MC for the two plots are shown in 
Fig.~\ref{fig:OrigETotalDiff} and \ref{fig:OptETotalDiff}. 

\begin{figure}[tbp] 
  \begin{center}
    \includegraphics[width=3.5in]{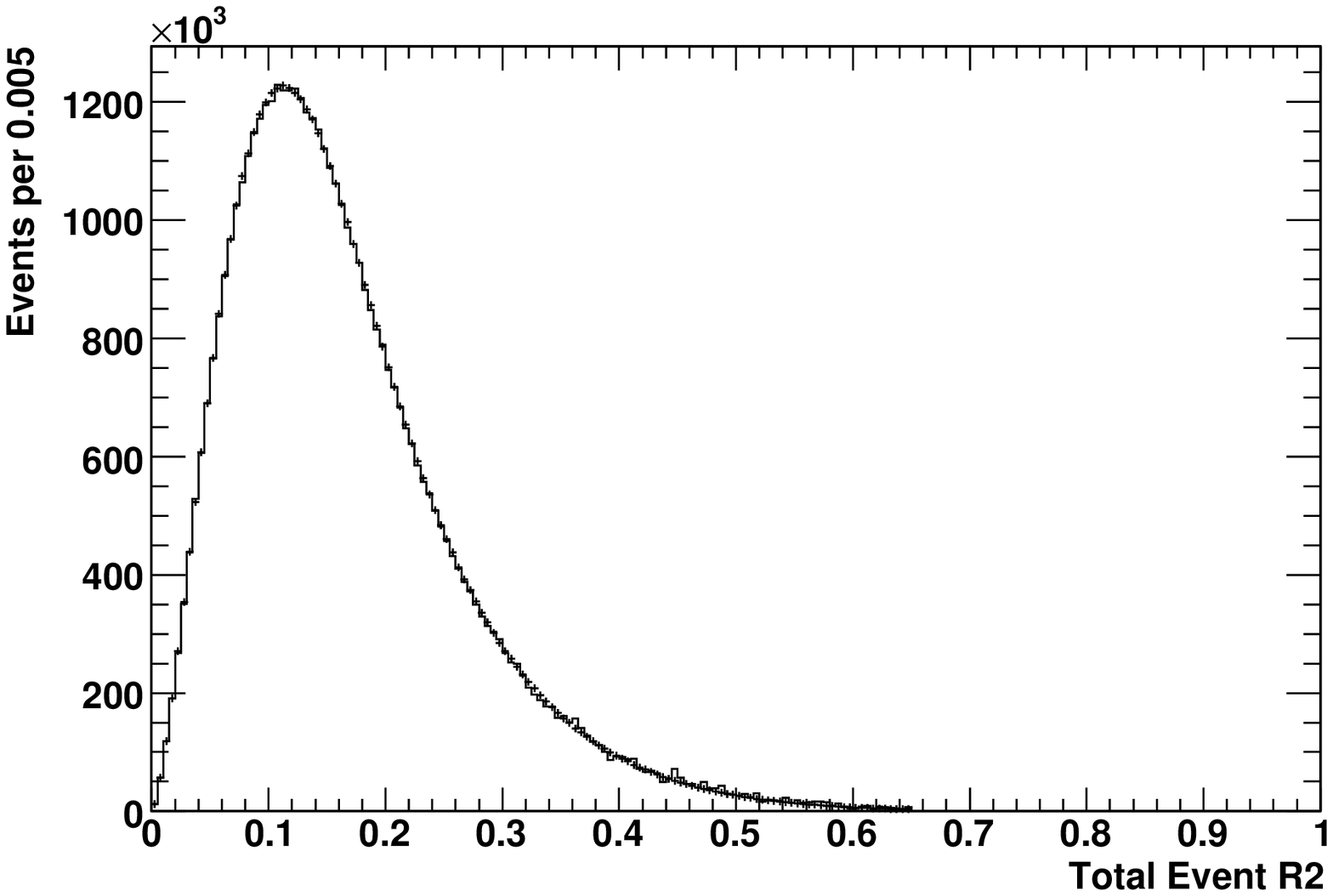}
    \caption[R2 for \BB data overlaid with revised MC.]{R2 for \BB data overlaid with \BB Monte Carlo for the  revised MC R2 cut. 
The solid histogram  represents data (off-peak subtracted from on-peak) and the dotted (+) histogram 
is from MC. The histograms are normalised to have the same area from 0.0--0.5.}\label{fig:OptR2}
  \end{center}
\end{figure}

\begin{figure}[tbp] 
  \begin{center}
    \includegraphics[width=3.5in]{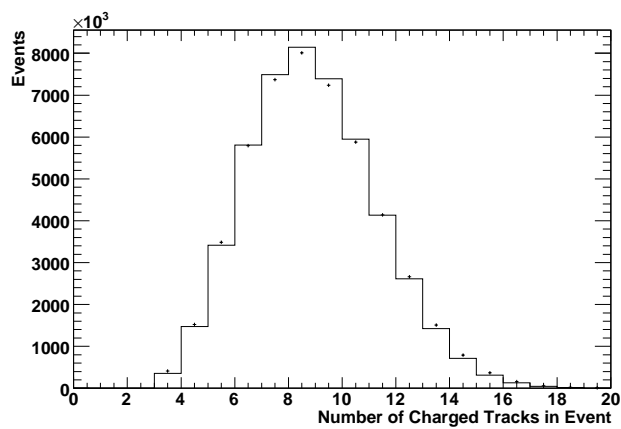}
    \caption[nTracks for \BB data overlaid with revised MC.]{nTracks for \BB data overlaid with \BB Monte Carlo for the  MC with revised
tracking probability. The solid histogram  represents data (off-peak subtracted from on-peak) and the dotted (+) histogram 
is from MC. The histograms are normalised to have the same area.}\label{fig:OptnTracks}
  \end{center}
\end{figure}

\begin{figure}[tbp] 
  \begin{center}
    \includegraphics[width=3.5in]{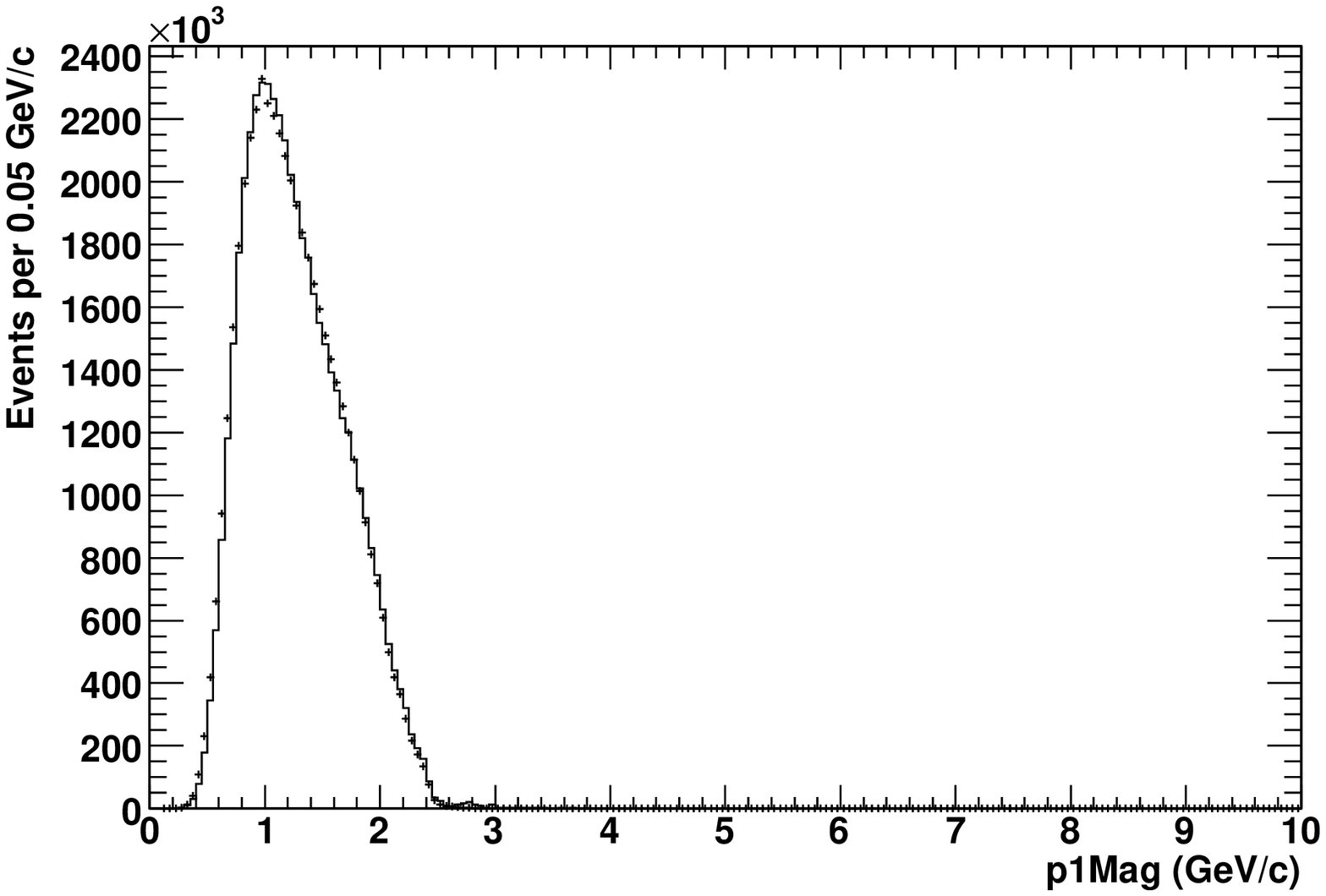}
    \caption[p1Mag for \BB data overlaid with revised MC.]{p1Mag for \BB data overlaid with \BB Monte Carlo for the  revised MC p1Mag cut. 
The solid histogram  represents data (off-peak subtracted from on-peak) and the dotted (+) histogram 
is from MC. The histograms are normalised to have the same area from 0.0--5.0.}\label{fig:Optp1Mag}
  \end{center}
\end{figure}

\begin{figure}[tbp] 
  \begin{center}
    \includegraphics[width=3.5in]{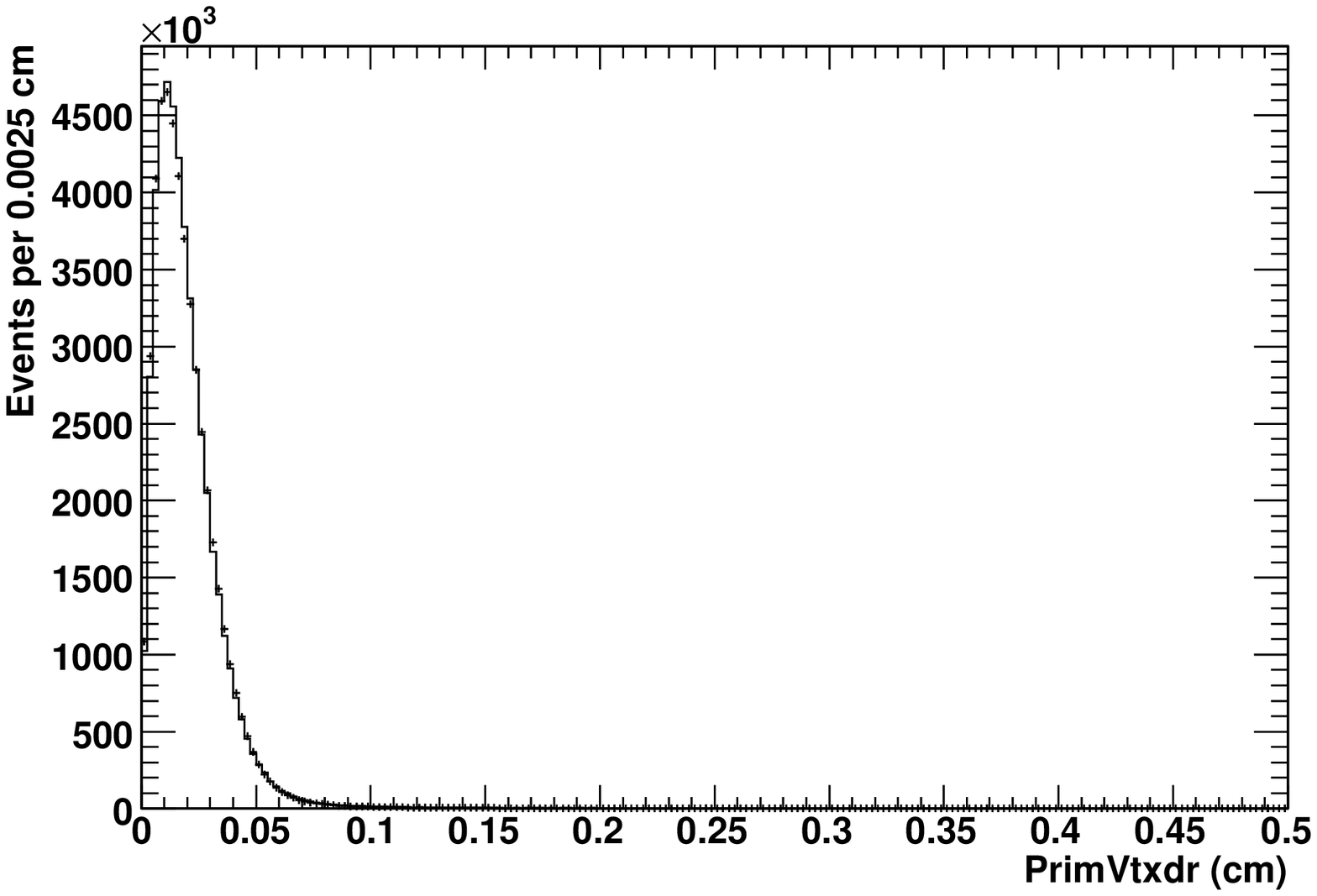}
    \caption[PrimVtxdr for \BB data overlaid with revised MC.]{PrimVtxdr for \BB data overlaid with \BB Monte Carlo for the  revised MC PrimVtxdr cut. 
The solid histogram  represents data (off-peak subtracted from on-peak) and the dotted (+) histogram 
is from MC. The histograms are normalised to have the same area.}\label{fig:OptPrimVtxdr}
  \end{center}
\end{figure}

\begin{figure}[tbp] 
  \begin{center}
    \includegraphics[width=3.5in]{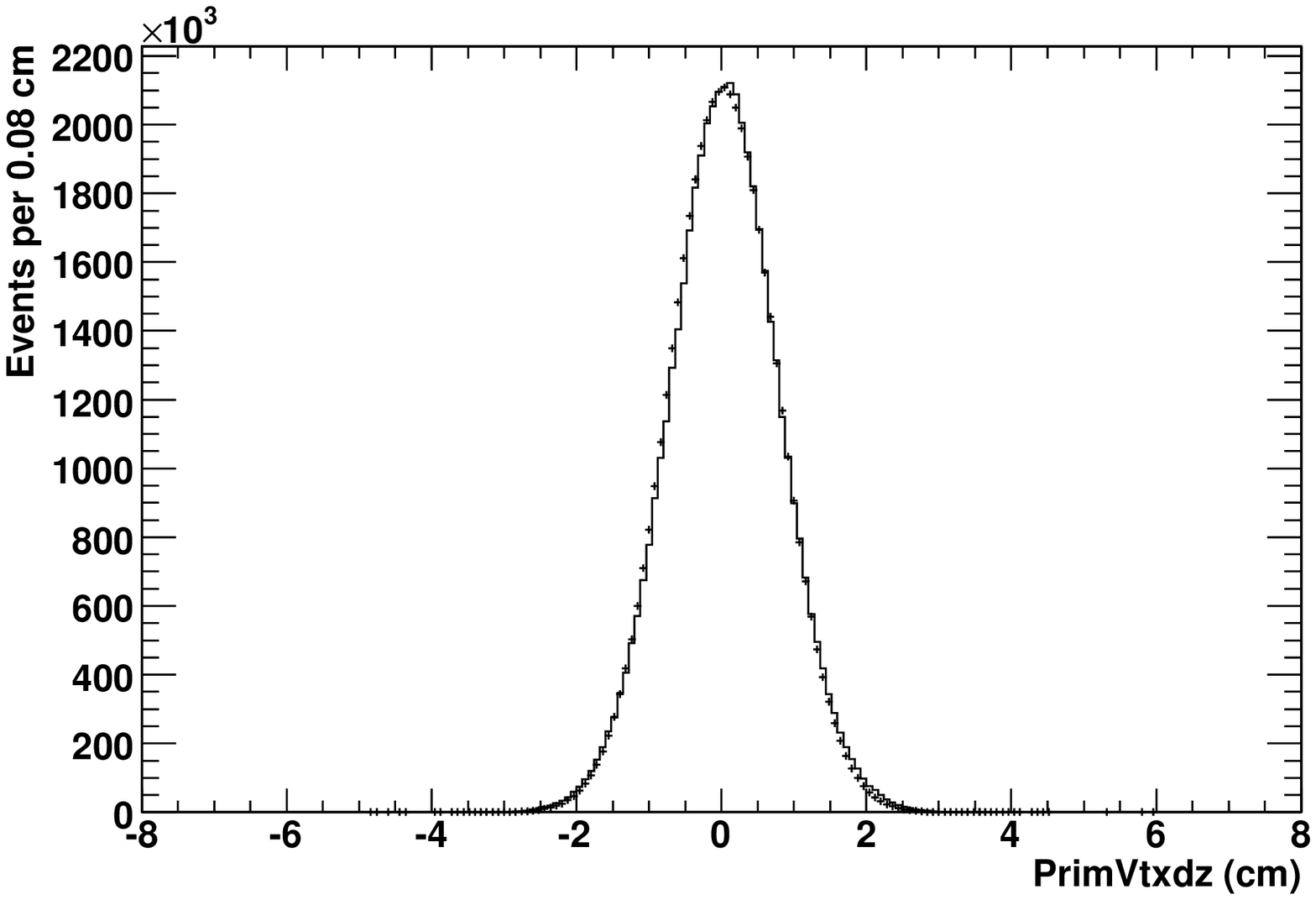}
    \caption[PrimVtxdz for \BB data overlaid with revised MC.]{PrimVtxdz for \BB data overlaid with \BB Monte Carlo for the  revised MC PrimVtxdz cut. 
The solid histogram  represents data (off-peak subtracted from on-peak) and the dotted (+) histogram 
is from MC. The histograms are normalised to have the same area.}\label{fig:OptPrimVtxdz}
  \end{center}
\end{figure}

The change in R2, p1Mag and PrimVtxdz are negligible in comparison to ETotal, and the variation
is comparable to the statistical uncertainty. The plots of data and revised MC for these variables are shown in Fig.~\ref{fig:OptR2} -- \ref{fig:OptPrimVtxdz}.

To estimate $\Delta\varepsilon_B$ we treat each efficiency difference as an independent systematic uncertainty and add them in quadrature. The result is:
\begin{equation}
\Delta\varepsilon_B = 0.404\%.
\end{equation}

\section{Summary of Systematic Uncertainties}
We can now summarize by estimating the total \B Counting systematic uncertainty for the new proposed cuts by combining the independent uncertainties in $\varepsilon_B$ 
and $\kappa$. Note that in Chapter \ref{chap:opt} an optimal cutset was found by minimizing the time-variation, measured by the expression (\ref{eqn:bigBUncert}). 
This cannot be used when estimating the actual systematic uncertainty in the number of \B mesons, since by treating all MC variation as independent, 
it double-counts much of the time-variation. For the true estimate of the systematic uncertainty, we instead only consider the uncertainties of
those quantities explicitly appearing in the \B Counting formula, i.e. $\kappa$ and $\varepsilon_B$. The statistical
and systematic uncertainties on these quantities are summarized in Table \ref{table:UncerSummary}.

\begin{table}[ht]
  \caption[Summary of \B Counting statistical and systematic uncertainties.]{
    \label{table:UncerSummary}Summary of \B Counting statistical and systematic uncertainty components
    of $\kappa_{\mu}$, 
    $\kappa_X$ and $\varepsilon_B$. }
  \centering
  \begin{tabular}{|l|l|c|c|}
    \hline
        &\ Uncert. &\ Value   &\ Contrib. to $\Delta N^0_{B}$ \\
    \hline
     $\Delta\kappa_{\mu}$ &\ Statistical      &\ 0.044\%  &\  0.10\%  \\
                   &\ Time var. of mu-pair eff.  &\ 0.032\%  &\ 0.07\%  \\
	           &\ Var. of $\kappa_{\mu} R_{\mu}$  &\ 0.047\%  &\ 0.11\% \\
	\hline
   &\             \bf{Total}	    &\           0.07\%   &\  0.16\% \\              	
    \hline
    \hline
     $\Delta\kappa_{X}$ &\ Statistical      &\ 0.016\%    &\ 0.04\%  \\
                   &\ Time var. of continuum eff.  &\ 0.096\%   &\  0.21\%  \\
                   &\ ISR/$2\gamma$ contribution  &\ 0.020\%    &\  0.04\%   \\
    \hline 
   &\       \bf{Total}      &\           0.10\%  &\ 0.22\%   \\
     \hline
     \hline
     $\Delta\varepsilon_B$ &\ Statistical      &\ 0.004\%  &\ 0.004\%   \\
                   &\ Low track-multiplicity events  &\ 0.360\%  &\   0.36\%   \\
                   &\ \BB data/MC comparison  &\ 0.404\%    &\  0.40\%  \\
  \hline   
    &\        \bf{Total}     &\           0.54\%   &\  0.54\%   \\
    \hline
    \hline
    $\Delta N^0_B$ &\ \bf{Total} &  \multicolumn{2}{|c|}{0.6\%} \\
\hline
  \end{tabular}

\end{table}

Combining the uncertainties (systematic and statistical) on $\varepsilon_{B}$ in quadrature, we have:  $\varepsilon_{B} = 0.9405 \pm 0.0051$. Similarly, we have the value of
$\kappa =  0.9979 \pm 0.0012$.  The total uncertainty on the number of \B mesons in a sample can now be found by the 
standard method of error propagation. 
From the \B Counting formula  (\ref{eqn:nb})  we can write:

\begin{equation}\label{eqn:ErrorProp}
\left( \Delta N^0_{B}\right)^2 = \left( \frac{\partial N^0_{B}}{\partial \kappa}\right)^2 \left(\Delta \kappa\right)^2 +
\left( \frac{\partial N^0_{B}}{\partial \varepsilon_{B}}\right)^2 \left(\Delta \varepsilon_{B}\right)^2.
\end{equation}

We have
\begin{equation}
\frac{\partial N^0_{B}}{\partial \kappa} = \frac{-1}{\varepsilon_{B}}( N_{\mu}\cdot R_{o\!f\!\!f}),
\end{equation}
and
\begin{equation}
\frac{\partial N^0_{B}}{\partial \varepsilon_{B}} =  \frac{-1}{\varepsilon_{B}}N^0_{B}.
\end{equation}

With this, we are able to make an estimate on the total systematic uncertainty on the number of \BB events, 
$\Delta N^0_{B}$. By considering the two terms of (\ref{eqn:ErrorProp}) separately we can determine how the uncertainties on
$\kappa$ and $\varepsilon_B$ propagate into uncertainties on $N^0_{B}$. 

The uncertainties on $\kappa_{\mu}$ and $\kappa_X$ combined in quadrature lead to a $\Delta\kappa$ value of 
$\pm 0.12\%$. This  corresponds to a $\pm 0.27\%$ uncertainty
in $N^0_{B}$  while the $\Delta\varepsilon_B$ value of $\pm 0.54\%$ corresponds to an 
uncertainty of  $\pm 0.54\%$ in $N^0_{B}$. These, added in quadrature via (\ref{eqn:ErrorProp}) give the total
uncertainty of 0.6\% on the number of B mesons.

The number of \BB events counted 
in the \babar\ dataset is $(464.8 \pm 2.8)\times 10^6$. The equivalent number given by the existing \B Counting code
is $(465.0 \pm 5.1)\times 10^6$. The number counted in each Run is shown in 
Table \ref{table:BCountSummary2}.

    \begin{table}
      \caption[Number of \B meson events in Runs 1--6 with revised uncertainty.]{
	\label{table:BCountSummary2}Number of \B mesons events with revised overall uncertainty 
	in Runs 1--6.}
        \centering
      \begin{tabular}{|l||c|c|}
	\hline
        &\  \bf{Run 1} &\     \bf{Run 2}  \\
       	\hline
	$N^0_B$ (new cutsets)  &\  $(22.4 \pm 0.1) \times 10^6$   &\  $(67.6 \pm 0.4) \times 10^6$   \\
	\hline
	$N^0_B$ (existing code)   &\ $(22.4 \pm 0.3) \times 10^6$ &\ $(67.4 \pm 0.7) \times 10^6$    \\
	\hline
	\hline
      &\  \bf{Run 3} &\     \bf{Run 4}  \\
       	\hline
	$N^0_B$ (new cutsets)  &\  $(35.7 \pm 0.2) \times 10^6$   &\  $(110.7 \pm 0.7) \times 10^6$   \\
	\hline
	$N^0_B$ (existing code)   &\ $(35.6 \pm 0.4) \times 10^6$ &\ $(110.5 \pm 1.2) \times 10^6$    \\
	\hline
	\hline
      &\  \bf{Run 5} &\     \bf{Run 6}  \\
       	\hline
	$N^0_B$ (new cutsets)  &\  $(146.2 \pm 0.9) \times 10^6$   &\  $(82.2 \pm 0.5) \times 10^6$   \\
	\hline
	$N^0_B$ (existing code)   &\ $(147.2 \pm 1.6) \times 10^6$ &\ $(82.0 \pm 0.9) \times 10^6$    \\
	\hline
	\hline
      &\  \bf{Total} &\      \\
       	\hline
	$N^0_B$ (new cutsets)  &\  $(464.8 \pm 2.8) \times 10^6$   &\    \\
	\hline
	$N^0_B$ (existing code)   &\ $(465.0 \pm 5.1) \times 10^6$ &\   \\
	\hline

	\end{tabular}

    \end{table}

%% file: chapter10.tex
\chapter{Summary}\label{chap:summary}

We have presented a new proposed set of cuts designed to  improve
the precision of \B Counting at the \babar\ Experiment. The overall uncertainty
in the number of \B mesons counted has been reduced from 1.1\% to approximately 0.6\%. 

BGFMultiHadron and nTracks are the largest contributors to uncertainty  due to inaccuracies in 
simulating low track-multiplicity events $(\approx  0.36\%)$ and from the differences between
the ETotal distributions of \BB data and MC $(\approx  0.40\%)$ .

Algorithms implementing these new  \B Counting selection criteria are being run in the Release 24 reprocessing 
of the \babar\ dataset. This ensures analysts running on the most recent version of the data will be able 
to count \B mesons with this improved method. To allow cross-checking, the old tagbits isBCMuMu and 
isBCMultiHadron are also recalculated during reprocessing.

%% file: appendix.tex
\chapter{Uncertainties of a Multi-Variable Function}\label{appendix:Uncert}

Let $f(x_1, \cdots, x_m)$ be an infinitely differentiable function of $m$ variables ($x_i$), each with an uncertainty $\Delta x_i$.  
The Taylor  series of $f$ expanded around the point $(a_1, \cdots, a_m)$ is:


\begin{eqnarray}
T(x_1,\cdots,x_m) & = &\sum_{n_1=0}^{\infty} \cdots \sum_{n_m=0}^{\infty} \frac{\partial^{n_1}}{\partial x_1^{n_1}}\cdots \frac{\partial^{n_m}}{\partial x_m^{n_m}}
\frac{f(a_1,\cdots,a_m)}{n_1!\cdots n_m!}\cdot  \nonumber \\ 
& & \quad \quad \quad \quad \quad \quad \quad \cdot (x_1-a_1)^{n_1}\cdots (x_m-a_m)^{n_m}.
\end{eqnarray}

Assume each $x_i$ comes from an independent  statistical distribution with mean  $\hat{x}_i$ and standard deviation $\Delta x_i$. If the standard
deviations are sufficiently small, we can approximate $f$ by evaluating the series to first order around $(\hat{x}_1,\cdots,\hat{x}_m)$. The 
uncertainty of the function can then be estimated by the equation:
\begin{equation}
[\Delta f(x_1, \cdots, x_m)]^2 = \sum_{i=1}^m \left(\frac{\partial f}{\partial x_i} \Delta x_i\right)^2,
\end{equation}
where the differential is evaluated at $(\hat{x}_1, \cdots, \hat{x}_m)$.  The zeroth order terms are constant and do not affect the uncertainty.

The expression for $\kappa_{\mu\mu}$ has the same form as
\begin{equation}
f(x_1,x_2) = \frac{x_1}{x_2},
\end{equation}
so in this case:
\begin{equation}\label{app:KappaMu}
(\Delta f)^2 = \left( \frac{\Delta x_1}{\hat{x}_2}\right)^2 + \left(\frac{-\hat{x}_1 \Delta x_2}{x_2^2}\right)^2.
\end{equation}
Similarly, $\kappa_{cont.}$ has the form
\begin{equation}
g(x_1, x_2, x_3, x_4) = \frac{x_1 + x_2 }{x_3 + x_4}
\end{equation}
and
\begin{eqnarray}\label{app:KappaCont}
(\Delta g)^2 &  = & \left( \frac{\Delta x_1}{\hat{x}_3 + \hat{x}_4}\right)^2 +  \left( \frac{\Delta x_2}{\hat{x}_3 + \hat{x}_4}\right)^2 
 + \left( \frac{ -(\hat{x}_1 + \hat{x}_2)\Delta x_3}{(\hat{x}_3 + \hat{x}_4)^2} \right)^2  +  \nonumber \\
& & + \left( \frac{- (\hat{x}_1 + \hat{x}_2)\Delta x_4}{(\hat{x}_3 + \hat{x}_4)^2} \right)^2.
\end{eqnarray}

\chapter{\ThreeS Counting}
Between 22 December 2007 and 29 February 2008 (Run 7), the \babar\ detector recorded events provided by PEP-II at the \ThreeS resonance. Data was provided at the resonance
peak with a CM energy of 10.35 GeV (on-peak data) and also  at approximately 30 MeV below this (off-peak data). 

The author and Chris
Hearty performed an analysis of \ThreeS Counting on this dataset. 
The methods used were closely based on the work presented in this thesis 
and \cite{Hearty:2000}, and are described in detail in a \babar\ Analysis Document
\cite{McGregor:2008}.

During Run 7, 121.9  million \ThreeS mesons were produced. These were counted to within a systematic uncertainty of 
1.0\%. 
Some  results utilizing this work have been published, including a 
significant observation of the bottomonium ground state $\eta_{b}(1S)$ 
through the decay $\ThreeS \to \gamma\eta_{b}(1S)$ \cite{Aubert:2008}.